\input harvmac
\input rotate
\input epsf
\input xyv2

\def\B{{\cal B}}
\font\teneurm=eurm10 \font\seveneurm=eurm7 \font\fiveeurm=eurm5
\newfam\eurmfam
\textfont\eurmfam=\teneurm \scriptfont\eurmfam=\seveneurm
\scriptscriptfont\eurmfam=\fiveeurm
\def\eurm#1{{\fam\eurmfam\relax#1}}
 \font\teneusm=eusm10 \font\seveneusm=eusm7 \font\fiveeusm=eusm5
\newfam\eusmfam
\textfont\eusmfam=\teneusm \scriptfont\eusmfam=\seveneusm
\scriptscriptfont\eusmfam=\fiveeusm
\def\eusm#1{{\fam\eusmfam\relax#1}}
\font\tencmmib=cmmib10 \skewchar\tencmmib='177
\font\sevencmmib=cmmib7 \skewchar\sevencmmib='177
\font\fivecmmib=cmmib5 \skewchar\fivecmmib='177
\newfam\cmmibfam
\textfont\cmmibfam=\tencmmib \scriptfont\cmmibfam=\sevencmmib
\scriptscriptfont\cmmibfam=\fivecmmib
\def\cmmib#1{{\fam\cmmibfam\relax#1}}
\writedefs

\noblackbox\input rotate
\let\includefigures=\iftrue
\includefigures
\message{If you do not have epsf.tex (to include figures),}
\message{change the option at the top of the tex file.}
\def\figin{\epsfcheck\figin}\def\figins{\epsfcheck\figins}
\def\epsfcheck{\ifx\epsfbox\UnDeFiNeD
\message{(NO epsf.tex, FIGURES WILL BE IGNORED)}
\gdef\figin##1{\vskip2in}\gdef\figins##1{\hskip.5in}
\else\message{(FIGURES WILL BE INCLUDED)}%
\gdef\figin##1{##1}\gdef\figins##1{##1}\fi}
\def\DefWarn#1{}

\def\underarrow#1{\vbox{\ialign{##\crcr$\hfil\displaystyle
{#1}\hfil$\crcr\noalign{\kern1pt\nointerlineskip}\rightarrowfill\crcr}}}


\def\figinsert{\goodbreak\midinsert}
\def\ifig#1#2#3{\DefWarn#1\xdef#1{fig.~\the\figno}
\writedef{#1\leftbracket fig.\noexpand~\the\figno}%
\figinsert\figin{\centerline{#3}}\medskip\centerline{\vbox{\baselineskip12pt
\advance\hsize by -1truein\noindent\footnotefont{\bf
Fig.~\the\figno:} #2}}
\bigskip\endinsert\global\advance\figno by1}
\else
\def\ifig#1#2#3{\xdef#1{fig.~\the\figno}
\writedef{#1\leftbracket fig.\noexpand~\the\figno}%
\global\advance\figno by1} \fi \noblackbox
\input amssym.tex
\def\hat{\widehat}
%
\overfullrule=0pt

\def\tilde{\widetilde}
\def\bar{\overline}

%

\def\EUBB{\cmmib B}

\def\tilde{\widetilde}
\def\bar{\overline}

\def\Tr{{\rm Tr}}

%
\def\tilde{\widetilde}
\def\bar{\overline}
\def\Z{{\Bbb{Z}}}
\def\R{{\Bbb{R}}}
\def\C{{\Bbb{C}}}

\font\zfont = cmss10 
\font\litfont = cmr6 
\def\bigone{\hbox{1\kern -.23em {\rm l}}}
\def\ZZ{\hbox{\zfont Z\kern-.4emZ}}
\def\half{{\litfont {1 \over 2}}}


\def\ZZ{{\Bbb{Z}}}

\def\L{{\cal L}}

\font\zfont = cmss10 
\font\litfont = cmr6 
\def\bigone{\hbox{1\kern -.23em {\rm l}}}
\def\ZZ{\hbox{\zfont Z\kern-.4emZ}}
\def\half{{\litfont {1 \over 2}}}

\font\zfont = cmss10 
\font\litfont = cmr6 
\def\bigone{\hbox{1\kern -.23em {\rm l}}}
\def\ZZ{\hbox{\zfont Z\kern-.4emZ}}
\def\half{{\litfont {1 \over 2}}}

\def\N{{\EUN}}
\def\g{{\frak g}}

\def\Bcc{{\cal B}_{cc}}
%
\def\tilde{\widetilde}
\def\bar{\overline}

\font\zfont = cmss10 
\font\litfont = cmr6 
\def\bigone{\hbox{1\kern -.23em {\rm l}}}
\def\ZZ{\hbox{\zfont Z\kern-.4emZ}}
\def\half{{\litfont {1 \over 2}}}


\let\includefigures=\iftrue
\def\g{{\frak g}}

\def\Tr{{\rm Tr}}

\def\a{\alpha}
\def\b{\beta}

\def\g{\gamma}

\def\k{\kappa}

\def\tilde{\widetilde}
\def\hat{\widehat}
\def\bar{\overline}

\def\CN{{\cal N}}

\def\K{\eurm K}

 \lref\kostant{B. Kostant, ``Line Bundles And The Prequantized
Schrodinger Equation,'' Coll. Group Theoretical Methods in Physics
(Marseilles, 1972) 81.}

\lref\souriau{J.-M. Souriau, {\it Quantification Geometrique},
Comm. Math. Phys. {\bf 1} (1966) 374.}

\lref\KO{
  A.~Kapustin and D.~Orlov,
  ``Remarks on $A$-branes, Mirror Symmetry, and the Fukaya Category,''
  J.\ Geom.\ Phys.\  {\bf 48}, 84 (2003)
  [arXiv:hep-th/0109098].
}

\lref\KW{
  A.~Kapustin and E.~Witten,
  ``Electric-Magnetic Duality and the Geometric Langlands Program,''
 Comm. Number Theory and Physics {\bf 1} (2007) 1-236 [arXiv:hep-th/0604151].
}

\lref\bayen{
  F.~Bayen, M.~Flato, C.~Fronsdal, A.~Lichnerowicz and D.~Sternheimer,
  ``Deformation Theory And Quantization. 1. Deformations Of Symplectic
  Structures,''
  Annals Phys.\  {\bf 111}, 61 (1978).
}

\lref\kontsevich{M. Kontsevich, ``Deformation Quantization Of
Poisson Manifolds,''  Lett. Math. Phys.  {\bf 66}  (2003) 157--216
[arXiv:q-alg/9709040].}

\lref\kontsevichtwo{M. Kontsevich, ``Deformation Quantization Of
Algebraic Varieties,'' Lett. Math. Phys. {\bf 56} 271-294
[arXiv:math/0106006].}

\lref\cf{
  A.~S.~Cattaneo and G.~Felder,
  ``A Path Integral Approach to the Kontsevich Quantization Formula,''
  Commun.\ Math.\ Phys.\  {\bf 212}, 591 (2000)
  [arXiv:math/9902090].
}

\lref\fw{E. Frenkel and E. Witten, ``Geometric Endoscopy And
Mirror Symmetry,'' [arXiv:0710.5939].}

\lref\barwit{
  D.~Bar-Natan and E.~Witten,
  ``Perturbative Expansion of Chern-Simons Theory with Noncompact Gauge
  Group,''
  Commun.\ Math.\ Phys.\  {\bf 141}, 423 (1991).
}

\lref\Brylinski{R.~Brylinski, ``Geometric Quantization of Real
Minimal Nilpotent Orbits,''
 Diff. Geom. Appl. {\bf 9} (1998) 5.}

\lref\Brylinskii{R.~Brylinski, ``Quantization of the 4-Dimensional
Nilpotent Orbit of ${\rm SL}(3,  R)$,'' Canad. J. Math. {\bf 49}
(1997) 916.}

\lref\Brylinskiii{R.~Brylinski, ``Instantons and Kaehler Geometry of
Nilpotent Orbits,'' {\it Representation Theories and Algebraic
Geometry}, ed. A. Broer, NATO Adv. Sci. Inst. Ser. C Math. Phys.
Sci. {\bf 514} Kluwer Acad. Publ., Dordrecht (1998) 85.}

\lref\AZ{
  M.~Aldi and E.~Zaslow,
  ``Coisotropic Branes, Noncommutativity, and the Mirror Correspondence,''
  JHEP {\bf 0506}, 019 (2005)
  [arXiv:hep-th/0501247].
}

\lref\soib{P. Bressler and Y. Soibelman, ``Mirror Symmetry And
Deformation Quantization'' [arXiv:hep-th/0202128].}

\lref\oldkap{A. Kapustin, ``$A$-Branes And Noncommutative Geometry''
[arXiv:hep-th/0502212].}

\lref\gualto{M. Gualtieri, ``Branes On Poisson Varieties''
[arXiv:0710.2719].}

\lref\pestun{V. Pestun, ``Topological Strings In Generalized Complex
Space,'' [arXiv:hep-th/0512189].}

\lref\ohitchin{N. Hitchin, ``Generalized Calabi-Yau Manifolds''
[arXiv:math/0209099].}

\lref\cds{
  A.~Connes, M.~R.~Douglas and A.~S.~Schwarz,
  ``Noncommutative Geometry and Matrix Theory: Compactification on Tori,''
  JHEP {\bf 9802}, 003 (1998)
  [arXiv:hep-th/9711162].
}

\lref\ADW{
  S.~Axelrod, S.~Della Pietra and E.~Witten,
  ``Geometric Quantization Of Chern-Simons Gauge Theory,''
  J.\ Diff.\ Geom.\  {\bf 33}, 787 (1991).
}

\lref\witten{
  E.~Witten,
  ``Quantum Field Theory and the Jones polynomial,''
  Commun.\ Math.\ Phys.\  {\bf 121}, 351 (1989).
}

\lref\CN{
  A.~Abouelsaood, C.~G.~.~Callan, C.~R.~Nappi and S.~A.~Yost,
  Nucl.\ Phys.\  B {\bf 280}, 599 (1987).
}

\lref\SW{
  N.~Seiberg and E.~Witten,
  JHEP {\bf 9909}, 032 (1999)
  [arXiv:hep-th/9908142].
}

\lref\bb{A. Beilinson and J. Bernstein, ``Localisation de $\frak
g$-Modules,'' C. R. Acad. Sci. Paris Ser. I Math. {\bf 292} (1981)
15-18.}

\lref\anderson{J. E. Anderson,  ``Hitchin's Connection, Toeplitz
Operators and Symmetry Invariant Deformation Quantization''
[arXiv:math/0611126].}

\lref\gelf{I. M. Gelfand, M. I. Graev, and I. I.
Pyatetskii-Shapiro, {\it Representation Theory And Automorphic
Functions} (Academic Press, 1990).}

\lref\atiyah{M. F. Atiyah, {\it The Geometry And Physics Of Knots}
(Cambridge University Press, 1990).}

\lref\abott{M. F. Atiyah and R. Bott, ``Yang-Mills Equations Over
Riemann Surfaces,'' Phil. Trans. R. Soc. Lond. {\bf A308} (1983)
523-615.}

\lref\elitz{
  S.~Elitzur, G.~W.~Moore, A.~Schwimmer and N.~Seiberg,
  ``Remarks On The Canonical Quantization Of The Chern-Simons-Witten Theory,''
  Nucl.\ Phys.\  B {\bf 326}, 108 (1989).
}

\lref\RSW{
  T.~R.~Ramadas, I.~M.~Singer and J.~Weitsman,
  ``Some Comments on Chern-Simons Gauge Theory,''
  Commun.\ Math.\ Phys.\  {\bf 126}, 409 (1989).
}

\lref\fw{D. Freed and E. Witten, ``Anomalies In String Theory With
$D$-Branes,'' Asian J. Math. {\bf 3} (1999) 819-52
[arXiv:hep-th/9907189].}

\lref\faltings{G. Faltings, ``Stable $G$-Bundles And Projective
Connections,'' J. Alg. Geom. {\bf 2} (1993) 507-68.}

\lref\TUY{A. Tsuchiya, K. Ueno, and Y. Yamada, ``Conformal Field
Theory On Universal Family Of Stable Curves With Gauge Symmetries,''
in {\it Integrable Systems In Quantum Field Theory And Statistical
Mechanics}, Adv. Stud. Pure Math {\bf 19} 495-566 (Academic Press,
Boston, 1989).}

\lref\hitch{N. Hitchin, ``Flat Connections And Geometric
Quantization,'' Comm. Math. Phys. {\bf 131} (1990) 347-380.}

 \lref\hitchin{N. Hitchin,
``The Self-Duality Equations On A Riemann Surface,'' Proc. London
Math. Soc. (3) {\bf 55} (1987) 59-126.}

\lref\BW{D. Bar-Natan and E. Witten, ``Perturbative Expansion Of
Chern-Simons Gauge Theory With Non-Compact Gauge Group,'' Commun.
Math. Phys. {\bf 141} (1991) 423-440.}

\lref\Ramified{S.~Gukov and E.~Witten, ``Gauge Theory,
Ramification, And The Geometric Langlands Program''
[arXiv:hep-th/0612073].}

\lref\gualttwo{M. Gualtieri, ``Generalized Complex Geometry''
[arXiv:math/0401221].}

\lref\tianyau{G. Tian and S.-T. Yau, ``Complete Kahler Manifolds
With Zero Ricci Curvature, I'' J. Amer. Math. Soc. {\bf 3} (1990) 579-609.}

\lref\nadzas{D. Nadler and E. Zaslow, ``Constructible Sheaves And
The Fukaya Category'' [arXiv:math/0604379].}

\lref\kostanttwo{B. Kostant, ``On Laguerre Polynomials, Bessel
Functions, Hankel Transform
  and a Series in the Unitary Dual of the Simply-Connected Covering Group
  of $SL(2,{\bf R})$,''  Represent. Theory  {\bf 4}  (2000), 181--224. }

\lref\gonch{V. V. Fock and A. Goncharov, ``Dual Teichmuller and
Lamination Spaces'' [arXiv:math/05120312].}

\lref\CRybnikov{A.~Chervov, L.~Rybnikov, ``Deformation quantization of
submanifolds and reductions via Duflo-Kirillov-Kontsevich map,''
EJTP {\bf 4}, No. 15 (2007) 71-90 [arXiv:hep-th/0409005].}

\lref\KontsevichBK{A.~Belov-Kanel, M.~Kontsevich,
``Automorphisms of the Weyl algebra,''
Lett. Math. Phys. {\bf 74} (2005) 181-199 [arXiv:math/0512169].}

 \noindent
 \Title{} {\vbox{ \centerline{
Branes And Quantization} }}
\smallskip
\centerline{Sergei Gukov}
\smallskip
\centerline{\it{Department of Physics, University of California}}
 \centerline{\it{Santa Barbara, CA 93106}}\centerline{and}
 \centerline{\it {Department of Physics, Caltech}}
 \centerline{\it Pasadena, CA 91125}
\bigskip \centerline{and}
\bigskip \centerline{Edward Witten}
\smallskip
\centerline{\it{School of Natural Sciences, Institute for Advanced
Study}} \centerline{\it{Princeton, New Jersey 08540}}
\bigskip\bigskip
\noindent The problem of quantizing a symplectic manifold
$(M,\omega)$ can be formulated in terms of the $A$-model of a
complexification of $M$.  This leads to an interesting new
perspective on quantization.  From this point of view, the Hilbert
space obtained by quantization of $(M,\omega)$ is the space of
$(\Bcc,\B')$ strings, where $\Bcc$ and $\B'$ are two $A$-branes;
$\B'$ is an ordinary Lagrangian $A$-brane, and $\Bcc$ is a
space-filling coisotropic $A$-brane. $\B'$ is supported on $M$,
and the choice of $\omega$ is encoded in the choice of $\Bcc$.  As
an example, we describe from this point of view the
representations of the group $SL(2,\R)$.  Another application is
to Chern-Simons gauge theory. \vskip .5cm
\noindent\Date{September, 2008} \listtoc \writetoc

\newsec{Introduction}\seclab\intro
\subsec{The Problem}

According to textbooks, the passage from classical mechanics to
quantum mechanics is made by replacing Poisson brackets with
commutators.  However, this is an unrealistically simple
description of the situation, even for a basic example such as the
classical phase space $\R^2$, with canonically conjugate variables
$x$ and $p$. One can associate a quantum operator $\CO_f$ to a
classical function $f(x,p)$, but not in a completely unique way,
because of what textbooks call the operator ordering problem.
Regardless of how one defines $\CO_f$, the map from classical
functions $f$ to quantum operators $\CO_f$ does not map Poisson
brackets to commutators. Only if one restricts oneself to
functions that are at most quadratic in $x$ and $p$ does one have
the simple relation \eqn\simprel{[\CO_f,\CO_g]=-i\hbar
\CO_{\{f,g\}}.}

The notion of a function being at most quadratic in $x$ and $p$ is
not invariant under canonical transformations.  Quantizations of
$\R^2$ with different choices of what one means by linear or
quadratic functions are  not the same.  One cannot conjugate one
such quantization to another by a unitary map between the two
Hilbert spaces that transforms the operators $\CO_f$ constructed
in one quantization to their counterparts $\tilde\CO_f$ in another
quantization. The order $\hbar^2$ corrections to \simprel\ are
simply different in the two quantizations.

The fact that quantization is ambiguous locally also means that it
is not clear how to carry out quantization globally.  Suppose that
$M$ is a $2n$-dimensional classical phase space that we wish to
quantize. (And suppose that we are given on $M$ an additional
structure known as a prequantum line bundle
\refs{\kostant,\souriau}; this will enter our story shortly.) Even
if one can locally identify $M$ with $\R^{2n}$, this does not
automatically tell us how to quantize $M$, even locally, since the
quantization of $\R^{2n}$ is not unique, as we have just explained.
If we make random local choices in quantizing $M$, we cannot expect
them to fit together to a sensible global quantization.  There is
also no good framework for trying to fit the pieces together,
because there is no general notion of restricting a quantization of
$M$ to a quantization of an open subset of $M$, which would be a
prerequisite for trying to quantize $M$ by gluing together
quantizations of open subsets.

One cannot expect to be able to quantize $M$ without some
additional structure beyond its classical symplectic structure
(and prequantum line bundle). There is no known general recipe for
what this additional structure should be.  As a result, there is
no general theory of quantization of classical phase spaces.

In practice, quantization is a somewhat informal notion, which
refers to a collection of loosely related procedures. The most
important example in which we know what quantization should mean is
$\R^{2n}$ with a given choice of affine structure, that is, a choice
of what one means by linear functions. This can be quantized in a
way that requires no further choices.  (In the usual procedure, one
splits the linear functions into coordinates and momenta, which are
then taken to act by multiplication and differentiation,
respectively. The resulting Hilbert space admits a natural action of
the symplectic group $Sp(2n,\R)$ or rather its double cover, and
thus does not really depend on the splitting between coordinates and
momenta.)  Another important example is a cotangent bundle $M=T^*U$
(with the standard symplectic structure), which can be quantized in
a natural way in terms of half-densities on $U$; similarly, there is
a natural procedure for quantization of Kahler manifolds by taking
holomorphic sections of the appropriate line bundle. Finally, if one
knows how to quantize $M$, and $G$ is a group that acts on $M$, then
(under some mild restrictions) one can define a quantization of the
symplectic quotient $M/\!/G$ by taking the $G$-invariant part of the
quantization of $M$.  There are various ways to combine the
procedures just mentioned.

There is no guarantee that the different procedures are
equivalent.  If $M$ is a cotangent bundle or a Kahler manifold in
more than one way or a symplectic quotient of one of these in more
than one way, or can be realized by more than one of these
constructions, there is no assurance that the different procedures
lead to equivalent quantizations.

\subsec{Quantization Via Branes}

In this paper, we offer a new perspective  on quantization, based
on two-dimensional sigma-models.  The goal is to get closer to a
systematic theory of quantization.  However, it is not clear to
what extent our perspective helps in computing new formulas.

\def\H{{\cal H}}
Our procedure is as follows.  We start with a symplectic manifold
$M$, with symplectic form $\omega$, that we wish to quantize. As
in geometric quantization \refs{\kostant,\souriau}, we assume that
$M$ is endowed with a prequantum line bundle $\L$; this is a
complex line bundle $\L\to M$ with a unitary connection of
curvature $\omega$.

For our purposes, saying that $Y$ is a complexification of $M$
simply means that (1) $Y$ is a complex manifold with an
antiholomorphic involution\foot{An involution is simply a symmetry
whose square is the identity.} $\tau:Y\to Y$, such that $M$ is a
component of the fixed point set of $\tau$; (2) the symplectic
form $\omega$ of $M$ is the restriction to $M$ of a nondegenerate
holomorphic two-form $\Omega$ on $Y$, such that
$\tau^*\Omega=\bar\Omega$; (3) the unitary line bundle $\L\to M$
can be extended to a unitary line bundle $\L\to Y$ with a
connection of curvature ${\rm Re}\,\Omega$, and moreover the
action of $\tau$ on $Y$  lifts to an action on $\L$, restricting
to the identity on $M$. These data are regarded as part of the
definition of $Y$.

The case of most interest in the present paper is the case that
$Y$ is an affine variety, which roughly means that it admits
plenty of holomorphic functions. More precisely, an affine variety
is defined by a finite set of polynomial equations for a finite
set of complex variables $x_1,\dots,x_s$, as opposed to a more
general algebraic variety which is obtained by gluing together
pieces which are each affine varieties.
 Our approach to quantization will be  based
on the $A$-model associated with the real symplectic form
$\omega_Y={\rm Im}\,\Omega$.  So we need a further condition on
$Y$ which ensures that this theory has a good $A$-model. A good
$A$-model is one in which the relevant correlation functions and
other observables are complex-valued, rather than being functions
of a formal deformation parameter. (For example, the most familiar
$A$-model observables are obtained from sums over worldsheet
instantons of different degrees.  Having a good $A$-model means
that such sums are not just formal power series but converge to
complex-valued functions.) $Y$ will have a good $A$-model if the
supersymmetric sigma-model with target $Y$, which can be twisted
to give the $A$-model, is well-behaved quantum mechanically; this
in turn should be true if $Y$ admits a complete hyper-Kahler
metric, compatible with its complex symplectic structure. For
instance, the example considered below and in more detail in
section 3 corresponds to the Eguchi-Hansen manifold, which is a
complete hyper-Kahler manifold. Having a good $A$-model should
imply that deformation quantization of $Y$ (which is part of the
$A$-model, as we discuss below) gives an actual deformation of the
ring of holomorphic functions on $Y$, with a complex deformation
parameter; a bad $A$-model merely leads to a formal deformation
over a ring of formal power series.

We require an actual deformation of the ring of functions, not
just a formal one, for our approach to quantization to make sense.
Interestingly, the conditions \kontsevichtwo\ under which
deformation quantization of an affine variety gives an actual
deformation of the ring of functions on $Y$ are very similar to
the conditions for $Y$ to admit a complete hyper-Kahler metric
along the lines of the complete Calabi-Yau metrics constructed in
\tianyau.

The most familiar $A$-branes are Lagrangian $A$-branes, supported
on a Lagrangian submanifold of $Y$; such a submanifold is of
middle dimension.  In general \KO, however, the $A$-model can also
admit $A$-branes whose support has a dimension greater than
one-half the dimension of $Y$.   The support of such a brane is a
coisotropic submanifold of $Y$ with certain somewhat special
properties.  In particular, the choice of the line bundle $\L\to
M$ with curvature ${\rm Re}\,\Omega$ determines, in the language
of \KW, a canonical coisotropic brane in the $A$-model of $Y$. Its
support is all of $Y$ and it will be one of the main ingredients
in the present paper.

Suggestions that the $A$-model is related to deformation
quantization (whose relation to quantization is discussed in
section 1.4) go back to \soib\  and \oldkap\ and have been
extended and made more precise in \refs{\pestun,\gualto}, partly
in the framework of generalized complex geometry
\refs{\ohitchin,\gualttwo}. The canonical coisotropic $A$-brane
was used in \KW\ to elucidate some of these matters.  For a
related approach to the $A$-model, see \nadzas.

We will also make use of ordinary Lagrangian $A$-branes.
 $M$ itself is a Lagrangian submanifold, so (if $M$ obeys a mild
topological condition) we can define a rank 1 $A$-brane supported
on $M$.  Let us pick such a brane (there are inequivalent choices
if $M$ is not simply-connected) and call it $\B'$.
 In this paper, we write $\Bcc$ for the
canonical coisotropic $A$-brane, $\B'$ for a Lagrangian $A$-brane,
and $\B$ for an $A$-brane of unspecified type.

Quantization of $M$ is now achieved by declaring that the Hilbert
space associated to $M$ is the space $\cal H$ of $(\Bcc,\B')$
strings.  This definition certainly gives a vector space associated
to the choice of $A$-brane $\B'$.  That the explicit construction of
this vector space is similar to quantization was originally shown by
Aldi and Zaslow in examples \AZ, and will be further discussed in
section 2.

To justify calling this process quantization, we need more
structure. For one thing, we want to associate to $\B'$ not just a
vector space but a Hilbert space. It is unusual to get a Hilbert
space structure in the topological $A$-model, but in the present
context, as explained in section 2.4,  $\H$ can be given a hermitian
metric by making use of the antiholomorphic involution $\tau$.  The
space $\cal H$, with its hermitian metric, depends only on the
choices of $Y$, $\L$, and $\B'$, and not on any additional data
(such as a metric on $Y$) that is used in defining the $A$-model. We
do not have a general proof that the hermitian metric on $\cal H$ is
positive definite, though this is true near the classical limit.  (A
generalization of the construction, involving an antiholomorphic
involution that maps $M$ to itself but does not leave $M$ fixed
pointwise, leads to a hermitian metric on $\cal H$ that is not
positive-definite near the classical limit.)

If our procedure is reasonably to be called quantization, we also
want to have a natural way to quantize a large class of functions
on $M$, that is to realize them as operators on $\H$. The
functions on $M$ that can be naturally quantized, in our approach,
are the functions that are restrictions of holomorphic functions
on $Y$ that have a suitable behavior at infinity. The details of
what is suitable behavior at infinity  are tied to the question of
what spaces have good $A$-models. However, for $Y$ an affine
variety (defined by a finite set of polynomial equations for a
finite set of complex variables $x_1,\dots,x_s$), a reasonable
condition is to allow only functions of polynomial growth (that
is, polynomials in the $x_i$). This gives a very large class of
holomorphic functions on such a variety, and it is for this class
of functions that (under certain restrictions on $Y$) deformation
quantization gives an actual rather than formal deformation.  So
this is the right class of functions to consider. For reasonable
$M\subset Y$, the restrictions of these functions are dense in the
space of smooth functions on $M$, and our procedure leads to
quantization in the sense of constructing a Hilbert space $\H$
with a map from a large class of functions on $M$ to operators on
$\H$. At the opposite extreme, if $Y$ is compact, the definition
of the Hilbert space $\H$ still makes sense, but we get no
operators acting on this Hilbert space; one might not want to call
this quantization.

To make this discussion a little more concrete, we will consider
an example (which will be explored more fully in section 3).  Let
$M=S^2$ be a two-sphere, and let $\omega$ be a symplectic form on
$M$ with $\int_M\omega=2\pi n$, $n\in\Z$. We expect quantization
to give a Hilbert space $\H$ of dimension $n$. The
infinite-dimensional group ${\rm Adiff}\,S^2$ of area-preserving
diffeomorphisms of $S^2$ acts on the classical phase space
$(M,\omega)$.  The group that acts on $\H$ is $U(n)$. There is no
natural way to map ${\rm Adiff}\,S^2$ to $U(n)$, so any approach
to quantization will involve some arbitrary choices.

One standard approach in this problem (which in geometric
quantization \refs{\kostant,\souriau} is known as picking a
complex polarization) is to pick a complex structure $J$ on $S^2$,
such that $\omega$ is of type $(1,1)$. The subgroup of ${\rm
Adiff}\,S^2$ that preserves $J$ is at most $SO(3)$, and it is
convenient to pick $J$ so that this subgroup is actually $SO(3)$.
If so, with some choice of coordinates, the Kahler metric on $S^2$
is a multiple of the round metric on the two-sphere
$x^2+y^2+z^2=1$. Quantization is now straightforward: one defines
$\H$ to be $H^0(S^2,\L)$, the space of holomorphic sections of
$\L$ in complex structure $J$.  Since the procedure of
quantization was $SO(3)$-invariant, the group $SU(2)$ (the
universal cover of $SO(3)$) acts naturally on $\H$, as does its
Lie algebra. The functions $x,y$, and $z$ generate via Poisson
brackets the action of this Lie algebra, so in this approach to
quantization, these functions naturally map to quantum operators.
One can then in a fairly natural way take polynomial functions in
$x,y$, and $z$ to act on $\H$ by mapping a monomial $x^ay^bz^c$ to
the corresponding symmetrized polynomial in the $\frak{so}(3)$
generators.

Clearly, we could have embedded $SO(3)$ in ${\rm Adiff}\,S^2$ in
many (conjugate) ways, so this approach to quantization depends on
an arbitrary choice.  Now let us discuss how one would quantize the
same example in our approach.  We are supposed to pick a suitable
complexification $Y$ of $S^2$.  We do this by again picking
coordinates in which $S^2$ is defined by the equation
\eqn\gzaz{x^2+y^2+z^2=1,} and we introduce $Y$ by simply regarding
$x,y,$ and $z$ as complex variables.  Thus $Y$ is an affine variety;
it admits a complete hyper-Kahler metric (the Eguchi-Hansen metric),
so we expect it to have a good $A$-model (the relevant deformation
of the ring of functions on $Y$ is explicitly described in section
3.1, and involves a complex parameter, not a formal variable). The
allowed holomorphic functions are polynomials in $x,y$, and $z$
subject to the relation \gzaz. On $Y$, there is a holomorphic
two-form $\Omega$ which restricts on $M$ to the properly normalized
symplectic form $\omega$; it is simply $\Omega=n\,dx\wedge dy/2z$.
The holomorphic functions on $Y$ with polynomial growth at infinity
are simply the polynomials in $x,y$, and $z$, so the functions that
we can quantize are those polynomials, just as in the previous and
more standard approach. In the standard approach, the special role
of $x$, $y$, and $z$ is that they generate via Poisson brackets
symmetries of the complex structure that is used in quantization. In
our approach, what is special about $x$, $y$, and $z$ is that they
generate the ring of holomorphic functions on $Y$ with polynomial
growth at infinity.

\def\tx{\tilde x}
\def\ty{\tilde y}
\def\tz{\tilde z}
It is illuminating to consider an alternative complexification of
$S^2$ that does not work well.  We can define the two-sphere by
the equation\foot{To show that this equation defines a two-sphere,
observe that if $\tx,\ty,$ and $\tz$ are real numbers obeying
$\tx^4+\ty^4+\tz^4=1$, there is a unique positive $t$ such that
$(x,y,z)=t(\tx,\ty,\tz)$ obey $x^2+y^2+z^2=1$.  This map gives an
isomorphism between the space of solutions of $x^2+y^2+z^2=1$ and
the space of solutions of $\tx^4+\ty^4+\tz^4=1$.}
 $\tx^4+\ty^4+\tz^4=1$, for
real variables $\tx,\ty,\tz$, so  one can define a
complexification $\tilde Y$ of $S^2$ by letting $\tx,\ty$, and
$\tz$ be complex variables obeying the same equation. But
deformation quantization of $\tilde Y$ is only a formal procedure
according to \kontsevichtwo, and the construction of \tianyau\
does not endow $\tilde Y$ with a complete Calabi-Yau metric.
Rather, $\tilde Y$ admits an incomplete Calabi-Yau metric, which
can be compactified to give a K3 surface. We expect that to give
$\tilde Y$ a good $A$-model, one must compactify it and consider
the $A$-model of the K3-surface; then there are no holomorphic
functions and no natural interpretation in terms of quantization.

\subsec{Comparison To Geometric Quantization}\subseclab\defqu

Here and in section 1.4, we will compare our approach to some
standard approaches to quantization.

In geometric quantization \refs{\kostant,\souriau}, the first
step, given $(M,\omega)$, is to pick a prequantum line bundle
$\L\to M$, that is, a unitary line bundle with a connection of
curvature $\omega$.  This is also an initial step in our approach,
as explained above, and probably (explicitly or implicitly) in any
approach to quantization. The second step is then to pick a
polarization (typical examples being a realization of $M$ as a
cotangent bundle $ T^*U$ for some $U$, or a choice of Kahler
structure on $M$), after which quantization is carried out via
half-densities on $U$ or holomorphic sections of $\L\otimes
K^{1/2}\to M$ ($K^{1/2}$ is a square root of the canonical bundle
of $M$).

This  second step has some drawbacks.  A global polarization may
not exist, even for phase spaces that should be quantizable.
Moreover, if a polarization exists, there are many possible
polarizations. It is not clear when quantization carried out with
two different polarizations gives equivalent results.

Our approach has analogous drawbacks.  Given $(M,\omega)$, it is
not clear whether a suitable $Y$ exists, or whether different
choices of $Y$ will give equivalent results. (We do not know of
any examples in which this is the case.) Our approach is therefore
particularly useful if there is a natural $Y$ (or at least a
natural class of $Y$'s with some special relationship), while
geometric quantization is particularly useful if there is a
natural polarization.

The problem of when geometric quantization with two different
polarizations gives equivalent results is vexing.  The most
important example is quantization of $ \R^{2n}$.  Once one picks an
affine structure on $\R^{2n}$ (a notion of what one means by linear
functions), a polarization can be picked by choosing a maximal
Poisson-commuting  set of linear functions $q^1,\dots,q^n$, which we
declare to be the coordinates (as opposed to the momenta).
Quantization is then carried out by introducing a Hilbert space of
functions (actually half-densities) $\Psi(q^1,\dots,q^n)$.  We may
call a choice  of this kind a linear polarization. It is a classic
result that quantizations with different linear polarizations
(compatible with the same affine structure) are equivalent.  The
usual proof uses the action of the symplectic group $Sp(2n,\R)$ (or
rather its double cover), generated by quadratic functions of the
coordinates and momenta.

This fundamental example has others as corollaries.  For $G$  a
subgroup of the symplectic group, consider the quantization of the
symplectic quotient $M=\R^{2n}/\negthinspace/G$. Any $G$-invariant
polarization of $\R^{2n}$ descends to a polarization of $M$, and
geometric quantization of $M$ with two polarizations that descend
from $G$-invariant linear polarizations of $\R^{2n}$ will be
equivalent. This statement, which follows from the equivalence of
linear polarizations of $\R^{2n}$, also has an analog for $n=\infty$
in the case of Chern-Simons gauge theory \ADW.

An example of a well-motivated procedure of quantization that is
awkward to describe in geometric quantization is the case that
$M=\R^{2n}/\negthinspace/G$ is the symplectic quotient of
$\R^{2n}$ (or some other space that can be quantized by geometric
quantization) by a subgroup $G\subset Sp(2n,\R)$ such that there
is no $G$-invariant polarization of $\R^{2n}$.  It is natural to
define quantization of $M$ by taking the $G$-invariant part of the
quantization of $\R^{2n}$, but this definition is not related in
any obvious way to what one can get from a polarization of $M$.

In   our approach to quantization via branes, near the
semiclassical limit, one may define the $A$-model of $Y$ by
picking a suitable metric on $Y$.  This is the analog of a
polarization in our approach. One illuminating and important case
is that the metric on $Y$ is a complete hyper-Kahler metric and
$M$ is a complex submanifold in one of the complex structures,
which we will call $J$. We will say that a metric of this kind
gives a hyper-Kahler polarization of the pair $(Y,M)$. In this
case, $J$ defines a complex polarization of $M$ in the sense of
geometric quantization, and the vector space $\H$ that is defined
in our procedure (but not in general its hermitian inner product)
agrees with what one would naturally define in geometric
quantization, as we explain in section 2.3.

The main advantage of our approach may be that the question of
what can be varied without changing the quantization is perhaps
clearer than it is in geometric quantization. As we have stressed,
our answer to the question ``Upon what additional data does a
quantization of $(M,\omega)$ depends?'' is ``It depends on the
choice of the complexification $Y$ with antiholomorphic involution
$\tau$, holomorphic two-form $\Omega$ and line bundle $\L$.''

The fact that different linear polarizations of $\R^{2n}$ lead to
equivalent quantizations is a special case of our statement. Given
$\R^{2n}$ with real-valued linear coordinates $x^1, \dots,x^{2n}$
and symplectic structure $\omega=\sum_{i<j}\omega_{ij}dx^i\wedge
dx^j$, we define $Y,\Omega$ without breaking the $Sp(2n,\R)$
symmetry by complexifying the $x^i$ and setting
$\Omega=\sum_{i<j}\omega_{ij}dx^i\wedge dx^j$. We define $\tau$ to
act by $x^i\to\bar x^i$. Since $Y$ is contractible, $\L$ exists
and is unique up to isomorphism, so in our approach, quantization
of $\R^{2n}$ endowed with an affine structure is natural.  The
group $Sp(2n,\R)$ of symmetries of the structure (and in fact, its
inhomogeneous extension to include additive shifts of the
coordinates) must therefore act, at least projectively.

\subsec{Comparison To Deformation Quantization}

Deformation quantization \bayen\ is another matter. Unlike
quantization, deformation quantization is a systematic procedure.
Starting with a symplectic manifold $M$ -- or more generally, any
Poisson manifold --  deformation quantization produces a
deformation of the ring of functions on $M$, depending on a formal
parameter $\hbar$. This can be done in a way that, up to a natural
automorphism, does not depend on any auxiliary choice (such as the
choices that are needed in quantization). The theory of
deformation quantization has led to beautiful results \kontsevich\
that can be expressed in terms of two-dimensional quantum field
theory  \cf, somewhat like our approach to quantization.

Since deformation quantization is a formal procedure, it makes
sense for complex manifolds.  In other words, if $Y$ is a complex
symplectic manifold (such as an affine variety) that admits many
holomorphic functions, one can apply deformation quantization to
deform the ring of holomorphic functions on $Y$ to an associative
but noncommutative algebra \kontsevichtwo.  Deformation
quantization of the ring of holomorphic functions on $Y$ requires
no arbitrary choices (beyond the structure of $Y$ as a complex
symplectic manifold) but quantization does.

However, deformation quantization is not quantization.
Generically, it leads to a deformation over a ring of formal power
series (in the formal variable $\hbar$), not a deformation with a
complex parameter. It does not lead to a natural Hilbert space
$\H$ on which the deformed algebra acts. In our earlier example of
quantizing a two-sphere whose area is $2\pi n$, quantum mechanics
requires that $n$ (which becomes the dimension of $\H$) should be
an integer, while in deformation quantization, $\hbar=1/n$ is
treated as a formal variable and there is no special behavior when
$\hbar^{-1}$ is an integer.

Generally speaking, physics is based on quantization, rather than
deformation quantization, although conventional quantization
sometimes leads to problems that can be treated by deformation
quantization. For a well-known example, see \cds. Our approach to
quantization does have a relationship to deformation quantization.
The relation is that deformation quantization of $Y$ produces an
algebra that then acts in the quantization of a real symplectic
submanifold $M\subset Y$.  (See
\refs{\Brylinski,\Brylinskii,\Brylinskiii} for a similar
perspective in the context of representation theory.)  This will
be explained in the next subsection.

As already noted, in our framework, the existence of a good
$A$-model for $Y$ is supposed to ensure that deformation
quantization of $Y$ produces an actual deformation of the algebra
of holomorphic functions, depending on a complex parameter $\hbar
$ (or $1/n$), not just a formal deformation depending on a formal
parameter $\hbar$.

\subsec{The Inverse Problem}\subseclab\inverse

In describing our approach to quantization, we followed tradition
and began with a symplectic manifold $(M,\omega)$ that one wishes
to quantize.  The solution to the problem involves picking a
suitable complexification $(Y,\Omega)$.

There is an alternative approach in which the starting point is a
complex symplectic manifold $(Y,\Omega)$, together with a unitary
line bundle $\L\to Y$ of curvature ${\rm Re}\,\Omega$.  The
following discussion is most interesting if $Y$ has plenty of
holomorphic functions.  This is so if $Y$ is an affine variety,
such as the variety $x^2+y^2+z^2=1$ that featured in the example
that we discussed previously.

Then one considers the $A$-model of $Y$ in symplectic structure
$\omega_Y={\rm Im}\,\Omega$.  The choice of $\L$ enables us,
following \KO\ and \KW, to define a coisotropic $A$-brane $\Bcc$,
whose support is all of $Y$.  For any $A$-brane $\B$, the space of
$(\B,\B)$ strings is a $\Z$-graded associative algebra.  In the
present case, additively, the space of $(\Bcc,\Bcc)$ strings is
just the space of holomorphic functions on the complex manifold
$Y$. However, in the $A$-model, the commutative ring of
holomorphic functions is deformed.  The first order deformation is
by the Poisson bracket, and the higher order corrections (which
can be computed in sigma-model perturbation theory, somewhat as in
\cf) are controlled by associativity.  Thus the space of
$(\Bcc,\Bcc)$ strings is an associative but noncommutative algebra
$\CA$ that we can think of as arising from deformation
quantization of $Y$.  As we have stressed, if $Y$ has a good
$A$-model, this deformation involves an actual complex parameter,
not a formal one.
Moreover, if $Y$ admits a good $A$-model, its symmetries that
preserve a coisotropic $A$-brane $\Bcc$ will act on $\CA$ as automorphisms.
In our approach, these are precisely the symplectomorphisms of
the complex symplectic manifold $(Y,\Omega)$.
This is closely related to what one finds in deformation
quantization of $Y$ when it produces an actual deformation of
the algebra of holomorphic functions, not just a formal deformation
(see \KontsevichBK\ for a detailed discussion of the affine space).

\def\H{{\cal H}}
If we are interested in quantization, as opposed to deformation
quantization, we need something smaller that $\CA$ acts on.  For
this, we note first that if $\B$ is any other $A$-brane, then by
general principles $\CA$ acts on the space of $(\Bcc,\B)$ strings.
Now, pick $\B$ to be a conventional Lagrangian $A$-brane,
supported on a Lagrangian submanifold $M\subset Y$.  We denote
this brane as $\B'$. Then the space $\H$ of $(\Bcc,\B')$ strings
admits a natural action of the algebra $\CA$.

Now suppose further that $M$ has been chosen so that ${\rm
Re}\,\Omega$ remains nondegenerate when restricted to $M$.
Analysis of the $(\Bcc,\B')$ strings, as first considered in
examples in \AZ, relates $\H$ to quantization of $M$ with
symplectic structure ${\rm Re}\,\Omega$. But how can we get in
this framework a hermitian inner product -- usually understood as
one of the main points of quantization? For this, we need one more
piece of data: an antiholomorphic involution $\tau:Y\to Y$ that
obeys $\tau^*\Omega=\bar\Omega$, maps $M$ to itself, and lifts to
an action on $\L$.  With the help of $\tau$ (and more standard
ingredients, such as CPT symmetry), one can define a hermitian
inner product on $\H$, with the property that holomorphic
functions on $Y$ that obey $\tau(\bar f)=f$  act on $\H$ as
hermitian operators. Near the classical limit, the hermitian inner
product is positive definite if and only if $\tau$ leaves $M$
fixed pointwise.

A noteworthy point here is that the algebra $\CA$ only depends on
the input data $Y,\Omega,\L$, and not on $\tau$.  If $\tau$ does
not exist (or $M$ is not a component of the fixed point set of any
$\tau$), then everything that we have said goes through, except
that $\H$ is not endowed with a natural hermitian metric.

Alternatively, $Y$ may admit several different antiholomorphic
involutions, say $\tau$ and $\tau'$.  Let $M$ and $M'$ be components
of the fixed points sets of $\tau$ and $\tau'$ (and suppose that
${\rm Re}\,\Omega$ is nondegenerate when restricted to either one).
Then we can quantize either $M$ or $M'$ by the above procedure,
giving Hilbert spaces $\H$ and $\H'$. The same algebra $\CA$ will
act in either case.  Functions that are real when restricted to $M$
will be hermitian as operators on $\H$, and functions that are real
when restricted to $M'$ will be hermitian as operators on $\H'$.

There are far more choices if we are not interested in a hermitian
metric.  Then $M$ can be the support of any rank 1 $A$-brane, and
the same algebra $\CA$ acts on the space $\H$ that we obtain by
quantizing $M$, regardless of what $M$ we pick.

\bigskip\noindent{\it The $A$-Model}

So far we have emphasized two points of view about this subject.

In the first approach, the starting point is the real symplectic
manifold $(M,\omega)$ that we wish to quantize.  The problem is
solved by complexifying $M$ to a complex symplectic manifold
$(Y,\Omega)$ that has a good $A$-model for symplectic structure
$\omega_Y={\rm Im}\,\Omega$.

In the second approach, the starting point is the complex
symplectic manifold $(Y,\Omega)$.  Picking a suitable coisotropic
brane $\Bcc$, assumed to be an $A$-brane with respect to
$\omega_Y={\rm Im}\,\Omega$, we deform the algebra of holomorphic
functions on $Y$ to a noncommutative algebra $\CA$.  Then picking
another $A$-brane, we get a module for $\CA$.

A third approach, and the most natural one from the point of view
of two-dimensional topological quantum field theory, is to
emphasize the $A$-model of $Y$, regarded as a real symplectic
manifold with symplectic structure $\omega_Y$.  There may be many
inequivalent choices of space-filling coisotropic $A$-brane $\Bcc$
on $Y$ -- corresponding to different choices of a complex
structure $I$ on $Y$ for which there is a holomorphic two-form
$\Omega$ with $\omega_Y={\rm Im}\,\Omega$. For each choice of such
a brane $\Bcc$, we get a noncommutative algebra $\CA$ that acts on
the space of $(\Bcc,\B')$ strings, for any other $A$-brane $\B'$.
If $M$ is the support of $\B'$, the space of $(\Bcc,\B')$ strings
gives a quantization of $M$ whenever ${\rm Re}\,\Omega$ is
nondegenerate when restricted to $M$.  Thus, the same $A$-model
can lead to quantization of $M$ in different symplectic
structures.

\subsec{Organization Of This Paper}\subseclab\orgo

In section 2, we will describe in more detail our $A$-model
approach to quantization.

Section 3 is devoted primarily to analyzing in more depth the
example related to $M=S^2$.  This example is surprisingly rich and
related to representation theory of $SL(2,\R)$ as well as $SU(2)$.
In this paper, we consider primarily the case of those groups, but
actually, as we explain at the end of section 3, the example has a
generalization in which $M$ is a coadjoint orbit of a semi-simple
real Lie group of higher rank, and $Y$ is the corresponding orbit
of its complexification. This leads to a perspective on the
representations of semi-simple real Lie groups, similar to that of
R. Brylinski \refs{\Brylinski,\Brylinskii,\Brylinskiii}.

Finally, in section 4, we discuss from the present point of view
one of the few known examples in which the subtleties of
quantization are actually important for quantum field theory. This
is three-dimensional Chern-Simons gauge theory.

\newsec{Basic Construction}\seclab\setup

\subsec{The $A$-Model And The Canonical Coisotropic Brane}

We begin with a complex symplectic manifold $Y$, that is, a
complex manifold endowed with a nondegenerate holomorphic two-form
$\Omega$.  Though we will not assume that $Y$ has a hyper-Kahler
structure, it is convenient to use a notation that is suggested by
the hyper-Kahler case.  We write $I$ for the complex structure of
$Y$, and we denote the real and imaginary parts of $\Omega$ as
$\omega_J$ and $\omega_K$: \eqn\locy{\Omega=\omega_J+i\omega_K.}
Since $\Omega$ is of type $(2,0)$, we have $I^t\Omega=i\Omega$, or
\eqn\ocy{I^t\omega_J=-\omega_K,~I^t\omega_K=\omega_J.} (We regard
$I$ as a linear  transformation of tangent vectors; $I^t$ is the
transpose map acting on one-forms. $\Omega$ and $I^t\Omega$ are
maps from tangent vectors to one-forms.)

In this paper, we view $Y$ as a real symplectic manifold with
symplectic structure $\omega_Y=\omega_K$, and we study the
associated $A$-model.   The most familiar branes of the $A$-model
are Lagrangian branes, supported on a Lagrangian submanifold that
necessarily is of middle dimension.  However \KO, in general it is
also possible to have an $A$-brane supported on a coisotropic
submanifold $Z\subset Y$ whose dimension exceeds half the
dimension of $Y$.  For our purposes, we are interested in a rank 1
coisotropic $A$-brane whose support is simply $Z=Y$.  Like any
rank 1 brane, such a brane is endowed with a unitary line bundle
$\L$ with a connection whose curvature we call $F$.  The necessary
condition for such a brane to be an $A$-brane is that
$I=\omega_Y^{-1}F$ should square to $-1$, in which case one can
show that $I$ is an integrable complex structure.

This is a rather special condition, but there is a simple way to
obey it that was important in \KW\ and will also be important in
the present paper.  We simply set $F=\omega_J$, in which case
$\omega_Y^{-1}F=\omega_K^{-1}\omega_J$, which coincides with $I$
according to eqn. \ocy.

Thus, starting with the complex symplectic manifold $(Y,\Omega)$,
for any choice of a unitary line bundle $\L$ with a connection of
curvature $\omega_J={\rm Re}\,\Omega$, we get an $A$-brane in the
$A$-model of symplectic structure $\omega_Y$.  We call this
$A$-brane the canonical coisotropic brane and denote it as $\Bcc$.

To make the $A$-model of symplectic form $\omega_Y$ concrete, it
is usual to introduce an almost complex structure with respect to
which $\omega_Y$ is positive and of type $(1,1)$.  This enables
one to develop a theory of pseudoholomorphic curves in $Y$,
leading to an $A$-model that depends only on $\omega_Y$ and not on
the chosen almost complex structure.  There is no need for the
almost complex structure to be integrable.  In the present case,
since the symplectic structure of the $A$-model is
$\omega_Y=\omega_K$, it is natural to write $K$ for the almost
complex structure that is used to define the $A$-model.

To write the sigma-model action, one also uses the associated
metric $g=-\omega_K K$.  Furthermore, it is always possible to
pick $K$ so that $IK=-KI$, implying that $J=KI$ is also an almost
complex structure.  $J$ will be useful in the quantization.

We stress that we make no assumption that $J$ and $K$ are
integrable.  A $K$ with the stated properties ($\omega_K$ is of
type $(1,1)$, and $IK=-KI$) always exists, and moreover the space
of choices for $K$ is contractible. To see this, let $Y$ be of
real dimension $4n$. Let $Sp(2n)$ be the compact form of the
symplectic group acting on $\C^{2n}$, and $Sp(2n)_\C$ its
complexification. The choice of $I,\Omega$ reduces the structure
group of $Y$ from $GL(4n,\R)$ to $Sp(2n)_\C$.  The further choice
of $K$ reduces this group to $Sp(2n)$. (We have $Sp(2n)=U(2n)\cap
Sp(2n)_\C$, where $U(2n)$ is the subgroup of $GL(4n,\R)$ that
commutes with $K$.)  As the quotient space $Sp(2n)_\C/Sp(2n)$ is
contractible, a global choice of $K$ can be made, and there is no
topology in the choice of $K$.

Though $J$ and $K$ need not be integrable, certainly the nicest
case is that $Y$ admits a complete hyper-Kahler metric in which
the three complex structures are $I$, $J$, and $K$ and the metric
is $g=-\omega_KK$. (We call this a hyper-Kahler polarization.) In
general, we cannot assume this, but many standard facts about the
hyper-Kahler case are true in greater generality. For instance, it
follows from $IK=-KI$ that $\omega_J$ is of type
$(2,0)\oplus(0,2)$ with respect to $K$. Indeed, the fact that
$\omega_K$ is of type $(1,1)$ with respect to $K$ can be written
$K^t\omega_KK=\omega_K$, which by \ocy\ is the same as
$-K^tI^t\omega_JK^t=\omega_K$, or $I^tK^t\omega_JK=\omega_K$. With
$I^2=-1$ and using \ocy\ again, this is $K^t\omega_JK=-\omega_J$,
which is equivalent to saying that $\omega_J$ is of type
$(2,0)\oplus(0,2)$ with respect to $K$. It now follows from $J=KI$
that $\omega_J$ is of type $(1,1)$ with respect to $J$. (We have
$J^t\omega_JJ=I^tK^t\omega_JKI=\omega_J$, since $\omega_J$ is of
type $(2,0)\oplus(0,2)$ with respect to both $I$ and $K$.)

The relation between $J$ and $K$ is completely symmetrical;
instead of beginning with $K$, and then defining $J$, we could
have begun by introducing an almost complex structure $J$,
constrained so that $IJ=-JI$ and $\omega_J$ is of type $(1,1)$
with respect to $J$. The same argument as above would show
existence of $J$, and then we could define $K=IJ$, arriving at the
same picture.  As an example of this viewpoint, suppose we are
given a middle-dimensional submanifold $M\subset Y$ such that
$\omega_J$ is nondegenerate along $M$, and such that the tangent
space $TY$ to $Y$, when restricted to $M$, has a decomposition
$TY=TM\oplus I(TM)$, where $TM$ is the tangent space to $M$. Then
we can constrain $J$ along $M$ so as to  preserve the
decomposition $TY=TM\oplus I(TM)$; we
 simply define $J$ on $TM$ by picking an almost complex
structure on $M$ such that $\omega_J$ is of type $(1,1)$, and then
define $J$ on $I(TM)$ to ensure $IJ=-JI$.  Having defined $J$
along $M$, there is no topological obstruction to extending it
over $Y$, again because $Sp(2n)_\C/Sp(2n)$ is contractible.

\subsec{Space of $(\Bcc,\Bcc)$ Strings}

Our next problem is to identify the algebra $\CA$ of $(\Bcc,\Bcc)$
strings.

First we describe the space of $(\Bcc,\Bcc)$ strings additively.
The space of $(\Bcc,\Bcc)$ strings in the $A$-model is the same as
the space of operators that can be inserted in the $A$-model on a
boundary of a string world-sheet $\Sigma$ that ends on the brane
$\Bcc$. So let us determine this.

We write $X$ for the bosonic fields in the sigma-model with target
$Y$, and $\psi_-$, $\psi_+$ for left- and right-moving fermionic
fields.  A boundary operator  must be invariant under the
supersymmetry (or BRST) symmetry of the $A$-model.  The general
$A$-model transformation law of $X$ is \eqn\urtu{\delta X =
(1-iK)\psi_+ + (1+iK) \psi_-.} Here we use an arbitrary almost
complex structure $K$ (relative to which $\omega_K$ is of type
$(1,1)$ and positive) in defining the $A$-model of symplectic
structure $\omega_K$.  A simple type of local operator is an
operator $f(X)$ derived from a complex-valued function $f:Y\to
\C$. For such an operator, inserted at an interior point of
$\Sigma$, to be invariant under \urtu, $f$ must be constant.

However, we are interested in boundary operators, rather than bulk
operators, and for this we must consider the boundary condition
obeyed by the fermions.  For a general space-filling rank 1 brane,
this boundary condition is \eqn\zno{\psi_+ = (g-F)^{-1}(g+F)
\psi_-.} In the present case, with $F=\omega_J$, we have
$(g-F)^{-1}(g+F)= J$. So for boundary operators, \urtu\ collapses to
$\delta X=\left((1-iK)J+(1+iK)\right)\psi_-$, which is equivalent to
\eqn\zoon{\delta X= (1+iI)(1+J)\psi_-.} If we decompose $\delta X$
as $\delta^{1,0}X+\delta^{0,1}X$, where the two parts of the
variation are of type $(1,0)$ and type $(0,1)$ with respect to $I$,
then \eqn\tycno{\eqalign{\delta^{1,0}X& = 0 \cr
                              \delta^{0,1}X& = \rho, \cr}}
where $\rho=(1+iI)(1+J)\psi_-$.  Clearly, since the square of the
topological symmetry of the $A$-model vanishes, \tycno\ implies
that \eqn\zycno{\delta\rho=0.}

{}From \tycno, we see that for a string ending on the canonical
coisotropic brane, a boundary operator $f(X)$ preserves the
topological symmetry of the $A$-model if and only if the function
$f$ is holomorphic in complex structure $I$. More generally, all
boundary observables of the $A$-model can be constructed from $X$
and $\rho$, which have dimension 0, since other fields have
strictly positive dimension or vanish at the boundary.  If we pick
local complex coordinates on $Y$ corresponding to fields $X^i$,
then a general operator of $q^{th}$ order in $\rho$ takes the form
$\rho^{\bar i_1}\rho^{\bar i_2}\dots\rho^{\bar i_q}f_{\bar i_1\bar
i_2\dots\bar i_q}(X,\bar X)$ and has charge $q$ under the ghost
number symmetry of the $A$-model.  By interpreting $\rho^{\bar i}$
as $d\bar X^{\bar i}$, we can interpret such an operator as a
$(0,q)$-form on $Y$. Then it follows from \tycno\ and \zycno\ that
the topological supercharge of the $A$-model corresponds to the
$\bar\partial$ operator of $Y$.

So the observables of the $A$-model correspond additively to the
graded vector space $H_{\bar
\partial}^{0,\star}(Y)$ (where $Y$ is viewed as a complex manifold
with complex structure $I$). For our purposes in this paper, we
are mainly interested in the ghost number zero part of the ring of
observables. Additively, this corresponds simply to the
holomorphic functions on $Y$. However, the multiplicative
structure is different.  Classically, the holomorphic functions on
$Y$ generate a commutative ring, but in the $A$-model, in the
special case of a brane of type $\Bcc$, this ring is deformed to a
noncommutative ring that we call $\CA$. The deformation
corresponds to deformation quantization using the Poisson brackets
derived from the holomorphic symplectic two-form $\Omega$. One
explanation of how this deformation comes about is given in
section 11.1 of \KW.

\subsec{Lagrangian Brane And Quantization}\subseclab\lbrane

\ifig\opst{An open string with $(\Bcc,\B')$ boundary conditions.}
{\epsfxsize2.5in\epsfbox{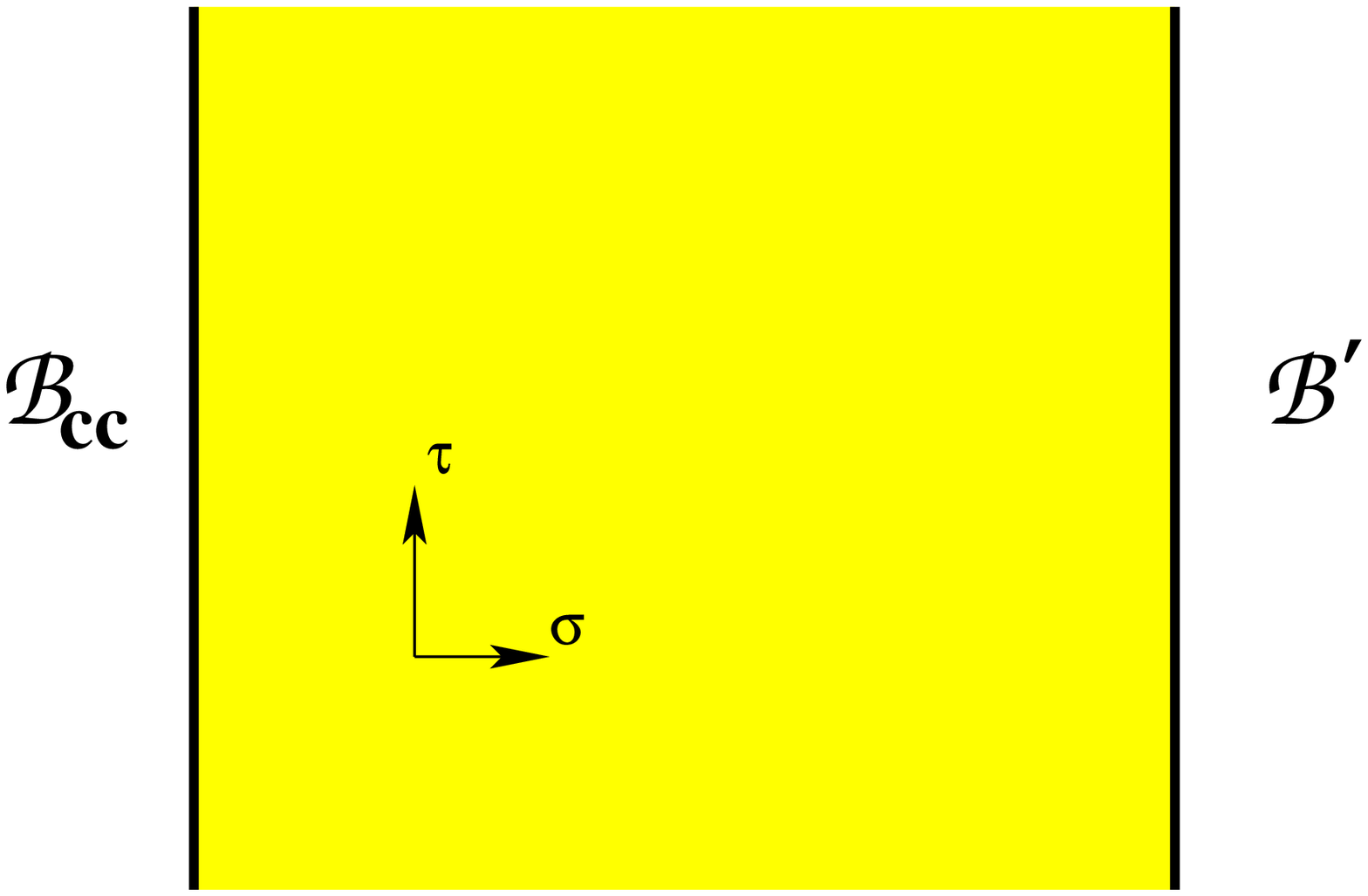}}

So far, we have obtained an algebra $\CA$ of $(\Bcc,\Bcc)$
strings. Now we want to find something (other than itself) that
$\CA$ can act on.  The immediate answer is to introduce a second
$A$-brane $\B'$. Then $\CA$ acts naturally on the space of
$(\Bcc,\B')$ strings (\opst).

\def\CT{\cal L'}
For this paper, we will consider $\B'$ to be an $A$-brane of the
simplest possible kind: a Lagrangian $A$-brane of rank 1.  Thus,
$\B'$ is supported on a Lagrangian submanifold $M$ and endowed
with, roughly speaking, a flat line bundle $\CT$. The natural
objects of study in what follows are branes, and it is important
to bear in mind some subtleties about the relation between
Lagrangian submanifolds and branes.  First, because of disc
instanton effects, not every pair $(M,\CT)$ defines an $A$-brane.
Second, in the absence of such effects, if two pairs $(M,\CT)$ and
$(M',\CT')$ differ by Hamiltonian isotopy, then the associated
$A$-branes are equivalent. Finally, the interpretation of $\CT$ as
a flat line bundle is oversimplified, because of the relation of
branes to $K$-theory and the role of the $B$-field.  Each of these
effects will play some role later.

To relate branes to quantization, we will impose a condition on
$M$ that has no analog in the usual theory of the $A$-model.  We
assume that $\omega_J$ is nondegenerate when restricted to $M$.
This is a mild condition in the sense that, acting on the tangent
space to a given point in $Y$, $\omega_J$ is nondegenerate when
restricted to a generic even-dimensional plane.  Hence if the
condition  is true for a submanifold $M$, it is true for any
sufficiently nearby submanifold.  (The opposite case that
$\omega_J$ is zero when restricted to $M$ was investigated in
section 11 of \KW, and leads to ${\cal D}$-modules. We return to
this in section 3.9.)

If $\omega_J$ is nondegenerate when restricted to $M$, then the
pair $(M,\omega_J)$ is a symplectic manifold, and this is the
symplectic manifold that we will quantize.

As explained in section 2.1, to define the $A$-model, we pick an
almost complex structure $K$, with respect to which $\omega_K$ is
of type (1,1), and such that $I$, $K$, and $J=KI$ obey the algebra
of quaternions.  Once $M$ is given, it is convenient to further
constrain $K$ such that the tangent bundle $TM$ is $J$-invariant.
(That one can do so was explained at the end of section 2.1.)

We now want to show that quantization of $(\Bcc,\B')$ strings leads
to quantization of the symplectic manifold $(M,\omega_J)$. To see
this, we describe the sigma-model on a Riemann surface $\Sigma$, as
in section 2.2, by bosonic fields $X$ that describe the sigma-model
map $\Phi:\Sigma\to Y$, and left and right-moving fermionic fields
$\psi_-$ and  $\psi_+$. In general, the boundary conditions for
fermions at the end of an open string are $\psi_+=R\psi_-$, for some
matrix $R$. In the case of a space-filling brane of rank 1, such as
$\Bcc$, \eqn\zelx{R=(g-F)^{-1}(g+F).} Here $g$ is the metric of $Y$
and $F$ is the curvature of the Chan-Paton line bundle. For the
brane $\Bcc$, we have $F=\omega_J$ and $R=J$, so the boundary
condition is \eqn\elx{\psi_+=J\psi_-.}

For a Lagrangian brane, supported on a submanifold $M$, $F=0$ but
this does not mean that $R=1$.  Rather, the boundary condition is
\eqn\belx{\psi_+={\cal R}\psi_-,} where ${\cal R}:TY|_M\to TY|_M$
 is a
reflection that leaves fixed the tangent bundle $TM$ to $M$, and
acts as $-1$ on the normal bundle to $M$ (here $TY|_M$ is the
tangent bundle of $Y$ restricted to $M$).

For our problem, we are principally interested in $(\Bcc,\B')$
strings, that is strings that couple to $\Bcc$ on the left and to
$\B'$ on the right.  The boundary conditions are thus
\eqn\zurx{\eqalign{\psi_+(0) &=J\psi_-(0) \cr \psi_+(\pi) &={\cal
R}\psi_-(\pi), \cr}} where $0$ and $\pi$ are the endpoints of the
string.  These boundary conditions do not allow $\psi$ to be
constant along the string.  For example, if $\psi$ is tangent to $M$
and also constant, the combination of the two boundary conditions
gives $\psi_+=J\psi_+$, which since $\psi_L$ is real and $J^2=-1$
implies that $\psi_+=0$.  Similarly, if $\psi$ is normal to $M$ and
constant, we get $\psi_+=-J\psi_+$, again implying that $\psi_+=0$.

Now let us discuss the bosonic fields $X(\sigma,\tau)$, where
$\sigma$ and $\tau$ are the worldsheet space and time coordinates.
The bosonic fields do have zero modes, because the boundary
condition $*d X=R dX$ is consistent with constant $X$; likewise
the classical equations of motion are obeyed if $X$ is constant.
The usual zero mode structure of the bosonic string is
$X=x+p\tau+\dots$, where $x$ is the zero mode, $p$ is its
canonical momentum,  and $\dots$ are the nonzero modes.  Here,
usually $x$ is canonically conjugate to $p$.  In the present case,
there is a constant $x$ term, but there is no $p\tau$ term in the
expansion, because as soon as $X$ becomes $\tau$-dependent, the
boundary condition $*dX=RdX$ at $\sigma=0$ does not permit $X$ to
be independent of $\sigma$.  Usually, in sigma-model perturbation
theory, this effect can be treated perturbatively because $R$ is
close to 1; but for a boundary associated with the brane $\Bcc$,
we have $R^2=-1$, so there is no way to expand around $R=1$.  The
result is that the $p\tau$ term is missing in the expansion, which
reads \eqn\jux{X=x+\dots,} where the omitted terms are nonzero
modes. The nonzero modes are, of course, related by worldsheet
supersymmetry to the nonzero modes of $\psi$.  For a detailed
explanation of this expansion in an example, see \AZ.

In the context of quantizing the $(\Bcc,\B')$ strings, in addition
to the boundary condition at $\sigma=0$, there is also a boundary
condition $X(\pi)\in M$ at the other end.  Because of this
boundary condition, the zero modes in \jux\ take values in $M$.

\def\N{{\cal N}}
So it must be that the components of $x$ are canonically conjugate
to each other.  The reason that this happens is that the action of
a string ending on a brane with Chan-Paton connection $A$ contains
a boundary term $\int_{\partial\Sigma}A_\mu\,dX^\mu$. In the case
of $(\Bcc,\B')$ strings, the Chan-Paton bundle $\L'$ of the brane
$\B'$ is flat, while the Chan-Paton bundle of $\Bcc$ is the
unitary line bundle $\L$ of curvature $\omega_J$. We write $A$ and
$A'$ for the connections on these two line bundles. Classically,
the action for the zero modes is $\int d\tau
(A_{\mu}-A'_\mu)dx^\mu/d\tau$, where the two terms come from the
left and right endpoints of the string. We define a line bundle $
\N=\L\otimes (\L')^{-1}$ over $M$. $\N$ is a unitary line bundle
with a connection $B=A-A'$ of curvature $\omega_J$. The action for
the zero modes is, in this approximation, \eqn\zelz{\int
B_\mu{dx^\mu\over d\tau}d\tau.}

Formally speaking, to quantize the zero modes with this action
(which actually has an important ``quantum'' correction that we will
describe shortly) amounts to quantizing $M$ with prequantum line
bundle $\N$.   Just knowing this does not give any general solution
to the problem of quantization. All we learn is that, if the
$A$-model of $Y$ exists, then the space of $(\Bcc,\B')$ strings can
be understood as the result of quantizing $M$ with prequantum line
bundle $\N$. If the $A$-model exists, and the two boundary
conditions associated with branes $\Bcc$ and $\B'$ also exist, then
the space of $(\Bcc,\B')$ strings exists even if it is hard to
describe this space explicitly.

Usually, given two $A$-branes or $B$-branes $\B_1$ and $\B_2$, one
can go to a large volume limit and describe the space of
$(\B_1,\B_2)$ strings in terms of classical geometry.  The delicate
structure of the coisotropic brane $\Bcc$ prevents us from doing
this successfully for the $(\Bcc,\B')$ strings, in general.

But we can use general properties of the $A$-model to learn
general properties of quantization.  One can also get some
information from classical geometry, as we explain next.

\bigskip\noindent{\it Branes Of Type $(A,B,A)$}

There is an important special case in which we can describe
explicitly the space of $(\Bcc,\B')$ strings.  This is the case that
we are given a hyper-Kahler polarization of $(Y,M)$.  This is a
hyper-Kahler structure on $Y$, extending its complex symplectic
structure $(I,\Omega)$, such that $M$ is a complex submanifold in
complex structure $J$.

Under these conditions, the branes $(\Bcc,\B')$ are both branes of
type $(A,B,A)$, that is, $A$-branes for the $A$-models with
symplectic forms $\omega_I$ or $\omega_K$, and $B$-branes for the
$B$-model of complex structure $J$. For example, $\Bcc$ is a
$B$-brane of type $J$ because the curvature form $\omega_J$ of its
Chan-Paton line bundle is of type $(1,1)$ in complex structure
$J$. Similarly, $\B'$ is a $B$-brane because $M$ is a complex
submanifold and the Chan-Paton bundle of $\B'$ is flat.

We can now look at the space of $(\Bcc,\B')$ strings in two
different ways.  We defined $\H$ to be the space of $(\Bcc,\B')$
strings in the $A$-model of $\omega_Y=\omega_K$.  Similarly, we
can define $\tilde\H$ to be the space of $(\Bcc,\B')$ strings in
the $B$-model of complex structure $J$.   As long as $M$ is
compact or wavefunctions are required to vanish sufficiently
rapidly at infinity, the two spaces are the same, since they both
can be described as the space of zero energy states of the
sigma-model with target $Y$ (compactified on an interval with
boundary conditions at the two ends determined by $\Bcc$ and
$\B'$).

One qualification is that the equivalence of $\H$ and $\tilde \H$
does not respect their grading. $\H$ and $\tilde \H$ are both
$\Z$-graded, because the $A$-model of type $\omega_K$ and the
$B$-model of type $J$ are both $\Z$-graded by ``ghost number.''
The gradings are different, but conjugate.  In fact, the
sigma-model of target $Y$, with boundary conditions set by two
branes of type $(A,B,A)$, has an $SU(2)$ group of $R$-symmetries
that we call $SU(2)_R$; the two ghost number symmetries are
conjugate but different $U(1)$ subgroups of $SU(2)_R$.  In
practice, however, one usually studies quantization in a situation
in which the grading is trivial; this is so if $\N$ is very ample
as a line bundle in complex structure $J$.

To use the equivalence between $\H$ and $\tilde\H$, we need to
have a way to determine $\tilde\H$.  For this, we simply observe
that in the $B$-model, the choice of Kahler metric is irrelevant.
So we can rescale the metric of $Y$ by a factor $t\gg 1$, reducing
to a situation in which sigma-model perturbation theory is valid.
In this limit, by a standard argument, we describe $\tilde\H$ by
$\bar\partial$ cohomology: \eqn\zolto{\tilde\H=\oplus_{i=0}^{{\rm
dim}_\C M}H^i(M,K^{1/2}\otimes \N).} Since $\H\cong\tilde\H$, we
can express this as a statement about $\H$:
\eqn\zoltox{\H=\oplus_{i=0}^{{\rm dim}_\C M}H^i(M,K^{1/2}\otimes
\N).}
 Here, roughly speaking,
$K^{1/2}$ is the square root of the canonical line bundle $K$ of
$M$.  (A more precise explanation is given presently.) For very
ample $\N$, the cohomology vanishes except for $i=0$ and its
$\Z$-grading is trivial.

The description  \zoltox\ of   $\H$ has an important limitation,
beyond the problem with the grading. It describes $\H$ as a vector
space, but it does not lead to a natural description of the
Hilbert space structure of $\H$, when there is one.  In section
2.4, we will describe the conditions under which $\H$ has a
hermitian inner product that can be defined in the $A$-model.
Analogous but different conditions\foot{To define a hermitian
metric in the $B$-model, one uses an involution $\tau$ of $Y$ that
maps $M$ to itself and reverses the sign of $J$, while in the
$A$-model, $\tau$ reverses the sign of $\omega_Y,$ as explained in
section 2.4.}
 lead to a natural
hermitian inner product on $\tilde \H$ in the $B$-model. The two
hermitian structures are different (when they both exist) and the
equivalence between $\H$ and $\tilde\H$  does not map a natural
hermitian structure of the $A$-model to a natural structure of the
$B$-model. (This identification does preserve a third hermitian
product, the one that $\H$ and $\tilde \H$ get from their
interpretation as the space of physical ground states in the
sigma-model with hyper-Kahler metric. This one is not natural in the
$A$-model or the $B$-model and is not visible in the large volume
limit that leads to \zolto.) Hence, when applicable, \zoltox\
describes $\H$ as a vector space, not as a vector space with a
hermitian inner product.

The description of $\H$ that we have just given has an obvious
resemblance to a standard statement in geometric quantization.  In
that context, the choice of an integrable complex structure $J$ on
$M$, such that the symplectic form of $M$ becomes a Kahler form, is
known as a complex polarization. Eqn. \zoltox\ then defines
quantization with a complex polarization.

A hyper-Kahler polarization of the pair $(Y,M)$, which we used in
the above derivation, plays an analogous role in our approach.
Our statement is that for any choice of hyper-Kahler polarization,
the space of $(\Bcc,\B')$ strings can be described as in \zoltox.

There are two primary differences between our statement and the
analogous statement in geometric quantization:

(1) In our framework, the space of $(\Bcc,\B')$ strings is an
$A$-model invariant and therefore is independent of the choice of
a hyper-Kahler polarization.  In geometric quantization, there is
no general statement about when the right hand side of \zolto\ is
independent of the choice of complex polarization $J$.

(2) In our framework, \zolto\ is a statement about vector spaces,
while in geometric quantization, one usually endows the right hand
side of \zolto\ with a hermitian structure.  For example, in the
very ample case, one defines a Hilbert space norm by
$|\psi|^2=\int_M (\omega^n/n!)\bar\psi\psi$, where $\omega$ is the
symplectic form of $M$, $2n$ is the real dimension of $M$, and
$\psi\in H^0(M,K^{1/2}\otimes \N)$.  This is certainly a natural
formula in Kahler geometry (it describes the hermitian metric that
arises in the sigma-model after rescaling the metric of $Y$ by a
factor $t\gg 1$), but it is not a natural $A$-model inner product.
A somewhat related statement is that in our framework, \zolto\
does not describe the natural $\Z$-grading of $\H$, but a
conjugate one.

\bigskip\noindent{\it Relation To $K$-Theory}

\def\T{{\cal T}}

A point that still remains to be clarified is the meaning of the
symbol $K^{1/2}$ in the above analysis.  In general, $M$ may not be
a spin manifold, so a line bundle $K^{1/2}$ may not exist, and $M$
may not be simply-connected, so that if $K^{1/2}$ exists, it may not
be unique up to isomorphism.

The resolution of this point depends upon the relation of branes to
$K$-theory. The following are general statements about branes,
independent of any specialization to an $A$-model or a $B$-model.
Consider a brane supported on a submanifold $N\subset Y$ and endowed
with a rank 1 Chan-Paton bundle $\T$.  Naively, $\T$ is a complex
line bundle, but actually, because of an anomaly in the sigma-model
\fw, $\T$ can be more precisely described as a ${\rm Spin}_c$
structure on the normal bundle to $N$ in $Y$.

For the space-filling brane $\Bcc$, $N$ is equal to $Y$, so the
normal bundle to $N$ is trivial.  Hence the Chan-Paton bundle $\L$
of $\Bcc$ is an ordinary complex line bundle.  For the Lagrangian
$A$-brane $\B'$, $N$ is the Lagrangian submanifold $M$.  The tangent
bundle and normal bundle to a Lagrangian submanifold are naturally
isomorphic (under multiplication by $\omega_Y=\omega_K$), so the
Chan-Paton ``line bundle'' $\L'$ of $\B'$ is really a choice of
${\rm Spin}_c$ structure on $M$.  If $\L'$ were actually a line
bundle, there would be no natural line bundle $(\L')^{-1}\otimes
K^{1/2}$, since $K^{1/2}$ may not exist and may not be unique.
However, for $\L'$ a ${\rm Spin}_c$ structure, there is a natural
line bundle $(\L')^{-1}\otimes K^{1/2}$; the two factors in this
tensor product are not naturally defined as complex line bundles,
but the tensor product is.

Now the meaning of \zolto\ is clear. With $\N=\L\otimes
(\L')^{-1}$, and $\L$ an ordinary complex line bundle, there is no
problem in defining ${\N}\otimes K^{1/2}$ as a complex line
bundle, though the two factors separately do not have this status.

In this discussion, we did not assume that the rank 1 brane
supported on $N$ is supposed to be an $A$-brane or a $B$-brane. If
this brane is supposed to be an $A$-brane supported on a
Lagrangian submanifold $M$, then $\L'$ must be a flat ${\rm
Spin}_c$ structure on $M$.  There is a topological obstruction to
having a ${\rm Spin}_c$ structure on $M$, and there is a further
obstruction to having a flat ${\rm Spin}_c$ structure. In general,
${\rm Spin}_c $ structures on $M$ are classified topologically by
the choice of a way of lifting the second Stieffel-Whitney class
$w_2(M)\in H^2(M,\Z_2)$ to an integral cohomology class $\zeta\in
H^2(M,\Z)$. Flat ${\rm Spin}_c$ structures are classified by a
choice of a lift $\zeta$ such that $\zeta$ is a torsion element of
$H^2(M,\Z)$.  In general, even if $M$ is ${\rm Spin}_c$, it may
not admit a flat ${\rm Spin}_c$ structure, since it may be
impossible to pick the lift $\zeta$ to be a torsion class.  A
symplectic manifold that does not admit a flat ${\rm Spin}_c$
structure cannot be quantized in our sense.

This obstruction to quantization has been encountered in the
literature on representations of a semi-simple non-compact real Lie
group $G$. In that context, $M$ is a coadjoint orbit of $G$, and one
aims to obtain a representation of $G$ by quantization of $M$. (See
section 3 for more on this.)  This problem has been approached from
many different points of view.   In \Brylinski, which is perhaps the
closest to the approach in the present paper, a preliminary step to
quantizing $M$ is to, in effect, endow $M$ with a flat ${\rm
Spin}_c$ structure. For example, the minimal orbit of $SO(p,q)$ with
$p+q$ odd and $p,q\geq 4$  does not admit such a structure and
cannot be quantized by the methods of \Brylinski\ or of the present
paper.

\bigskip\noindent{\it General Shift By $K^{1/2}$}

Now let us return to the zero mode action \zelz,
 \eqn\belz{\int B_\mu{dx^\mu\over d\tau}d\tau,}
 dropping the assumption of a hyper-Kahler polarization.
In the original derivation, $B$ emerged as a connection on the
``complex line bundle'' $\N=\L\otimes (\L')^{-1}$.  However, as we
have just explained, in general $\N$ does not make sense as a
complex line bundle.

Simply to make sense of the formula, there must be a correction that
shifts $\N$ to $\hat \N=\L\otimes (\L')^{-1}\otimes K^{1/2}$ (or
something similar to this), where $K$ is the canonical line
bundle\foot{The metric on $Y$ that is used to define the $A$-model,
even if it is not hyper-Kahler, restricts to a metric on $M$. The
metric on $M$, together with the symplectic structure, determines an
almost complex structure on $M$, and this almost complex structure
is used to define $K$.} of $M$.  The connection that appears in the
action must be a connection on $\hat \N$.

The way that this correction arises is as follows.  As  a step
toward quantizing the open strings with $(\Bcc,\B')$ boundary
conditions, one quantizes the worldsheet fermions, expanding around
a map from the string worldsheet to $Y$ that consists of a constant
map to a point $p\in M$.  In expanding around such a constant map,
there are no fermion zero modes, since they are all removed by the
boundary condition \zurx.  Hence, the space of ground states in the
fermion Fock space is a one-dimensional vector space $W_p$. As $p$
varies, $W_p$ varies as the fiber of a complex ``line bundle''
${\cal W}\to M$. ${\cal W}$ and the induced connection on it must be
included as an additional factor in quantizing the bosonic zero
modes.

 In fact, ${\cal W}$ is
isomorphic to $K^{1/2}$ (and is not quite well-defined as a line
bundle because of an anomaly in  the relevant  family of fermion
Fock spaces). One can show that ${\cal W}\cong K^{1/2}$ by using
standard methods to determine the quantum numbers of the Fock space
ground state.  This result is also needed to match the $B$-model
analysis in the case that one uses a hyper-Kahler polarization.

\subsec{Unitarity}

So far, to an $A$-brane $\B'$, we have associated a vector space
$\H$ consisting of $(\Bcc,\B')$ strings.  Quantum mechanics
usually involves Hilbert spaces, however, and the question arises
of how to define a hermitian inner product on $\H$.

For this, we want to make a general analysis of how a hermitian
inner product can appear in the topological $A$-model.  In this
general discussion, we are only concerned with the topological
$A$-model with symplectic structure $\omega_K$.  (Additional
structures such as $I$ and $\omega_J$ are not relevant.)

If $\B_1$ and $\B_2$ are two branes in the topological $A$-model,
there is always a natural duality between the space of
$(\B_1,\B_2)$ strings and the space of $(\B_2,\B_1)$ strings.  It
is given by the two point function on the disc. Consequently,
defining a hermitian inner product on the space of $(\Bcc,\B')$
strings is equivalent to finding a complex antilinear map from
$(\Bcc,\B')$ strings to $(\B',\Bcc)$ strings.

Let us start by assuming that our $A$-model is a twisted version of
a physically sensible, unitary, supersymmetric field theory. In
general, any such field theory has an antilinear CPT symmetry, which
we will denote $\Theta$. For any pair $\B_1,\B_2$, the
transformation $\Theta$ maps $(\B_1,\B_2)$ strings to $(\B_2,\B_1)$
strings. This gives an antilinear map from $(\Bcc,\B')$ strings to
$(\B',\Bcc)$ strings, but it cannot be the map we want, because it
is not a symmetry of the $A$-model. The definition of the $A$-model
depends on a choice of a differential $Q$, which is  a complex
linear combination of the supercharges.  CPT maps $Q$ to $Q^\dagger$
(its hermitian adjoint), which is the differential of a complex
conjugate $A$-model.

To get a symmetry of the $A$-model, we need to combine $\Theta$
with a transformation that maps the conjugate $A$-model back to
the original $A$-model.  We get such a transformation if $Y$
admits an involution $\tau$ (that is a diffeomorphism obeying
$\tau^2=1$) with the property that $\omega_K$ is odd under $\tau$:
\eqn\lefdo{\tau^*\omega_K=-\omega_K.} This property implies that
$\tau$ maps the $A$-model to the conjugate $A$-model.  In fact, we
can always pick a $\tau$-invariant metric $g$, such that
$\omega_K$ is of type $(1,1)$.  Then the almost complex structure
$K=g^{-1}\omega_K$ is odd under $\tau$.  So $\tau$ maps
pseudoholomorphic curves to pseudo-antiholomorphic curves, and
thus maps the $A$-model to the conjugate $A$-model.

So $\Theta_\tau=\tau\Theta$ is an antilinear map from the $A$-model
to itself.  This is a general statement about the $A$-model and
holds whether or not the $A$-model can be obtained by twisting an
underlying physical theory.  (The latter is possible if one can
choose $K$ to be integrable.)

Now if $\B_1$ and $\B_2$ are $\tau$-invariant $A$-branes, we can
use $\Theta_\tau$ to define a hermitian inner product on the space
$\H$ of $(\B_1,\B_2)$ strings.  If $(~,~)$ is the pairing between
$(\B_1,\B_2)$ strings and $(\B_2,\B_1)$ strings given by the
two-point function on the disc, then the inner product on $\H$ is
given by \eqn\gurf{\langle \psi,\psi'\rangle=(\Theta_\tau
\psi,\psi').}

Now let us apply this to our problem.  For the brane $\Bcc$ to be
$\tau$-invariant, we first of all need that the curvature
$\omega_J$ of the Chan-Paton bundle of this brane should be
$\tau$-invariant: \eqn\labk{\tau^*\omega_J=\omega_J.} We also
actually need a little more: the action of $\tau$ on $M$ should
lift to an action on the Chan-Paton line bundle $\L$, whose
curvature is $\omega_J$, and we must pick such a lift. Since
$\omega_K$ is odd under $\tau$ while $\omega_J$ is even, it
follows that $\tau$ is antiholomorphic from the point of view of
the complex structure $I=\omega_J^{-1}\omega_K$:
\eqn\zodoc{\tau^*I=-I.} It follows from this that each component
of the fixed point set of $\tau$ is middle-dimensional.  With
$\Omega=\omega_J+i\omega_K$, we also have
\eqn\zabk{\tau^*\Omega=\bar\Omega.} So more briefly, we can
summarize the above conditions by saying that $\tau$ is an
antiholomorphic involution of the complex symplectic manifold
defined by the data $(Y,I,\Omega)$, with a lift to $\L$.

For the Lagrangian $A$-brane $\B'$ to be $\tau$-invariant, its
support $M$ (and its Chan-Paton line bundle $\L'$) must be
$\tau$-invariant, and again, we need a lift of $\tau$ to act on
$\L'$.  In this case, we get, finally, a hermitian form on the
space $\H$ of $(\Bcc,\B')$ strings.  Note that $\tau$ need not
leave $M$ fixed pointwise, but it must map $M$ to itself.  If
$\tau$ acts trivially on $M$, the lift to $\L'$, if it exists, can
be uniquely specified by saying that $\tau$ acts trivially on the
restriction of $\L'$ to $M$.  If $\tau$ acts nontrivially on $M$,
its lift to $\L'$ may involve a subtle choice.

The hermitian inner product $\langle~,~\rangle$ is not necessarily
positive definite (which means as usual that
$\langle\psi,\psi\rangle>0$ for all nonzero $\psi\in\H$), but it
is always nondegenerate. Nondegeneracy means that given $\psi\in
\H$, there is always $\chi\in\H$ such that
$\langle\chi,\psi\rangle\not=0$.  This follows from nondegeneracy
of the underlying topological inner product $(~,~)$ and the fact
that $\Theta_\tau^2=1$.  Picking $\chi_0$ such that
$(\chi_0,\psi)\not=0$ and setting $\chi=\Theta_\tau\chi_0$, we
have $\langle\chi,\psi\rangle\not=0$.

 Near the classical
limit, the norm of a state $\psi\in \H$ is roughly
$\langle\psi,\psi\rangle =\int_M \bar\psi(\tau x)\psi(x)$. Such a
form can be positive-definite only if $\tau$ acts trivially on
$M$. So for $\langle~,~\rangle$ to be positive near the classical
limit, we require that $\tau$ should act trivially on $M$. Since
$M$ (being Lagrangian) is middle-dimensional, while the fixed
point set of the antiholomorphic involution $\tau$ is also
middle-dimensional, this means that $M$, assuming it is connected,
is a component of the fixed point set.   (In section 3.8, we give
an example far from the classical limit in which
$\langle~,~\rangle$ is positive even though $\tau$ acts
nontrivially on $M$.)

Conversely, if $M$ is a component of the fixed point set of
$\tau$, then, as $\omega_K$ is odd under $\tau$, $M$ is
automatically Lagrangian for $\omega_K$.  Moreover, $\omega_J$ is
automatically nondegenerate when restricted to $M$.  (To prove
this, start with the fact that, if $M$ is held fixed by an
antiholomorphic involution $\tau$, then the restriction to $M$ of
the tangent bundle to $Y$ decomposes as $TY|_M=TM\oplus I(TM)$.
$\tau$ acts as 1 and $-1$ on the two summands, and, being
$\tau$-invariant, $\omega_J$ is the sum of a nondegenerate
two-form on $TM$ and one on $I(TM)$.)

Thus, the case of our construction that leads to unitarity near the
classical limit is much more specific.  Introducing the space of
$(\Bcc,\Bcc)$ strings, we carry out deformation quantization of the
complex symplectic manifold $Y$, constructing an associative algebra
$\CA$. Then, to get a Hilbert space $\H$ on which $\CA $ acts, we
pick an antiholomorphic involution $\tau$ (with a lift to the line
bundle $\L\to Y$) and a component $M$ of the fixed point set
supporting a $\tau$-invariant $A$-brane $\B'$. $\H$ is defined as
the space of $(\Bcc,\B')$ strings, and $\CA$ acts on $\H$.

The operation $\Theta_\tau$ acts on a function $f$ on $Y$,
defining a $(\Bcc,\Bcc)$ string, as the composition of $\tau$ with
complex conjugation.  So if ${\cal O}_f:\H\to\H$ is the operator
associated to $f$, then the hermitian adjoint of ${\cal O}_f$ is
associated with the function $\tau(\bar f)$.  In particular, if
$\tau$ leaves $M$ fixed pointwise and $f$ is real when restricted
to $M$, then $\tau(\bar f)=f$ and ${\cal O}_f$ is hermitian.

\newsec{Branes And Representations}

\subsec{An Example}

To make these ideas more concrete, we will consider in more depth an
example that was introduced in section 1.2.   This is the case that
$Y$ is an affine algebraic variety defined by the equation
$x^2+y^2+z^2=\mu^2/4$ in complex variables $x,y,z$. Here $\mu^2$ is
a complex constant.  $Y$ admits the action of a group $SO(3,\C)$,
rotating $x,y$, and $z$.  In quantization, we will encounter the
double cover of $SO(3,\C)$, which is $SL(2,\C)$.

We make $Y$ a complex symplectic manifold by introducing the
holomorphic two-form \eqn\oosely{\Omega=h^{-1}{dy\wedge dz\over
x},} where $h$ is another complex constant. $h$ could be
eliminated (at the cost of changing $\mu$) by rescaling $x,y$, and
$z$, and we will eventually set $h=1$.

$\Omega$ is $SO(3,\C)$-invariant.  One way to show this is to
observe that $Y$ is defined by the equation $f=0$ in $\C^3$, where
$f=x^2+y^2+z^2-\mu^2/4$.  The meromorphic differential
$2h^{-1}\,dx\wedge dy\wedge dz/f$ is manifestly
$SO(3,\C)$-invariant, with a pole precisely at $f=0$.  The residue
of this pole is $\Omega$.

\def\hx{\hat x}\def\hy{\hat y}\def\hz{\hat z}\def\smu{\sqrt{\mu^2}}
Consequently, the $A$-model of $Y$ with symplectic structure
$\omega_Y={\rm Im}\,\Omega$ has $SO(3,\C)$ symmetry. If we
consider only the usual Lagrangian $A$-branes, we do not get much
from the $SO(3,\C)$ action. That action becomes interesting,
however, if we introduce coisotropic $A$-branes.

As a preliminary, we compute the cohomology class of $\Omega$.  We
set $x=\hx \smu/2,$ $y=\hy\smu/2$, $z=\hz\smu/2$.  The
two-dimensional homology of $Y$ is generated by the real two-cycle
$S$ defined by $\hx^2+\hy^2+\hz^2=1$.  We compute that
\eqn\zonk{\int_S{\Omega\over 2\pi}= h^{-1}\mu,} where
$\mu=\pm\sqrt{\mu^2}$. The sign depends on the orientation of $S$.

Hence the cohomology class of ${\rm Re}\,\Omega/2\pi$ is integral
if $n={\rm Re}\,(h^{-1}\mu)$ is an integer.  Under this condition,
we can construct a unitary line bundle $\L$ with a connection of
curvature ${\rm Re}\,\Omega$.  This determines a space-filling
coisotropic brane $\Bcc$ in the $A$-model with symplectic
structure $\omega_Y={\rm Im}\,\Omega$.  The first Chern class of
$\L$, integrated over $S$, is equal to $n$:
\eqn\lozg{\int_Sc_1(\L)=n.}

Now let us describe the ring $\CA$ of $(\Bcc,\Bcc)$ strings.
Classically, as explained in section 1.2, it is the commutative ring
of polynomial functions in $x,y, $ and $z$, modulo the commutativity
relation \eqn\zelg{[x,y]=[y,z]=[z,x]=0} and the geometrical relation
\eqn\elg{x^2+y^2+z^2=\mu^2/4.} In the context of the coisotropic
brane, we have to consider possible corrections to those relations.
Corrections involve the Poisson brackets, which are proportional to
$\Omega^{-1}=h z (dx\wedge dy)^{-1}$, and higher order terms
involving higher powers of $\Omega^{-1}$ and its derivatives.

It is simple to determine the corrections to the classical
relations, because they are severely constrained by (i) $SO(3,\C)$
symmetry, (ii) holomorphy, and (iii) an approximate scaling
symmetry.  Of these points, $SO(3,\C)$  symmetry requires little
explanation. By holomorphy, we mean holomorphy in $h$ and $\mu^2$
near $h=\mu^2=0$.  Holomorphy in $h$ is manifest in sigma-model
perturbation theory.  At $\mu^2=0$,  $Y$ develops a singularity at
the origin, but the ring $\CA$ is nevertheless holomorphic in
$\mu^2$ since   the corrections to the classical ring structure can
be computed in a region of field space far away from the singularity
of $Y$. Finally, if $\mu^2=0$, then $Y$ has a scaling symmetry
$(x,y,z)\to (tx,ty,tz)$, $t\in \C^*$.

For $\mu^2\not=0$, this is only an asymptotic symmetry, valid for
$(x,y,z)\to\infty$. Even if $\mu^2=0$, the algebra $\CA$ does not
have the scaling symmetry, because  $\Omega$ does not possess this
symmetry.  Still, the scaling symmetry heavily constrains the
corrections to the classical algebra generated by $x,y$, and $z$.
Under the scaling symmetry, we have $\mu^2\to t^2\mu^2$, $h\to
th$. The crucial point is that $\mu^2$ and $h$ both scale with
positive powers of $t$.

The most general deformation of \zelg\ that is allowed by holomorphy
and $SO(3,\C)$ symmetry is $[x,y]=f(\mu^2,h)z$ (and cyclic
permutations) for some function $f$.  Under scaling, $f$ must be of
degree 1, and holomorphy then implies that it is linear in $h$ and
independent of $\mu^2$.  But the term linear in $h$ can be computed
simply from the classical Poisson bracket, giving $[x,y]=h z$, and
cyclic permutations.

As for the relation \elg, holomorphy, $SO(3,\C)$ symmetry, and
scaling symmetry dictate that $h$-dependent corrections can only
take the form of a constant multiple of $h^2$.  Thus the relation
must take the form $x^2+y^2+z^2=\mu^2/4+ch^2$, for some complex
constant $c$.   This constant can be computed by going to second
order in sigma-model perturbation theory, but instead we will use an
indirect method to show in section 3.2 that $c=-1/4$.

Because of the scaling symmetry, we lose nothing if we set $h=1$.
The form of the algebra is then that $x,y,$ and $z$ obey the
$SO(3,\C)$ commutation relations
\eqn\torf{[x,y]=z,~[y,z]=x,~[z,x]=y,} and that the quadratic Casimir
operator $J^2=x^2+y^2+z^2$ is \eqn\nory{J^2={\mu^2-1\over 4}.}
For an alternative way to compute this algebra, from a different
starting point, see \CRybnikov.

The preceding statements imply that if $\B'$ is any $A$-brane,
then the space of $(\Bcc,\B')$ strings has a natural action of the
Lie algebra $\frak{so}(3,\C)$ or $\frak{sl}(2,\C)$  (generated by
$x,y,z$). The value of the quadratic Casimir is $(\mu^2-1)/4$,
independent of the choice of $\B'$.  We will see examples below;
with different choices of $\B'$, we will get different
$\frak{sl}(2,\C)$-modules, all with the same value of the
quadratic Casimir.

To get a natural group action (as opposed to an action of the Lie
algebra) requires more. The reason for this is that if $G$ is a
noncompact group, then a representation of the Lie algebra $\frak g$
of $G$ does not automatically exponentiate to an action of $G$. The
brane $\Bcc$ is $SL(2,\C)$-invariant, since its Chan-Paton curvature
${\rm Re}\,\Omega$ is invariant under $SL(2,\C)$.  A Lagrangian
$A$-brane $\B'$ will not be $SL(2,\C)$-invariant. If $\B'$ is
invariant under a subgroup $G$ of $SL(2,\C)$, then $G$ acts
naturally on the space of $(\Bcc,\B')$ strings.  We use this
starting in sections 3.2 and 3.3 to construct representations of
$SU(2)$ and $SL(2,\R)$.

Actually, something less than $G$-invariance of $\B'$ suffices to
get an action of $G$ on the space of $(\Bcc,\B')$ strings.  What we
need is only that, for all $g\in G$, the brane $(\B')^g$ obtained by
transforming $\B'$ by $g$ should be equivalent to $\B'$ in the
$A$-model. Hamiltonian isotopies of $Y$ with compact support, and
more generally those that rapidly approach the identity at infinity,
act trivially in the $A$-model. So if the $h$ action on $\B'$ can be
compensated by a Hamiltonian isotopy that has compact support, or
that approaches the identity fast enough at infinity, then $(\B')^g$
is equivalent to $\B'$ as an $A$-brane. If this holds for all $g\in
G$, then $G$ acts on the space of $(\Bcc,\B')$ strings.  For
examples, see sections 3.7-8.

\def\I{{\cal I}}
The functions $x,y$, and $z$ transform in the adjoint representation
of $SO(3,\C)$.  The polynomial functions in $x,y,$ and $z$ modulo
the commutation relations generate the universal enveloping algebra
${\cal U}$ of $SO(3,\C)$.  The ring $\CA$ is a quotient ${\cal
U}/{\cal I}$, where $\I$ is the ideal generated by
$x^2+y^2+z^2-(\mu^2-1)/4$.

Now let us discuss hermitian structures.  As explained in section
2.4, to get a hermitian structure from branes, we need an
antiholomorphic involution $\tau:Y\to Y$ that maps $\Omega$ to
$\bar\Omega$. There are essentially two choices.  The obvious choice
\eqn\zolg{\tau:(x,y,z)\to (\bar x,\bar y,\bar z)} has all the
necessary properties and breaks $SO(3,\C)$ to the compact subgroup
$SO(3)$.  Quantization of a $\tau$-invariant brane will therefore
lead to unitary representations of $SO(3)$, or its double cover
$SU(2)$.  The alternative, up to a change of coordinates, is
\eqn\nolg{\tilde\tau:(x,y,z)\to (-\bar x,-\bar y,\bar z).} This
breaks $SO(3,\C)$ to $SO(1,2)$, whose double cover is $SL(2,\R)$.
Quantization of $\tilde\tau$-invariant branes will lead to unitary
representations of $SL(2,\R)$.

\bigskip\noindent{\it Parameters Of The $A$-Model}

A nice property of this model is that $Y$ actually admits a
hyper-Kahler metric, extending the complex symplectic structure
$(I,\Omega)$.  In fact, with such a hyper-Kahler structure, $Y$ is
known as the Eguchi-Hansen manifold.  We write $I,J,K$ for the
three complex structures and $\omega_I,\omega_J,\omega_K$ for the
corresponding Kahler forms.  The hyper-Kahler metric on $Y$ is
completely determined by the three periods \eqn\labely{\eqalign{
\int_S{\omega_I\over 2\pi} & = \alpha \cr \int_S{\omega_J\over
2\pi} & = \beta \cr \int_S{\omega_K\over 2\pi} & = \gamma ,\cr}}
where $\alpha,\beta,\gamma$ are arbitrary real parameters.  These
parameters are uniquely determined up to an overall sign
$(\alpha,\beta,\gamma)\to (-\alpha,-\beta,-\gamma)$, which is
equivalent to a reversal of orientation of $S$.

The holomorphic two-form in complex structure $I$ is
$\Omega=\omega_J+i\omega_K$, so looking back to \zonk, we see that
the parameters are related by \eqn\bonk{\mu=\beta+i\gamma.} (As
before, $\mu=\pm \sqrt{\mu^2}$; the sign depends on the
orientation of $S$, which also affects the sign of $\beta$ and
$\gamma$.)

At the fixed point $\alpha=\beta=\gamma=0$, $Y$ develops an $A_1$
singularity.  In fact, according to \bonk, at
$\alpha=\beta=\gamma=0$, the equation for $Y$ reduces to
$x^2+y^2+z^2=0$, with a singularity at the origin.  From the point
of view of complex structure $I$, turning on $\alpha$ resolves the
singularity, and turning on $\beta+i\gamma$ deforms it.

The supersymmetric sigma-model with target $Y$ has a fourth
parameter, a mode of the sigma-model $B$-field.  This parameter is
\eqn\onk{\eta=\int_S{B\over 2\pi}} and takes values in $\R/\Z$. A
sign change of $\alpha,\beta,\gamma$ must be accompanied by a sign
change of $\eta$, since it involves a reversal of orientation of
$S$.  The symmetry is therefore
\eqn\hurto{(\alpha,\beta,\gamma,\eta)\to
(-\alpha,-\beta,-\gamma,-\eta).} The parameter space of the model
is therefore ${\cal W}=(\R^3\times{S}^1)/\Z_2$.

The present paper is really based on the $A$-model for symplectic
structure $\omega_K$.  This $A$-model is independent of $\alpha$
and $\beta$ (which control complex structure $K$ rather than
symplectic structure $\omega_K$) and depends holomorphically on
\eqn\zurto{\lambda=\eta+i\gamma.} Indeed, $\gamma$ is a period of
$\omega_K$, according to eqn. \labely.  As usual, the $A$-model
parameters are obtained by complexifying the periods of the
symplectic form to include $B$-field periods.

Though the $A$-model is independent of $\alpha$ and $\beta$, some
$A$-branes are conveniently defined and studied for favorable values
of those parameters.  A case in point is the problem of defining a
space-filling $A$-brane $\Bcc$ of rank 1.  The Chan-Paton bundle of
such a brane has a curvature two-form $F$ such that
$(\omega_K^{-1}(F+B))^2=-1$.  (In discussing this equation in
section 2.1, we assumed that $B=0$.)  A convenient way to solve this
equation is to pick $\alpha$ and $\beta$, so as to determine a
hyper-Kahler metric, and then require that \eqn\polyo{F+B=\omega_J.}
Dividing by $2\pi$ and integrating over $S$, this implies that
\eqn\zolyo{\int_Sc_1(\L)+\eta=\beta.} Thus, to define a coisotropic
brane in this way, we must pick $\beta$ so that $\beta-\eta$ is
equal to the integer $n=\int_Sc_1(\L)$, which was already introduced
in eqn. \lozg. For any given $A$-model parameter $\eta$, this
construction can be made for any $n$, with $\beta$ adjusted
accordingly.  So the same $A$-model has a family of coisotropic
branes depending on the integer $n$, though it is hard to use a
single hyper-Kahler metric to construct all of them.

Comparing \zolyo\ with \bonk\ and \zurto, we see that for a
convenient choice of hyper-Kahler metric, the geometrical
parameter $\mu^2$ is related to the $A$-model parameter $\lambda$
by \eqn\onx{\mu=\lambda+n.}
 In most of our discussion, we
will set $\eta=0$, and we will pick a particular coisotropic brane
with $n$ fixed. This means that we set $\beta=n$.

Now let us determine how the involutions $\tau$ and $\tilde\tau$,
which are antiholomorphic in complex structure $I$, act on the
parameters. Of course, neither $\tau$ nor $\tilde\tau$ is a symmetry
unless $\mu^2$ is real. Thus, either $\gamma=0$ and $\mu^2=\beta^2$
is non-negative, or $\beta=0$ and $\mu^2=-\gamma^2$ is non-positive.

  In general, an antiholomorphic symmetry in complex
structure $I$ reverses the sign of the Kahler form $\omega_I$. Also,
$\tau$ and $\tilde \tau$ map $\Omega=\omega_J+i\omega_K$ to
$\bar\Omega$, so they leave fixed $\omega_J$ and reverse the sign of
$\omega_K$. To be symmetries of the sigma-model, $\tau$ and $\tilde
\tau$ (which are not supposed to be orientifold symmetries; they
should preserve the orientation of a string worldsheet) must
preserve the $B$-field. The action of $\tau$ or $\tilde\tau$ on the
parameters $(\alpha,\beta,\gamma,\eta)$ is therefore
\eqn\zono{(\alpha,\beta,\gamma,\eta)\to \pm
(-\alpha,\beta,-\gamma,\eta).} (The sign depends on whether $\tau$
or $\tilde\tau$ preserve the orientation of $S$, which as we see
later can depend on the values of the parameters.) Thus, if
$\beta\not=0$, we may have $\eta\not=0$ without spoiling $\tau$ or
$\tilde\tau$ symmetry, but we must set $\alpha=\gamma=0$. And if
$\gamma\not=0$, we may have $\alpha\not=0$, but we must set
$\beta=\eta=0$.

\subsec{Representations Of $SU(2)$}

We begin by considering unitary representations. To extract from
the $A$-model a space of strings with a Hilbert space structure,
we need to pick a second $A$-brane $\B'$ that is supported at the
fixed point set of $\tau$ or $\tilde\tau$.

We first consider the case of a brane $\B'$ supported at the fixed
point set of $\tau$. This fixed point set, which we will call $M$,
consists of real $x,y,z$, obeying $x^2+y^2+z^2=\mu^2/4$. In
particular, $\mu^2=(\beta+i\gamma)^2$ must be real and positive (or
$M$ is empty or collapses to a point, which is not a Lagrangian
submanifold), so $\beta\not=0$, $\gamma=0$. $M$ is
$SU(2)$-invariant, so quantization of $(\Bcc,\B')$ strings will lead
to unitary representations of $SU(2)$.

A Lagrangian brane supported on $M$ exists only if $\eta=0$. Indeed,
if $F'$ denotes the curvature of the Chan-Paton bundle of $\B'$,
then the Lagrangian condition is $(F'+B)|_M=0$.   Given \onk\ (and
the fact that $M$ coincides with $S$ in this example), this implies
that $\int_MB/2\pi=0$ mod $\Z$, or $\eta=0$. Since also $\gamma=0$,
\onx\ tells us that \eqn\zono{\mu=n,} where
\eqn\lobal{n=\int_Mc_1(\L).} As usual $\L$ is the Chan-Paton bundle
of $\Bcc$.

As $M$ is simply-connected, the Chan-Paton bundle of $\B'$ is
trivial, so the space of $(\Bcc,\B')$ strings is related to
quantization of $M$ with prequantum line bundle $\L|_M$.

An interpretation via quantization only exists if $\omega_J$ is
nondegenerate when restricted to $M$.  This implies that $n$ must be
nonzero.  We may as well orient $M$ so that $n>0$.

Quantization of $M$ with $n$ units of flux gives a Hilbert space
$\H$ of dimension $n$, furnishing an irreducible representation of
$SU(2)$ with $j=(n-1)/2$.  The quadratic Casimir operator
$J^2=j(j+1)$ thus equals $(n^2-1)/4$.  Since $n^2=\mu^2$ according
to eqn. \zono, this is equivalent to $J^2=(\mu^2-1)/4$.  This result
agrees with \nory, and accounts for the choice of constant in this
formula.

The generators $J_x,$ $J_y,$ and  $J_z$ of $SU(2)$ arise by
quantizing the functions $x$, $y,$ and $z$.  We diagonalize $J_z$,
finding a basis of eigenstates of $J_z$, each obeying
$J_z\psi_s=s\psi_s$ for some $s$. According to a standard
analysis, in an $n$-dimensional irreducible representation of
$SU(2)$, the values of $s$ are \eqn\gg{-{n-1\over 2},1-{n-1\over
2},2-{n-1\over 2},\dots,{n-1\over 2}.} The range of eigenvalues of
$J_z$ is in accord with the classical fact that on the brane
characterized by $x^2+y^2+z^2=n^2/4$ (with real $x,y,z$) $z$  is
bounded by $-n/2\leq z\leq n/2$. Because of quantum mechanical
fluctuations, the maximal value of $|J_z|$ in the quantum theory
(namely $(n-1)/2$) is slightly less than the classical upper bound
$n/2$.

\subsec{Discrete Series Of $SL(2,\R)$}

To construct unitary representations of $SL(2,\R)$, we proceed in
a similar way, but now we consider $\tilde\tau$-invariant branes.
It is convenient to make a change of coordinates $x\to ix$, $y\to
iy$, so that the equation defining $Y$ becomes
\eqn\orf{-x^2-y^2+z^2={\mu^2\over 4},} and $\tilde\tau$ acts
simply by \eqn\zorf{(x,y,z)\to(\bar x,\bar y,\bar z).}

We continue to assume that $\eta=0$, and to ensure that $Y$ admits
the symmetry \zorf, we take $\mu^2$ real.  We begin with the case
$\mu^2>0$, which means that $\gamma=0$, $\beta\not=0$.  Just as in
the last subsection, existence of the brane $\Bcc$  requires us to
take $\beta$ to be an integer.  This integer is
$n=\int_Sc_1(\L)$, where now $S\subset Y$ is given by imaginary
$x$ and $y$ and real $z$. As in section 3.2, we may as well take
$n\geq 0$.  However, in section 3.2, we required $n>0$ to ensure
that $\omega_J$ remains nondegenerate when restricted to $S$.  For
our present purposes, this is irrelevant (we will not consider a
brane wrapped on $S$), so we allow $n=0$.  For $n=0$, the equation
defining $Y$ reduces to $-x^2-y^2+z^2=0$, with an $A_1$
singularity at $z=y=z=0$, so we should expect that some unusual
behavior may occur at that point.

The fixed point set of $\tilde\tau$ consists of real $x,y,z$.  The
equation $z^2=\mu^2/4+x^2+y^2$ shows that this fixed point set has
two components $M_\pm$, given by \eqn\deuf{z=\pm
\sqrt{\mu^2/4+x^2+y^2}.} Each component is equivalent to the
complex upper half-plane with the usual action of $SL(2,\R)$. In
particular, the components $M_\pm$ are simply-connected, and each
supports a unique rank 1 $A$-brane $\B'_\pm$.

We quantize $M_\pm$ by taking the space of $(\Bcc,\B'_\pm)$ strings.
This should give a unitary representation $D_n^\pm$ of $SL(2,\R)$
with the same value of the quadratic Casimir that we found in
section 3.2, namely $J^2=(n^2-1)/4$.  (The value of the quadratic
Casimir must be the same because it is determined entirely by the
choice of the space $Y$ and the brane $\Bcc$, independent of the
choice of  a second $A$-brane $\B'$.) On the classical phase space
$M_\pm$, the range of values of $z$ is $n/2\leq z\leq \infty$ on
$M_+$ and $-\infty \leq z\leq -n/2$ on $M_-$.  Hence in the
representation $D_n^+$, $J_z$ should be unbounded above, but bounded
below by approximately $n/2$; and in the representation $D_n^-$,
$J_z$ should be unbounded below, but bounded above by approximately
$-n/2$.

Unitary representations of $SL(2,\R)$ with these properties do exist
and are known as the discrete series.  The precise set of values of
$J_z$ is \eqn\humbo{{n+1\over 2},~1+{n+1\over 2},~ 2+{n+1\over
2},\dots} for $D_n^+$ and \eqn\zumbo{-{n+1\over 2},~-1-{n+1\over
2},~-2-{n+1\over 2},\dots} for $D_n^-$. We will explain these
formulas  in section 3.5 after describing the principal series.

\subsec{Principal Series Of $SL(2,\R)$}

Now we consider the case $\mu^2<0$, that is $\beta=0$,
$\gamma\not=0$.  The fixed point set $M$ of $\tilde\tau$ is now
given by the equation \eqn\lson{x^2+y^2={\gamma^2\over 4}+z^2,}
with $x,y$, and $z$ real. It is connected, with the topology of
$\R\times S^1$ (for $\gamma\not=0$). $z$ is unbounded above and
below, so in a representation  obtained by quantizing $M$, $J_z$
is similarly unbounded.

The first Betti number of $M$ is 1, so there is a one-parameter
family $\B'_\delta$ of rank 1 $A$-branes supported on $M$; roughly
speaking, the monodromy of the Chan-Paton bundle of $\B'_\delta$
around the circle in $M$ is $\exp(2\pi i\delta)$. To be more
precise, we recall that the Chan-Paton bundle of an $A$-brane is
actually a flat ${\rm Spin}_c$ bundle (not a flat line bundle), so
we actually need to pick a spin structure in order to define
$\delta$ as a number. To define $\delta$ precisely, we declare
that $\delta=0$ corresponds to the Ramond spin structure (the one
that corresponds to a trivial real line bundle over a circle).

Topologically, $\R\times S^1$ is the same as the cotangent bundle
$T^*S^1$. We explain below how to make contact between the present
problem and quantization of $M=T^*S^1$ with its standard
symplectic structure. For now, we just proceed informally.  In
quantizing $M=T^*S^1$ via functions (actually half-densities) on
$S^1$, a wavefunction picks up a phase $\exp(2\pi i\delta)$ in
going around $S^1$.  As a result, the eigenvectors of $J_z$ are
exponentials $\exp(i(n+\delta)\theta)$, where $\theta$ is an
angular variable on $S^1$, and $n$ is any integer. The spectrum of
$J_z$ is thus of the form $\{\delta+n|n\in \Z\}$, and is unbounded
above and below, as expected.

To get a representation of $SL(2,\R)$, the eigenvalues of $J_z$
must   be integers or half-integers, and so $\delta$ must be 0 or
$1/2$. For generic $\delta$, we get a representation not of
$SL(2,\R)$, but of its universal cover, which we denote
$\widetilde{SL}(2,\R)$.

The representations obtained from this construction are known as
the principal series representations $P_{\gamma,\delta}$.  These
are representations of $SL(2,\R)$ (or its universal cover) with
quadratic Casimir \eqn\zolf{J^2=-{\gamma^2+1\over 4},} for real
$\gamma$. The spectrum of $J_z$ is $\{\delta+n|n\in\Z\}$, as we
described above, and we get a representation of $SL(2,\R)$ (as
opposed to a cover) precisely if $\delta=0$ or $1/2$.  The central
element ${\rm diag}(-1,-1)$ of $SL(2,\R)$, which corresponds to a
$2\pi$ rotation of the circle, acts by $\exp(2\pi i\delta)$.

A standard algebraic analysis (which we will essentially explain
in section 3.5) shows that $P_{\gamma,\delta}$ is irreducible
unless $\gamma=0$ and $\delta=1/2$.  In that case, one gets
\eqn\lopo{P_{0,\half}=D_0^+\oplus D_0^-,} where $D_0^\pm$ is the
$n=0$ case of the discrete series representations $D_n^\pm$ of
subsection 3.3.  Let us try to understand this decomposition from
the present point of view.

If we set $\gamma=0$, $M$ reduces to the cone $z^2=x^2+y^2$.  It
is a union of two components  with $z\geq 0$ and $z\leq 0$,
respectively.  These are the specialization to $n=0$ of the two
symplectic manifolds $M_+$ and $M_-$ whose quantization led to the
discrete series.  Topologically, each of $M_+$ and $M_-$ is a
disc, with a unique flat ${\rm Spin}_c$ structure that actually is
a spin structure (since $M_\pm$ are simply-connected).  On a
circle at infinity, this spin structure corresponds to the
bounding or Neveu-Schwarz spin structure, which we get at
$\delta=1/2$.  So it is only at $\delta=1/2$ that a brane
supported on $M$ decomposes (when we set $\gamma$ to zero) as the
sum of a brane supported on $M_+$ and one supported on $M_-$.  At
any other value of $\delta$, $M_+$ and $M_-$ are linked by a
monodromy at the origin.

\bigskip\noindent{\it The Singularity At $\gamma=0$}

At $\gamma=0$,  $Y$ is described by $x^2+y^2-z^2=0$, with an $A_1$
singularity at the origin.  To understand better this case, we
exploit the fact that the parameter $\alpha$ is trivial in the
$A$-model. We can turn on $\alpha$ and remove the singularity
without changing the $A$-model.  Moreover, since we here have
$\beta=0$, this does not disturb $\tilde\tau$ symmetry.

Turning on $\alpha$ has the effect of blowing up the singularity.
A convenient way to describe the blowup is as follows.  Consider a
space $\C^3$, with coordinates $(a,b,p)$, subject to a scaling
symmetry $(a,b,p)\to (ta,tb,t^{-2}p)$, $t\in \C^*$.  The ring of
$\C^*$-invariant functions is generated by
\eqn\xun{\eqalign{x+z&=a^2p,\cr x-z&=-b^2p,\cr y&=abp,\cr}}
subject to one relation $x^2+y^2-z^2=0$. Let $Z$ be the locus in
$\C^3$ consisting of points in which $a$ and $b$ are not both
zero.  The quotient $Y'=Z/\C^*$ has a natural map to $Y$, given by
\xun.  This map is one-to-one except that the inverse image of the
origin in $Y$ is a copy of $\Bbb{CP}^1$ given by $p=0$.  Thus,
$Y'$ is obtained (in complex structure $I$) by blowing up the
singularity of $Y$ at the origin. The holomorphic two-form of
$Y'$, which corresponds to $\Omega$ under the above map, is
$\Omega'=(a\,db-b\,da)\wedge dp-2p\,da\wedge db$.

The Lagrangian submanifold $M'$ that corresponds to $M$ is defined
by taking $a,b,p$ real.  To describe $M'$ explicitly, we use the
scaling symmetry to set $a^2+b^2=1$.  This leaves a freedom of
reversing the sign of $a$ and $b$, so we set  $a=\cos(\theta/2)$,
$b=\sin(\theta/2)$ for some angle $\theta$.  The symplectic
structure with which we need to quantize $M'$ is $\omega'_J={\rm
Re}\,\Omega'= dp\wedge d\theta$.  This is the standard symplectic
structure  of $T^*S^1$. Moreover, the action of $SL(2,\R)$ on
$T^*S^1$ is the natural action coming from the usual action of
$SL(2,\R)$ on $S^1\cong \Bbb{RP}^1$. Quantization leads to
half-densities on $S^1$, with the natural action of $SL(2,\R)$ on
the space of half-densities. If we take $\delta\not=0$, we get
half-densities twisted by a flat bundle of monodromy $\exp(2\pi
i\delta)$.  This leads to the spectrum of $J_z$ described above.
(In this description, the reducibility of the $SL(2,\R)$
representation at $\delta=1/2$ is not clear.)

When $\gamma\not=0$, $M$ is smooth and the blowup via $\alpha$ is
not necessary.  In this case, by setting
$(x,y,z)=\half\gamma(\cosh p \cos\theta,\cosh p\sin\theta,\sinh
p)$, we can identify $M$ with $T^*S^1$ with its usual symplectic
structure.  This identification commutes with the group of
rotations of the circle, so it can be used to determine the
spectrum of $J_z$, but it does not commute with $SL(2,\R)$, so it
does not directly determine the $SL(2,\R)$ representation.  It can
be shown that the representation $P_{\gamma,\delta}$ can be
interpreted as the space of densities of weight $1/2+i\gamma$
twisted by a flat line bundle with monodromy $\exp(2\pi i\delta)$,
with the natural action of $\tilde{SL}(2,\R)$ on this space.  One
way to show this is to use the fact that $M$ has an
$SL(2,\R)$-invariant map to $S^1$ with Lagrangian fibers; in other
words, $M$ admits an $SL(2,\R)$-invariant real polarization.  (In
fact, $M$ has two such polarizations.)

\subsec{Algebraic Description}

To better understand some things that we have already encountered
and in preparation for what follows, we will summarize some
standard facts about representations of $\frak{sl}(2)$ from an
algebraic point of view.

In our construction, quantization of the functions $x,y$, and $z$
gives operators $J_x,J_y,J_z$ that obey the usual $\frak{sl}(2)$
commutation relations such as $[J_x,J_y]=iJ_z$. As usual, it is
convenient to set $J_\pm = J_x\pm iJ_y$. For unitary
representations of $\frak{su}(2)$, $J_z$ is hermitian and $J_+$ is
the hermitian adjoint of $J_-$.  For unitary representations of
$\frak{sl}(2,\R)$, $J_z$ is hermitian and $J_+$ is minus the
adjoint of $J_-$.  We summarize the facts about $J_\pm$:
\eqn\zorfo{J_+^\dagger=\cases{ J_- & for $\frak{su}(2)$ \cr
                               -J_- & for $\frak{sl}(2,\R)$. \cr}}
We will consider both unitary representations and representations
that are not required to be unitary.  In the unitary case, since
$J_z$ is hermitian, it can be diagonalized with real eigenvalues,
but in any event, we only consider representations in which $J_z$
can be diagonalized.  We also assume that the quadratic Casimir
operator can be diagonalized, and we often write simply $J^2$ for
its eigenvalue.

\def\s{s}

Suppose that $\psi$ is an eigenvector of $J_z$, with
$J_z\psi=\s\psi$ for some $\s$.  Additional eigenvectors can be
constructed by acting with $J_+$ or $J_-$.  This process may
continue indefinitely in both directions, giving a representation
in which $J_z$ (or more exactly the real part of its eigenvalue)
is unbounded above and below. Alternatively, it may terminate in
either or both directions, if we find a highest weight vector
$\psi$ obeying \eqn\yert{J_z\psi=s_+\psi, ~J_+\psi=0,} or a lowest
weight vector $\tilde\psi$ obeying
\eqn\zert{J_z\tilde\psi=s_-\tilde\psi,~J_-\tilde\psi=0.}

In an irreducible representation, the quadratic Casimir operator
$J^2=J_x^2+J_y^2+J_z^2$ is a multiple of the identity.  It can be
written \eqn\gert{\eqalign{J^2=&J_z^2+J_z+J_-J_+\cr
=&J_z^2-J_z+J_+J_-.\cr}} If \yert\ is obeyed, we use the first
formula in \gert\ to deduce that \eqn\noble{J^2=s_+(s_++1),} and
if \zert\ is obeyed, we use the second formula to deduce that
\eqn\tobel{J^2=s_-( s_--1).} If \noble\ is obeyed for some value
of $s_+$, then \tobel\ is obeyed with \eqn\obel{s_-=s_++1.} Thus,
it is possible to have a representation of highest weight $s_+$ if
and only if it is possible for the same value of the quadratic
Casimir to have a representation of lowest weight $s_++1$. This
explains the relation between the bounds on $J_z$ for
finite-dimensional representations in \gg\ and the corresponding
bounds for discrete series representations in \humbo\ and \zumbo.

If we assume unitarity, we can learn a little more.  A vector
$\psi$ obeying $J_z\psi=s_+\psi$, where $s_+$ obeys \noble, must
be annihilated by $J_+$. Indeed, we have
$J_-J_+\psi=(J^2-J_z^2-J_z)\psi=0$.  So $0=(\psi,J_-J_+\psi)$.  In
the unitary case, the right hand side is $\pm (J_+\psi,J_+\psi)$,
and its vanishing implies that $J_+\psi=0$.  Similarly, if
$J_z\tilde\psi=s_-\tilde\psi$, and $s_-$ obeys \tobel, then in the
unitary case, it follows that $J_-\tilde\psi=0$.

If $\psi$ is a highest weight vector with $J_z\psi=s_+\psi$, then
$s_+\geq 0$ for a unitary representation of $\frak{su}(2)$, and
$s_+\leq 0$ for a unitary representation of $\frak{sl}(2,\R)$. The
two statements can be combined to $\epsilon s_+\geq 0$, where
$\epsilon=1$ for $\frak{su}(2)$ and $\epsilon=-1$ for
$\frak{sl}(2,\R)$. To see this, we note that $J_+\psi=0$ implies
that
$2s_+(\psi,\psi)=(\psi,2J_z\psi)=(\psi,[J_+,J_-]\psi)=(\psi,J_+J_-\psi)=\epsilon
(J_-\psi,J_-\psi)$, where \zorfo\ was used in the last step. This
indeed implies that $\epsilon s_+\geq 0$, with equality only if
$J_-\psi=0$, which implies that $\psi$ generates a one-dimensional
trivial representation.  A similar argument shows that a lowest
weight vector has $s_-\leq 0$ in the case of a unitary
representation of $\frak{su}(2)$, and $s_-\geq 0$ in the case of a
unitary representation of $\frak{sl}(2,\R)$.

A standard fact, which we will not review here, is that every
irreducible representation of $SU(2)$ has both a highest weight
vector $\psi$ and a lowest weight vector $\tilde\psi$. Moreover,
we can assume that $\tilde\psi=J_-^n\psi$ for some integer $n$. If
$J_z\psi=s_+\psi$, then $J_z\tilde \psi=(s_+-n)\tilde\psi$.  The
quadratic Casimir is $J^2=s_+(s_++1)=(s_+-n)(s_+-n-1)$, implying
that $s_+=n/2$ is an integer or a half-integer.

For the case of a non-trivial irreducible unitary representation of
$\frak{sl}(2,\R)$, it is impossible to have both a lowest weight
vector and a highest weight vector.  The $J_z$ eigenvalues would
have to be negative for the highest weight vector, and positive for
the lowest weight vector; but a highest weight vector has a higher
weight than any other vector in an irreducible representation. This
being so, we cannot make an argument like the one in the last
paragraph, and there is no way to show algebraically that the
eigenvalues of $J_z$ take values in $\Z/2$. When the eigenvalues are
valued in $\Z/2$, we get a representation of the group $SL(2,\R)$,
while the more general case leads to representations of its
universal cover. In sections 3.3-4, we have encountered unitary
representations of $SL(2,\R)$ with neither a highest weight vector
nor a lowest weight vector (the principal series), as well as
unitary representations with a lowest weight vector (the discrete
series $D_n^+$) or a highest weight vector ($D_n^-$).

\bigskip\noindent{\it Relaxing Unitarity}

Now let us consider representations that are not necessarily
unitary.  The quadratic Casimir  can act as a complex number,
which we simply call $J^2$, and likewise the eigenvalues of $J_z$
can be complex. Consider a representation that contains a vector
$\psi$ with $J_z\psi=s\psi$ for some $s$. If $s$ and $J^2$ are
generic, then there does not exist an integer $n$ such that $J^2$
is equal to $(s+n)(s+n+1)$. In this case, none of the states
$J_+^n\psi$ or $J_-^m\psi$ have eigenvalues obeying \noble\ or
\tobel, so these states are all nonzero.  So a representation with
generic $s$ and $J^2$ is infinite in both directions, like the
principal series representations.

For special values of $s$ and $J^2$, something special can happen.
If $J^2=s(s+1)$, it is possible to have a representation $R^-$
spanned by vectors $\psi_n$, $n=0,-1,-2,\dots$, with
$J_z\psi_n=(s+n)\psi_n$. This representation has a highest weight
vector $\psi_0$ with $J_z\psi_0=s\psi_0$.  It is also possible to
have a representation $R^+$ spanned by vectors $\psi_n$,
$n=1,2,3\dots$, again with $J_z\psi_n=(s+n)\psi_n$.  This
representation has a lowest weight vector $\psi_1$ with
$J_z\psi_1=(s+1)\psi_1$.

Now let us consider a general representation $R$ spanned by
vectors $\psi_n$, $n\in \Z$, obeying $J_z\psi_n=(s+n)\psi_n$.
Suppose that $J_+\psi_{n}=a_n\psi_{n+1}$,
$J_-\psi_{n+1}=b_n\psi_n$, for some complex constants $a_n,b_n$.
Again, write $J^2$ for the eigenvalue of the quadratic Casimir
operator. Using \gert, we deduce that for all $n$,
\eqn\zolf{a_nb_n=J^2-(s+n)(s+n+1).} This is a necessary and
sufficient condition to get a representation of the Lie algebra
$\frak{sl}(2,\C)$ with the assumed value of the Casimir. For every
$n$, we are free to redefine $\psi_n\to\lambda_n\psi_n$, for
$\lambda_n\in \C^*$, along with
\eqn\olg{a_n\to\lambda_n\lambda_{n+1}^{-1}a_n,~ b_n\to
\lambda_{n+1}\lambda_n^{-1}b_n.} As long as $J^2$ and $s$ are such
that $J^2-(s+n)(s+n+1)$ never vanishes, $a_n$ and $b_n$ are
nonzero and are uniquely determined, modulo a transformation of
the type \olg, by the equations \zolf. This gives an irreducible
representation for the assumed values of $J^2$ and $s$,
generalizing the principal series.

Suppose on the other hand that $J^2-(s+n)(s+n+1)=0$ for some value
of $n$. Then we have $a_nb_n=0$, leaving three choices:

(1) We may have $a_n=b_n=0$.  This gives a representation $R$ that
decomposes as a direct sum $R=R_1\oplus R_2$, where $R_1$ has a
highest weight vector $\psi_{n}$ and $R_2$ has a lowest weight
vector $\psi_{n+1}$.

(2) We may have $b_n=0$, $a_n\not=0$.  This gives a representation
$R$ that contains a subrepresentation $R_2$ spanned by vectors
$\psi_m$, $m\geq n+1$.  $R_2$ has a lowest weight vector
$\psi_{n+1}$. There is no complementary representation $R_1$, but
rather $R$ can be described as an extension: \eqn\onxop{0\to
R_2\to R\to R_1\to 0.} Here $R_1$ is spanned by $\psi_m$, $m\leq
n$, but is not a subrepresentation of $R$, since $a_n\not=0$ and
$\psi_n$ is not a highest weight vector.  However, $R_1$ can be
understood as a quotient representation $R/R_2$.

(3) Finally, we may have $a_n=0$, $b_n\not=0$.  This gives an
extension in the opposite direction: \eqn\ponxop{0\to R_1\to R\to
R_2\to 0.}  Here $R_1$ is the subrepresentation spanned by $\psi_m$,
$m\leq n$, and containing the highest weight vector $\psi_n$.  $R_2$
is the quotient representation $R/R_1$.

The phenomenon just described involving non-split extensions does
not occur for  representations of either real form of
$\frak{sl}(2)$ that admit a nondegenerate hermitian form, since in
that case $b_n=\pm \bar a_n$ in view of \zorfo.   What nonsplit
extensions mean in terms of branes will be described in section
3.7.

If we further specialize $s$ and $J^2$, the equation
$J^2=(s+n)(s+n+1)$ may have two integer solutions, say $n_1$ and
$n_2$.  In this case, we run into the two equations
$a_{n_1}b_{n_1}=0$ and $a_{n_2}b_{n_2}=0$, and we can
independently choose, in each of the two cases, which if either of
$a$ and $b$ is nonzero.

\bigskip\noindent{\it Hermitian Structure}

Now let us specialize this to the case of a representation of
$\frak{sl}(2,\R)$ compatible with a hermitian structure $(~,~)$.
The relations $J_+\psi_n=a_n\psi_{n+1}$, $J_-\psi_{n+1}=b_n\psi_n$
imply $(\psi_{n+1},J_+\psi_n)=a_n(\psi_{n+1},\psi_{n+1})$,
$(\psi_n,J_-\psi_{n+1})=b_n(\psi_n,\psi_n)$.  If
$J_-=-J_+^\dagger$, then $(\psi_n,J_-\psi_{n+1})$ is minus the
complex conjugate of $(\psi_{n+1},J_+\psi_n)$, so we get
\eqn\ppome{a_nb_n=-{|(\psi_{n+1},J_+\psi_n)|^2\over
(\psi_{n+1},\psi_{n+1})(\psi_n,\psi_n)}.}

For a unitary representation, $(\psi_n,\psi_n)>0$ for all $n$. In
this case, \ppome\ implies that $a_nb_n<0$ (unless $\psi_n$ is a
highest weight vector).  Let us apply this to a unitary
representation in which the set of $J_z$ eigenvalues is $\{s+n|n\in
\Z\}$ for some real $s$.  Together with the fact that $a_nb_n<0$,
\zolf\ implies that $(s+n)(s+n+1)>J^2$ for all $n$, or
\eqn\togo{(s+n+1/2)^2>J^2+1/4.} For example, let us consider
representations of $SL(2,\R)$ in which $s=0$ so that the central
element ${\rm diag}(-1,-1)$ acts trivially.  The condition that
$(n+1/2)^2>J^2+1/4$ for all integers $n$ is equivalent to $J^2< 0$.
We have already encountered the relevant unitary representations
with $J^2\leq -1/4$.  These are the principal series representations
with $J^2=-(\gamma^2+1)/4$.  The unitary representations with
$0>J^2>-1/4$ are known as the complementary series.  Algebraically,
the complementary series is simply the analytic continuation of the
principal series to $J^2>-1/4$ or imaginary $\gamma$.  (Eqn. \zolf\
with the equivalence relation \olg\ give a general description for
all $J^2$ and $s$.)  However, natural geometric realizations of the
hermitian structure undergo a sort of phase transition at
$J^2=-1/4$.  For an explanation of how this happens (in a standard
description of the representations via densities of suitable weight
on $S^1$), see chapter 1.3 of \gelf. In terms of branes, we
described the principal series in section 3.4, and we will describe
the complementary series in section 3.8.

A representation of $\frak{sl}(2,\R)$ with any real $J^2$ and $s$
can admit a nondegenerate hermitian structure, though not
necessarily a positive-definite one. Let us see what happens for
$s=0$ and $J^2>-1/4$.
  If there is an integer $n$
such that $J^2=n(n+1)$, the representation has a lowest weight or
highest weight vector, as we have seen above.  Let us suppose that
this is not the case.  Eqns. \zolf\ and \ppome\ imply that
\eqn\overto{(\psi_n,\psi_n)(\psi_{n+1},\psi_{n+1})=
\bigl(n(n+1)-J^2\bigr)\bigl|(\psi_{n+1},J_+\psi_n) \bigr|^2.} Let
$n_0$ be the largest integer such that $n_0(n_0+1)-J^2<0$. Then
the condition $n(n+1)-J^2<0$ is obeyed for $2n_0+2$ values of $n$,
namely $-n_0-1\leq n\leq n_0$.  From \overto, it follows that if
$n$ is in this range, then $(\psi_{n+1},\psi_{n+1})$ and
$(\psi_n,\psi_n)$ have opposite signs.  The total number of sign
changes is even, so if $(\psi_n,\psi_n)$ is positive for  large
positive $n$, then it is also positive for large negative $n$.
However, the $n_0+1$ states $\psi_{n_0-2k}$, $0\leq k\leq n_0$,
have negative norm.

A similar analysis can be made for other values of $s$.  For
brevity, we consider only the question of $s=1/2$, which is
associated to representations of $SL(2,\R)$ in which the central
element acts as $-1$.  For $s=1/2$, \togo\ tells us that $-1/4\geq
J^2$, which is the range covered by the principal series.  So
there are no new unitary representations to be had.  There are, of
course, representations with indefinite but nondegenerate
hermitian forms. For $J^2>-1/4$ and $s=1/2$, the number of sign
changes is odd, so if $(\psi_n,\psi_n)$ is positive for large
positive $n$, it is negative for large negative $n$.  In the
region near $n=0$, the signs alternate.

\subsec{Discrete Series Of $\tilde{SL}(2,\R)$}

It is now straightforward to complete the description of the
discrete series. For a unitary representation,
$\mu^2=(\beta+i\gamma)^2$ must be real, so either $\beta$ or
$\gamma$ vanishes.  We have already considered in section 3.4 the
case of $\beta=0$ with generic $\gamma$.  So here we take
$\gamma=0$ with generic $\beta$.

{}From \zolyo, we have $\beta=\eta+n$, where $n=\int_Sc_1(\L)$ is
an integer and $\eta$ is the world-sheet theta-angle.  Thus, to
get generic $\beta$, we must take $\eta$ to be nonzero.
 The quadratic Casimir is $J^2=(\beta^2-1)/4$,
just as in the case of integer $\beta$.
 As in section
3.2, we introduce a second $A$-brane $\B'_\pm$ supported on the
locus $M_\pm$ defined by $z=\pm\sqrt{\mu^2+x^2+y^2}$.  (The
Chan-Paton bundle of this brane is determined up to isomorphism by
the requirement that $F+B|_M=0$.)  Classically, on the support of
this brane, $z$ is bounded below or above by $\mu$.

Quantum mechanically, there must be a lowest weight or highest
weight vector.  The $J_z$ eigenvalue $s_-$ of a lowest weight
vector is determined by $s_-(s_--1)=J^2=(\beta^2-1)/4$ or
\eqn\yoffe{s_-={\beta+1\over 2}.}  The spectrum is exactly as in
\humbo, with $n$ replaced by $\beta$.  For non-integral $n$, $J_z$
is no longer a half-integer.  What we get this way is a unitary
representation of the universal cover $\tilde{SL}(2,\R)$ of
$SL(2,\R)$.  It is the generalization of the discrete series to
$\tilde{SL}(2,\R)$.  See \kostanttwo\ for more.

\subsec{Harish-Chandra Modules From Branes}

 Our next goal is to describe in terms of branes the not
necessarily unitary representations that were described
algebraically in section 3.5, for example in eqns. \zolf\ and
\olg.  These are representations in which the spectrum of $J_z$ is
of the form $s+n$, for some complex constant $s$ and for $n$
ranging over a sequence of consecutive integers that may be
infinite, semi-infinite, or finite.  Such representations are
known mathematically as Harish-Chandra modules.  (Sometimes one
considers only the case $2s\in \Z$, corresponding to
representations of $SL(2,\R)$ rather than a cover, but we will not
make this restriction.)

In general, Harish-Chandra modules are not unitary, so we will
have to describe them via branes that are not
$\tilde\tau$-invariant. Harish-Chandra modules do admit an action
of the group $SL(2,\R)$, though in general not a unitary action.
This suggests that we might describe them via $SL(2,\R)$-invariant
branes, but in fact, the full list of $SL(2,\R)$-invariant
Lagrangian submanifolds of $Y$ is very short and we have exhausted
it already. However, since sufficiently well localized Hamiltonian
isotopies act trivially in the $A$-model, it suffices for our
purposes to consider branes that are $SL(2,\R)$-invariant only
asymptotically. It turns out that there is a sufficient supply of
these.

Since $J_z$ can be naturally diagonalized in these modules, it is
reasonable to guess that one can choose the brane $\B'$ to be
invariant under the subgroup $\K\cong U(1)=SO(2)$ generated by
$J_z$.  $\K$ is a maximal compact subgroup of $SL(2,\R)$.

As we will see, it is possible to describe all $\K$-invariant
Lagrangian submanifolds of $Y$. It is convenient to use the fact
that $Y$ admits a family of hyper-Kahler metrics with parameters
$(\alpha,\beta,\gamma)$, as described in section 3.1.  The
hyper-Kahler metric is not $SL(2,\C)$-invariant, but it is
invariant under the maximal compact subgroup $SU(2)\subset
SL(2,\C)$ and in particular under the maximal compact subgroup
$\K=U(1)\subset SL(2,\R)$. The moment map for the $\K$ action
gives three natural functions on $Y$, which we denote as $\vec w$.
(Thus the components of $\vec w$ are the moment maps for $\K$ with
respect to the three symplectic structures
$\vec\omega=(\omega_I,\omega_J,\omega_K)$.) Mapping a point in $Y$
to the corresponding value of $\vec w$ gives a map $\pi:Y\to\R^3$,
which away from the fixed points of $\K$ is a fibration with $S^1$
fibers. The hyper-Kahler metric of $Y$ takes the form
\eqn\ghmetric{ ds^2 = H d \vec w \cdot d \vec w + H^{-1} (d \chi +
\vec a)^2 }
where  $\chi \cong \chi+2\pi$ is an angular parameter along the
fibers of $\pi$. The group $\K$ acts by translation of $\chi$ or
in other words by rotation of the fibers. $H$ is the harmonic
function \eqn\huryt{ H = {1 \over 2 |\vec w - \vec w^*|} + {1
\over 2 |\vec w + \vec w^*|}.} The moduli of the hyper-Kahler
metric are contained in the choice of point $\vec w^*\in \R^3$:
\eqn\gurt{\vec w^*=\half(\alpha,\beta,\gamma).}  This formula will
be explained shortly.
 At the two points $\vec w=\pm \vec w^*$,
$H^{-1}$ vanishes and the fibers of the map $Y\to\R^3$ collapse to
points. Those points are the fixed points of $\K$.  Away from
those two points, the map $Y\to \R^3$ is a circle fibration and
$\vec a$ is a connection on this fibration with curvature
\eqn\herft{ \vec \nabla \times \vec a = \nabla H.} What we have
just described is the Gibbons-Hawking form of the Eguchi-Hansen
metric.

\ifig\ghmetricfig{The space of $\vec w \in \R^3$. The
tri-holomorphic  action of $\K\cong U(1)$ has fixed points at the
two ``Taub-NUT centers,'' $\vec w = \pm \vec w^*$, where $\vec w^* =
\half(\a,\b,\g)$.  The inverse image in $Y$ of a straight line
connecting these points is a two-sphere $S$.}
{\epsfxsize2.5in\epsfbox{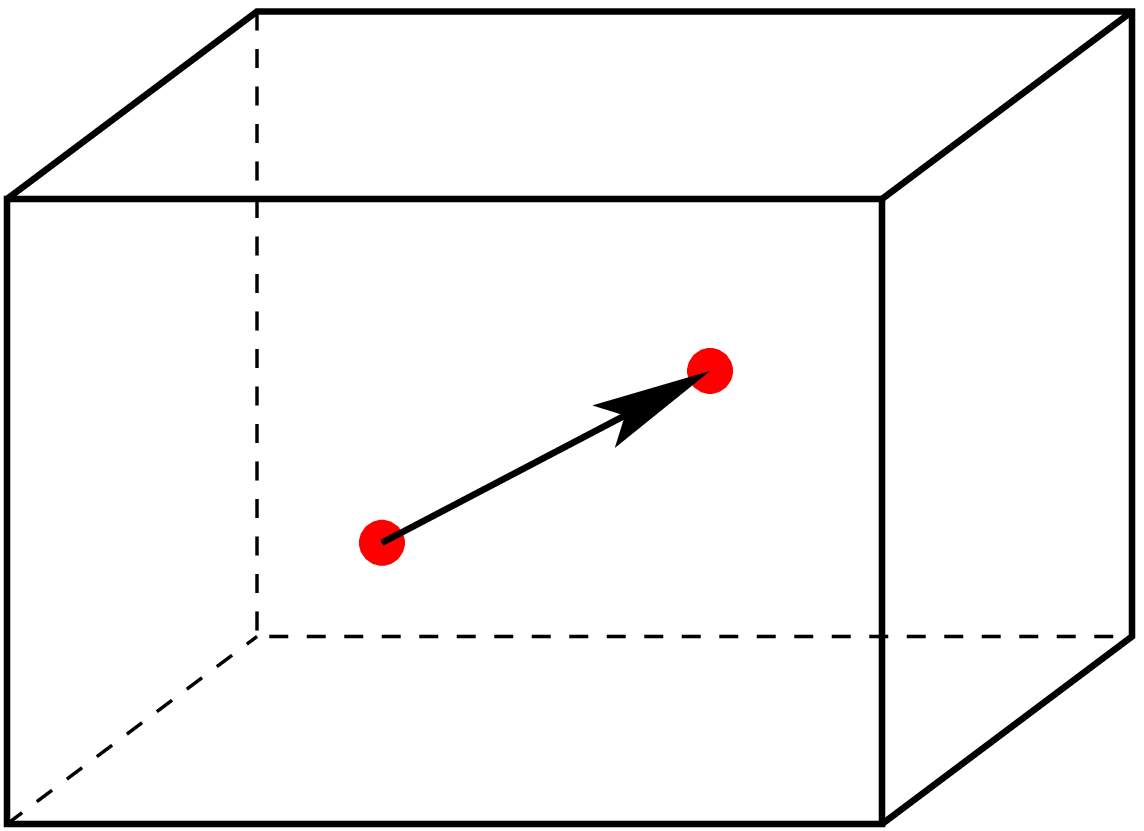}}

The three Kahler forms on $Y$ are
\eqn\ghomegaforms{ \vec \omega = (d \chi + \vec a\cdot d\vec
w)\wedge d \vec w - {1 \over 2} H d \vec w \times d \vec w }
where $(d \vec w \times d \vec w)_i = \epsilon_{ijk} d w_j \wedge
dw_k$.  As a check, we note that with the action of $\K$
corresponding to the vector field $\partial/\partial\chi$, the
moment map for $\K$ is indeed \eqn\zong{\vec\mu=\vec w,} as
promised. A convenient choice of a two-surface $S$ that generates
the second homology of $Y$ is (\ghmetricfig) the inverse image in
$Y$ of a straight line in $\R^3$ from $\vec w=-\vec w^*$ to $\vec
w=\vec w^*$.  A short calculation shows that
$\int_S\vec\omega/2\pi =2\vec w^*$, justifying the formula \gurt\
for the moduli.

Finally, we want to understand the action of $\tau$ and
$\tilde\tau$. For $\tau$ or $\tilde\tau$ to be a symmetry, they
must either (1) leave fixed the two points $\vec w=\pm \vec w^*$,
which occurs if $\alpha=\gamma=0$, or (2) exchange them, which
occurs if $\beta=0$.

$\tau$ and $\tilde\tau$ both commute with $\K$, and transform the
symplectic forms by $(\omega_I,\omega_J,\omega_K)\to
(-\omega_I,\omega_J,-\omega_K)$.  Since $\vec w$ is the moment
map, this means that $\tau$ and $\tilde\tau$ acts on the base of
$Y\to\R^3$ by $(w_1,w_2,w_3)\to (-w_1,w_2,-w_3)$. Consequently,
any $\K$-invariant $A$-brane that is pointwise $\tau$ or
$\tilde\tau$-invariant must project under $\pi:Y\to\R^3$ to the
$w_2$ axis, or a piece of it.

Given that $\tau$ and $\tilde\tau$ commute with rotations of the
fiber and square to 1, they can, roughly speaking, only act on
$\chi$ by $\chi\to\chi$ or $\chi\to \chi+\pi$.  Let us first
discuss case (2), with $\beta=0$.  This is the case that is
related to the principal series, and we know from section 3.4 that
there is a (pointwise) $\tilde\tau$-invariant $A$-brane with
topology $\R\times S^1$. This brane must project to the full $w_2$
axis. Hence, $\tilde\tau$ must leave $\chi$ fixed in this case. On
the other hand, for $\beta=0$, $\gamma\not=0$, there is no such
$\tau$-invariant $A$-brane, so $\tau$ must act by
$\chi\to\chi+\pi$.  Thus, for $\beta=0$, we have
\eqn\gnx{\eqalign{\tau:\,&\chi\to\chi+\pi\cr
                  \tilde\tau:\,&\chi\to\chi.\cr}}

On the other hand, consider the case $\alpha=\gamma=0$, $\beta>0$.
The two fixed points are at $(w_1,w_2,w_3)=\pm\half(0,\beta,0).$
{}From section 3.2, we know that there is an $SU(2)$-invariant
brane with topology $S^2$ on which $\tau$ acts trivially.  This
brane must project to the part of the $w_2$ axis connecting the
two fixed points. On the other hand, from section 3.3, we know
that there are two $\tilde\tau$-invariant branes with topology
$\R^2$. These must correspond to the regions $w_2\geq \beta/2$ and
$w_2\leq -\beta/2$ of the $w_2$ axis.  The action on $\chi$ of
$\tau$ and $\tilde\tau$ is therefore
\eqn\znx{\eqalign{\tau&:\chi\to\cases{\chi+\pi &
$|w_2|>\beta/2$\cr
                                       \chi     &
                                       $|w_2|<\beta/2$\cr}\cr
\tilde\tau&:\chi\to\cases{\chi & $|w_2|>\beta/2$\cr
                                       \chi+\pi     &
                                       $|w_2|<\beta/2$.\cr}\cr}}
For $\beta\to 0$, \znx\ and \gnx\ coincide, as expected.  In \znx,
we have not explained what happens for $|w_2|=\beta/2$; these are
the fixed point of $\K$ so the value of $\chi$ is immaterial
there.

Now let us consider $Y$ as a symplectic manifold with symplectic
form $\omega_Y=\omega_K=(d\chi+\vec a\cdot \vec w)\wedge dw_3
-H\,dw_1\wedge dw_2$, and try to describe the $\K$-invariant
Lagrangian submanifolds $M$.  $\K$-invariance means that $M$ is a
union of fibers of the projection $Y\to\R^3$.  So in fact, $M$
must be the inverse image in $Y$ of a one-dimensional curve $\ell$
in $\R^3$.  The restriction of $\omega_Y$ to such an $M$ is the
same as the restriction of $d\chi\wedge dw_3$, and vanishes if and
only if $\ell$ is at a constant value of $w_3$.   So to find $M$,
we specify a constant value of $w_3$, which we call $b$, and we
let $\R^2_b\subset \R^3$ be the plane defined by $w_3=b$. Then we
pick any curve $\ell\subset \R_b^2$, and take $M$ to be its
inverse image in $Y$.

 \ifig\abranefig{A $\K$-invariant Lagrangian submanifold
$M\subset Y$ is represented by a planar curve $\ell \in \R_b^2$.
Plotted is the case that $\R^2_b$ contains both  special points,
depicted here as solid dots.}
{\epsfxsize2.7in\epsfbox{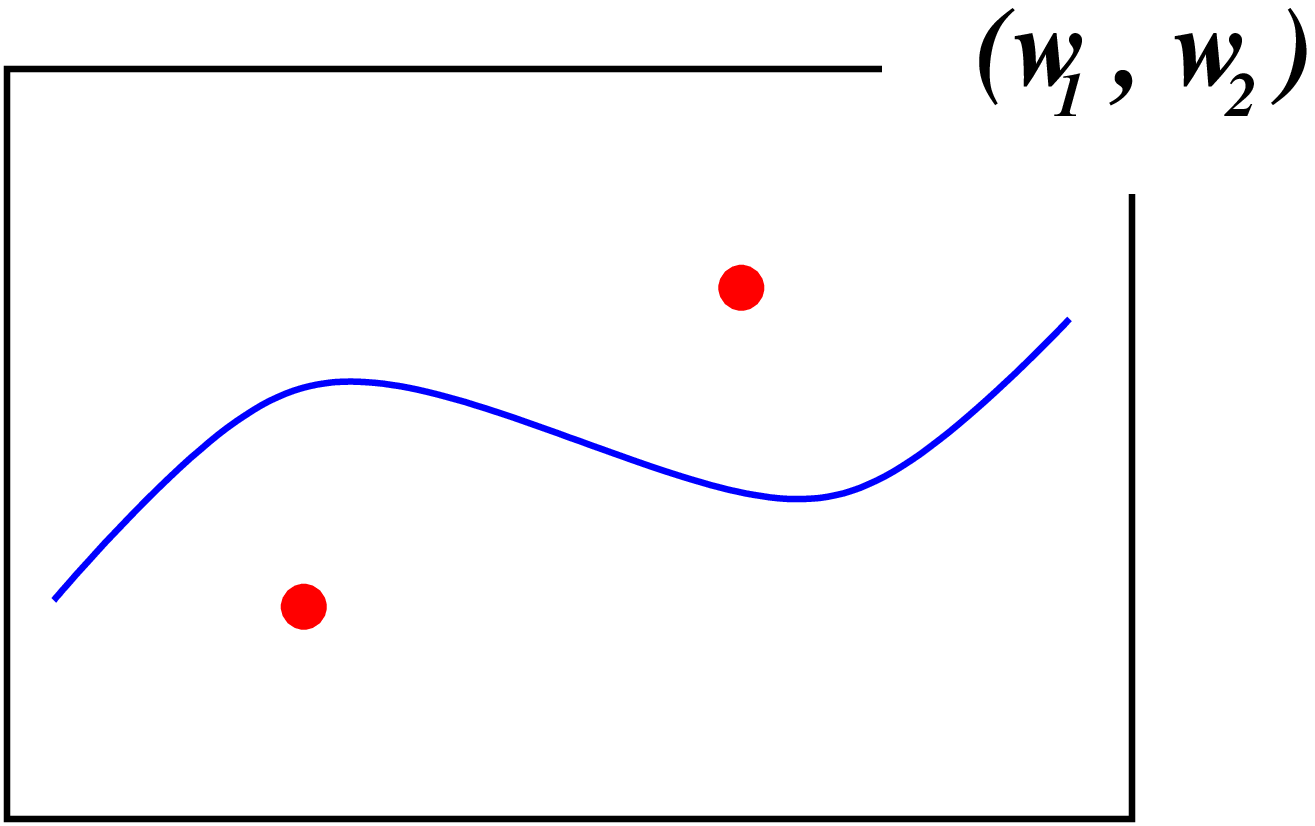}}

The plane $\R^2_b$ may contain zero, one, or two of the special
points $\vec w=\pm \vec w^*$. $\R^2_b$ contains one special point
if $\beta/2=\pm b$, and both special points if $b=\beta=0$. In
\abranefig, we sketch a case in which both of the special points
are contained in $\R^2_b$; in this example, $\ell$ extends to
infinity in both directions. Although a closed curve $\ell$ will
lead to a Lagrangian submanifold, it does not generally lead to an
$A$-brane. The reason is that a closed curve in $\R^2_b$ is the
boundary of a disc instanton (holomorphic in complex structure
$K$), so that such a Lagrangian submanifold is not likely to be
the support of an $A$-brane.\foot{Also, the space of $(\Bcc,\B')$
strings cannot have an interpretation in terms of quantization if
$\ell$ is a closed curve, because the condition for nondegeneracy
of $\omega_J$ that we discuss shortly cannot be satisfied for a
closed curve.} On the other hand, it is definitely possible for
$\ell$ to terminate at one or both ends at one of the special
points $\vec w=\pm \vec w^*$, where the fiber of the map
$Y\to\R^3$ collapses to a point.

\ifig\herm{If $\ell$ is semi-infinite, running from a fixed point
to infinity, then the corresponding Lagrangian submanifold is a
semi-infinite cigar, equivalent topologically to $\R^2$.}
{\epsfxsize3.0in\epsfbox{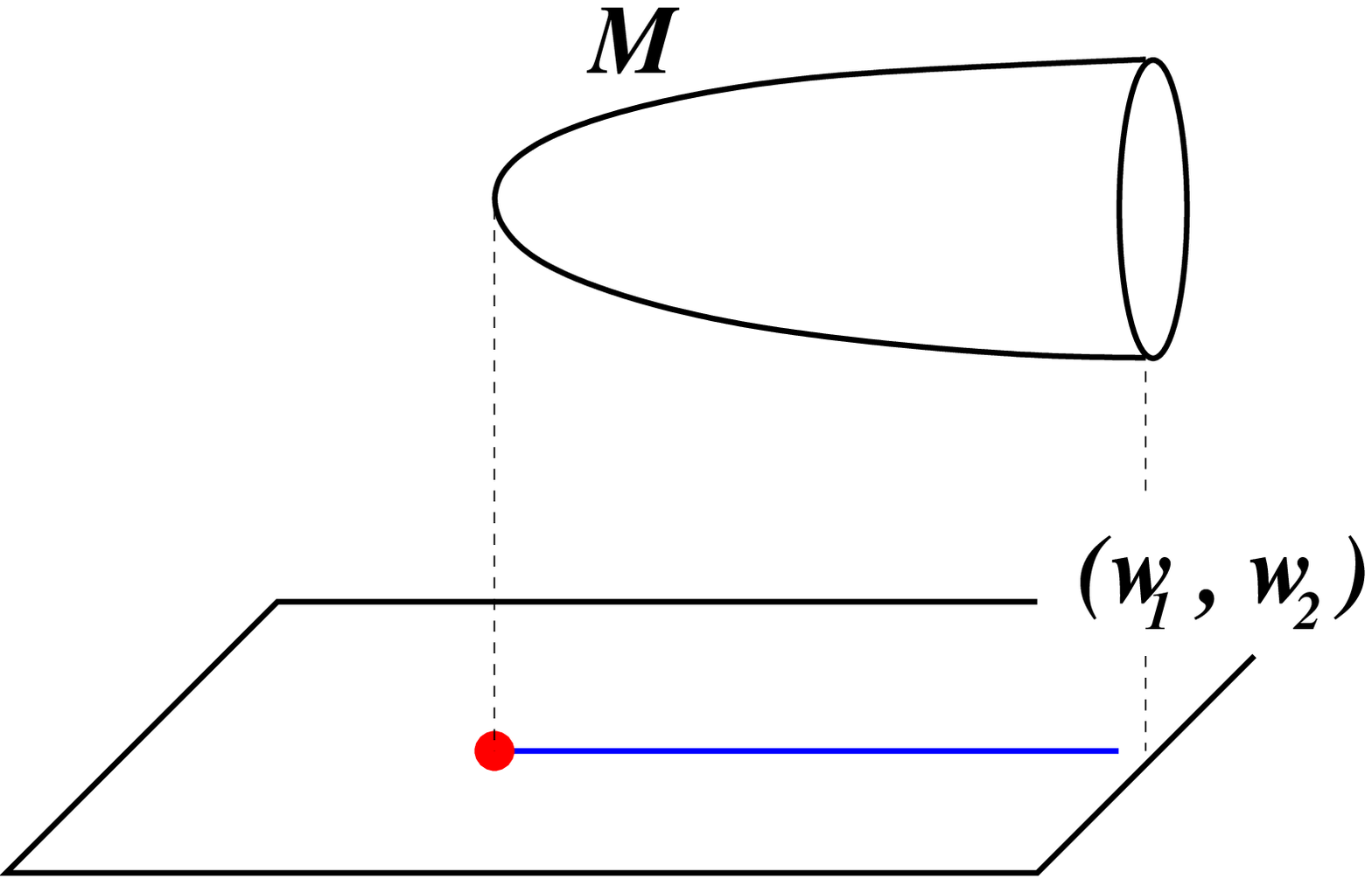}}

There are thus three topologies for $M$:

(1) If $\ell$ connects the two points $\vec w=\pm \vec w^*$, the
fiber collapses at both ends and $M$ is topologically a
two-sphere.

(2) If $\ell$ is infinite at one end, and ends at the other end at
one of the two special points, then the fiber collapses at one end
and $M$ is topologically $\R^2$.  (This is illustrated in \herm.)

(3) If $\ell$ is infinite at both ends, the fiber never collapses
and $M$ is topologically $\R\times S^1$.

We have already met $A$-branes of all three types.  In section
3.2, we associated finite-dimensional representations of $SU(2)$
with branes of type (1).  In section 3.3, we associated the
discrete series with branes of type (2).  And in section 3.4, we
associated the principal series with branes of type (3).  All of
these branes have their support on the $w_2$ axis, with
$w_1=w_3=0$.  Asymptotic $SL(2,\R)$ invariance means that at
infinity, if $\ell$ does get to infinity, $\ell$ must be
sufficiently close to the $w_2$ axis.  We do not know the right
criterion for ``sufficiently close,'' and our results will not
depend on this very much.  For simplicity, we will consider the
case that $w_1$ is bounded at infinity.

For the space of $(\Bcc,\B')$ strings to have an interpretation in
terms of quantization, $\omega_J$ must be nondegenerate when
restricted to $M$.  Since $\omega_J=(d\chi+\vec a\cdot \vec
w)\wedge dw_2-H dw_3\wedge dw_1$, and $dw_3=0$ along $\ell$, this
is equivalent to saying that $dw_2$ is everywhere nonzero along
$\ell$.  In other words,  $w_2$ is everywhere a good coordinate
along $\ell$, and $\ell$ can be described by an equation
$w_1=f(w_2)$, for some function $f$. Hamiltonian transformations
of $Y$ that commute with $\K$ can include, as a special case,
area-preserving transformations of $\R^2_b$ that leave invariant
the fixed points $\vec w=\pm \vec w^*$ (if either of these points
lies in $\R^2_b$). Modulo such a transformation, the only
invariants of $\ell$ are whether it ends at a fixed point and
whether it goes to the left or right of such a point.

Let us consider an $\ell$ that is infinite at both ends and
asymptotically parallel to the $w_2$ axis.  Thus, the Lagrangian
submanifold $M$ is topologically $\R\times S^1$.  $\ell$ may have
complicated wiggles, but these are irrelevant in the $A$-model. To
find the parameters in the $A$-model, we simply observe that as
the first Betti number of $M$ is 1, a rank 1 $A$-brane supported
on $M$ depends on a single complex parameter or a pair of real
parameters. One real parameter is the Wilson line -- the holonomy
of the Chan-Paton line bundle around $S^1$.  As in section 3.4, we
parametrize the holonomy of this bundle as $\exp(2\pi i\delta)$.
Then $\delta$ is one real parameter characterizing a rank 1
$A$-brane supported on $M$. The second real parameter corresponds
to a displacement of $M$ by the vector field
$V=\omega_Y^{-1}\zeta$, where $\zeta$ is a closed but not exact
one-form on $M$, and $\omega_Y=\omega_K$ is the symplectic form of
the $A$-model.  In the case at hand, we can take $\zeta=d\chi+\vec
a\cdot d\vec w$ (which is closed when restricted to $M$), in which
case $V$ is the vector field $\partial/\partial w_3$, which shifts
$b$.  Thus, the second modulus is simply $b$. Moreover, the
$A$-model is holomorphic in $\delta+ib$.

It is not hard to understand the meaning of the moduli. In
quantizing $M$ with respect to symplectic structure $\omega_J$,
the moment map for the $\K$ symmetry is $w_2$, which thus plays
the role of $z$ in sections 3.3-4.  $w_2$ is unbounded below along
$M$, since $\ell$ is asymptotically parallel to the $w_2$ axis. So
the spectrum of $J_z$ eigenvalues in quantization of $M$ is of the
form $\{s+n|n\in \Z\}$, for some complex number $s$ that is
defined modulo 1.  As in section 3.4, the real part of $s$ is
determined by the monodromy around $S^1$ of the Chan-Paton bundle
of $M\cong \R\times S^1$.  Thus ${\rm Re}\,s=\delta+{\rm
constant}$, so holomorphy of the $A$-model in $\delta+ib$ implies
that \eqn\golno{s=\delta+ib+{\rm constant}.}

Now let us discuss what sort of branes are associated with modules
that have a highest or lowest weight vector.  For this, $\ell$
should end on one of the fixed points $\vec
w^*=\pm\half(\alpha,\beta,\gamma)$, and hence (taking the positive
sign) we must have \eqn\ols{b=\gamma/2.} According to \bonk,
\eqn\kids{\mu=\beta+i\gamma=\eta+i\gamma+n,} with $n$ an integer.
Comparing the last three formulas, the imaginary part of $\mu$ is
twice the imaginary part of $s$, so
 via holomorphy, we have $\mu/2=s+n+{\rm constant}$.
The constant is $1/2$, since, as we explained in section 3.3, a
highest or lowest weight module with $\mu=\beta=\gamma=0$ has
$s=1/2$. So the quadratic Casimir operator is
\eqn\hixn{J^2=(\mu^2-1)/4=(s+n+1/2)^2-1/4=(s+n)(s+n+1),} which is
the expected result for a module with a lowest or highest weight
vector.

It is inevitable that there is an undetermined integer $n$ in this
formula, since $s$ (which is defined by saying that the
eigenvalues of $J_z$ are congruent to $s$ mod $\Z$) is only
determined modulo an additive integer.  Moreover, the derivation
involved no input that would distinguish a representation
containing a highest weight vector from one containing a lowest
weight vector. For a brane to have an interpretation in terms of
quantization, $w_2$ must be a monotonic function along $\ell$, so
either $\ell$ runs from $w_2=-\infty$ to the fixed point or it
runs from the fixed point to $w_2=+\infty$. Since the value of
$w_2$ is a classical approximation to $J_z$, the two types of
brane correspond to highest weight and lowest weight
representations, respectively. In either case, to determine the
precise $J_z$ value of the highest or lowest weight vector, we
must treat the quantization more precisely, or use an algebraic
argument such as that explained in section 3.5.

 If one end of $\ell$ is at a fixed point (\herm), the
corresponding Lagrangian submanifold $M$ is topologically $\R^2$.
The first Betti number of $M$ vanishes, so the corresponding brane
$\B'$ has no deformations as an $A$-brane. This is what we expect
for a brane associated to a highest or lowest weight
representation. The value of $s$ for such a brane is constrained
to obey \hixn\ for some integer $n$, so there is no possible
deformation.  ($J^2$ is a property of the space $Y$, not the brane
$\B'$, so $J^2$ will remain fixed in deformations of $\B'$ in the
$A$-model of $Y$.)

Now let us consider a pair of curves $\ell_1$ and $\ell_2$, both
ending at the fixed point.  We suppose that $\ell_1$ runs from
$-\infty$ to the fixed point and $\ell_2$ from the fixed point to
$+\infty$.  $\ell_1$ and $\ell_2$ correspond to $A$-branes $\B_1$
and $\B_2$ that are associated, respectively, to a highest weight
representation $R_1$  and a lowest weight representation $R_2$ of
$\frak{sl}(2,\R)$. $\B_1$ and $\B_2$ are supported on Lagrangian
submanifolds $M_1$ and $M_2$, each of which is topologically
$\R^2$. The two copies of $\R^2$ meet transversely. In terms of
some local complex coordinates $u$ and $v$ (which we can take to
be holomorphic in complex structure $J$), $M_1$ is defined by
$v=0$ and $M_2$ by $u=0$.

Consider the reducible $A$-brane $\B'=\B_1\oplus \B_2$, which is
associated to the reducible representation $R=R_1\oplus R_2$, in
which $J_z$ is bounded neither above nor below. It is supported on
$M=M_1\cup M_2$, which is defined by the equation \eqn\rork{uv=0.}
What are the deformations of $\B'$ as an $A$-brane? The space of
first order deformations is the space of $(\B',\B')$ strings of
ghost number 1, which for $\B'=\B_1\oplus \B_2$ decomposes as
$\oplus_{i,j=1}^2{\cal H}_{ij}$, where ${\cal H}_{ij}$ is the
space of $(\B_i,\B_j)$ strings of ghost number 1.  ${\cal H}_{11}$
and ${\cal H}_{22}$ are each trivial, since the first Betti
numbers of $M_1$ and $M_2$ are trivial.  On the other hand, as
$M_1$ and $M_2$ intersect transversely at a single point in $Y$,
which is of codimension 4, the spaces ${\cal H}_{12}$ and ${\cal
H}_{21}$ are each of complex dimension 1.  A general deformation
of $\B'$ is thus given by a pair of complex parameters $a\in {\cal
H}_{12}$ and $b\in {\cal H}_{21}$.

If $a=b=0$, $\B'$ remains as a direct sum.  If $a\not=0$, $b=0$,
or $a=0$, $b\not=0$, then $\B'$ is deformed to an extension in one
direction or the other, containing $\B_1$ or $\B_2$ as a
sub-brane. This gives us three representations, with the same
value of $s$, exactly in parallel with the situation that was
described in section 3.5, for example in eqns. \onxop\ and
\ponxop.  In fact, the parameters $a$ and $b$ in the present
derivation with branes correspond to $a_n$ and $b_n$ in the
previous purely algebraic analysis.

Finally, if $a$ and $b$ are both nonzero, the support of the brane
$\B'$ is deformed from $uv=0$ to $uv=\epsilon$, where
$\epsilon\sim ab$.   The support of the deformed brane is
topologically $\R\times S^1$, with first Betti number 1.  So it is
possible to turn on a monodromy parameter $\delta$.  Of the three
real parameters $\delta$, ${\rm Re}\,\epsilon$, and ${\rm
Im}\,\epsilon$, two (namely $\delta$ and, say, ${\rm
Im}\,\epsilon$) represent a deformation in $s$ away from its
initial value that obeyed eqn. \hixn. This deformation forces the
brane and the associated representation to become irreducible.
(Indeed, when $\epsilon\not=0$, the brane no longer passes through
the fixed point at $u=v=0$; when $\delta$ is varied, the different
components are linked by a monodromy.)  The third real parameter,
say ${\rm Re}\,\epsilon$, represents a displacement of
$\ell=\ell_1\cup\ell_2$ in the $w_1$ direction. This type of
displacement is irrelevant in the topological $A$-model, though it
is meaningful as a deformation in the underlying sigma-model of
$Y$.

One can go one step farther and consider a situation in which
$\ell$ terminates  on a fixed point at each end; in other words,
$\ell$ connects the two fixed points at $\vec w=\pm \vec w^*$.
This leads to the same derivation with $\vec w^*$ replaced by
$-\vec w^*$, and $s$ replaced by $-s$ in \hixn. Requiring that
\hixn\ should hold for both signs of $s$, we learn as usual that
$2s\in \Z$.  The associated representations are
finite-dimensional.  We studied them from the viewpoint of $SU(2)$
in section 3.2, and we will  re-examine them in section 3.8 from
the viewpoint of $SL(2,\R)$.

\subsec{$\tilde\tau$-Invariant Branes With Only Asymptotic
$SL(2,\R)$ Symmetry}

Here we will study a more general class of $\tilde\tau$-invariant
$A$-branes associated to $\frak{sl}(2,\R)$ representations that
admit a hermitian structure that may or may not be
positive-definite.  Thus, we will re-examine in terms of branes
questions that were considered algebraically at the end of section
3.5.

We continue to describe $Y$ in a manifestly $SL(2,\R)$-invariant
way by the equation \eqn\osnox{-x^2-y^2+z^2={\mu^2\over 4}.} In
this description, $\tau$ and $\tilde\tau$ act by
\eqn\bosnox{\eqalign{\tau:(x,y,z)&\to (-\bar x,-\bar y, \bar z)\cr
                     \tilde\tau:(x,y,z)&\to (\bar x,\bar y,\bar
                     z).\cr}}
We begin with the case $\mu^2>0$. For the compact cycle $S$ that
generates the second homology of $Y$, we can take the fixed point
set of $\tau$. In the present description, this fixed point set is
characterized by $(x,y,z)=(i\hat x,i\hat  y,z)$, where $\hat
x,\hat y,$ and $z$ are real and $\hat x^2+\hat y^2+z^2=\mu^2/4$.

Let $\B'$ be a rank one $A$-brane supported on $S$.  Since $\tau$
acts trivially on $S$, the space $\cal H$ of $(\Bcc,\B')$ strings
gives a unitary representation of $SU(2)$, a fact that we
exploited in section 3.2.

Here we will look at the same brane from the point of view of
$SL(2,\C)$ and eventually $SL(2,\R)$.  First of all, the set $S$
is not $SL(2,\C)$-invariant. Nevertheless, the $A$-brane $\B'$ is
$SL(2,\C)$-invariant, simply because $S$ is compact.  For $v\in
\frak{sl}(2,\C)$, let $h_v$ be the Hamiltonian function that
generates the vector field on $Y$ corresponding to $v$.  Let $\hat
h_v$ be any function on $Y$ of compact support that coincides with
$h_v$ in a neighborhood of $S$.  The action of $\hat h_v$ is
trivial in the $A$-model (since Hamiltonian isotopies of compact
support are trivial in the $A$-model) and this action on $\B'$
coincides with the action of $h_v$. (We could make a similar
argument concerning the action of an element of the group
$SL(2,\C)$ that is close to the identity.)  This argument still
goes through if $S$ is not compact but is asymptotically
$SL(2,\C)$-invariant. For example, we will presently consider a
Lagrangian submanifold that is $SL(2,\C)$-invariant on the
complement of a compact set. This is certainly an adequate
condition.

Since the brane $\B'$ is $SL(2,\C)$-invariant, the group
$SL(2,\C)$ must act on $\H$.  Indeed, in any finite-dimensional
representation of $SU(2)$, the representation matrices can be
analytically continued to give an action of $SL(2,\C)$.  (By
contrast, the group $SL(2,\C)$ will generally not act in a Hilbert
space that furnishes a representation of $SL(2,\R)$, even a
unitary one; such representations are generally associated to
branes whose support is non-compact in an essential way, and the
above argument does not go through.)

Though $SL(2,\C)$  acts on the Hilbert space $\H$ obtained in
quantization of $S$, it certainly does not preserve the
positive-definite hermitian structure of $\H$ that arises from
$\tau$ symmetry.  In fact, $SL(2,\C)$ does not preserve any
hermitian structure on $\H$.  However, $\H$ does admit an
$SL(2,\R)$-invariant hermitian structure (though not a positive
one).

The reason for this is simply that $S$ is $\tilde\tau$-invariant,
as well as $\tau$-invariant.  $\tilde \tau$ acts on $S$ by
\eqn\hgt{(\hat x,\hat y,z)\to (-\hat x,-\hat y,z).} $\tilde \tau$
does not leave $S$ fixed pointwise (as $\tau$ does), but it does
map $S$ to itself, and therefore maps $\B'$ to itself. So the
construction of section 2.4 can be applied using $\tilde\tau$
symmetry to get a manifestly $SL(2,\R)$-invariant hermitian
structure on $\H$.

By following the logic of eqn. \gurf, we can make this explicit.
If $(~,~)$ is the natural pairing between $(\Bcc,\B')$ strings and
$(\B',\Bcc)$ strings, then the $SU(2)$-invariant hermitian pairing
is $\langle \psi,\psi'\rangle=(\Theta_\tau \psi,\psi')$ and the
$SL(2,\R)$-invariant pairing is
$\langle\langle\psi,\psi'\rangle\rangle
=(\Theta_{\tilde\tau}\,\psi,\psi')$.  Here
$\Theta_\tau=\tau\Theta$ and
$\Theta_{\tilde\tau}=\tilde\tau\Theta$. Finally, $\tilde\tau$ is
the same as $h\tau$, where $h$ is the rotation in \hgt, which acts
on the Lie algebra $\frak{sl}(2,\C)$ the same way that it does on
the coordinates: \eqn\ngt{h(J_x,J_y,J_z)h^{-1}=(-J_x,-J_y,J_z).}
So the relation between $\langle\langle ~,~\rangle\rangle$ and
$\langle~,~\rangle$ is
\eqn\gt{\langle\langle\psi_1,\psi_2\rangle\rangle = \langle
h\psi_1,\psi_2\rangle.} Explicitly, we can now verify that
$\langle\langle~,~\rangle\rangle$ is $SL(2,\R)$-invariant.  Since
$J_x,J_y$, and $J_z$ are hermitian with respect to
$\langle~,~\rangle$, \ngt\ implies that $J_z$ is hermitian but
$J_x$ and $J_y$ are antihermitian with respect to
$\langle\langle~,~\rangle\rangle$, which is the condition for
$SL(2,\R)$-invariance.

The fact that the same representation admits hermitian forms
invariant under either $SU(2)$ or $SL(2,\R)$ is related to the
fact that these groups are ``inner real forms'' of $SL(2,\C)$.
This means the following.  $SL(2,\R)$ is the subgroup of
$SL(2,\C)$ characterized by $g=\bar g$ (where we regard $g\in
SL(2,\C)$ as a $2\times 2$ complex unimodular matrix), while
$SU(2)$ is characterized by $g=h\bar g h^{-1}$.  An ``outer form''
(there are none for $SL(2,\C)$) would be characterized by
$g=\phi(\bar g)$, where $g\to \phi(g)$ is an outer automorphism of
order 2.

Let us describe the hermitian form
$\langle\langle~,~\rangle\rangle$ more explicitly in a basis of
eigenstates of $J_z$.  First consider a representation of odd
dimension $n=2k+1$.  We diagonalize $J_z$ with
$J_z\psi_s=s\psi_s$, $s=-k,-k+1,\dots,k$. Eqn. \ngt\ determines
$h$ up to multiplication by a constant $c$, which for the moment
we will assume to be real: \eqn\nerf{h\psi_s =c(-1)^s\psi_s.} So
the sign of $\langle\langle\psi_s,\psi_s\rangle\rangle$ is
$c(-1)^s$. Thus the states $ \psi_s$ have alternating positive and
negative norms. We have actually seen this structure from an
algebraic point of view in section 3.5, in a problem  (see eqn.
\overto) that is closely related, as will become clear.

The value of $c$ depends upon how one lifts $\tilde\tau$ from an
automorphism of $Y$ to an automorphism of the relevant Chan-Paton
line bundles. When one has a hermitian form that is
positive-definite if the sign is chosen properly, that gives a
natural choice.  Otherwise, what is natural may depend upon the
problem.  One fairly natural way to pick a lift of $\tilde\tau$ is
to pick a fixed point $p\in S$ of $\tilde\tau$, assuming that
there is one (in the present case there are two fixed points
$p_\pm$ defined by  $(\hat x,\hat y,z)=(0,0,\pm \mu/2)$), and
require that $\tilde\tau$ acts trivially on the fiber at $p$ of
the relevant Chan-Paton bundles. This gives a definite recipe for
defining the hermitian form, but in general it depends on the
choice of $p$.

A further subtlety arises for an even-dimensional representation
of $SU(2)$. In this case, since $s$ is half-integral, in order to
make $\langle\langle \psi,\psi\rangle\rangle$ real-valued, we need
to pick $c$ imaginary (which will ensure that the eigenvalues of
$h$ are real).  Here it is fairly clear that there cannot be a
preferred choice between $c=i$ and $c=-i$.  Indeed, it can be
shown that in the recipe mentioned in the last paragraph, the sign
depends on the choice of fixed point $p_\pm$.

\bigskip\noindent{\it The Complementary Series }

We return to the framework of section 3.7 and we consider an
$A$-brane associated with a suitable curve $\ell$.  However, here
we will take $\ell$ to be simply the $w_2$ axis, $w_1=w_3=0$.  For
$\beta=0$, $\gamma\not=0$, this gives the brane, studied in
section 3.4, whose quantization leads to the principal series.
Here we will consider the opposite case $\alpha=\gamma=0$,
$\beta>0$.  We take $\B'$ to be a rank 1 $A$-brane supported on
$M=\pi^{-1}(\ell)$.

In this case, the two fixed points of $\K$ lie on $\ell$ at the
points $\pm \half(0,\beta,0).$  Hence, geometrically one can
divide $M$ into three pieces $M_+\cong\R^2$, $M_0\cong S^2,$ and
$M_-\cong\R^2$, respectively, corresponding  to $w_2\geq \beta/2$,
$\beta/2\geq w_2\geq -\beta/2$, and $w_2\leq -\beta/2$.
Corresponding to this, if the parameters are chosen correctly,
$\B'$ may be the direct sum of three $A$-branes $\B_+,\B_0$, and
$\B_-$.

For this, we need $\eta=0$ and $\beta$ equal to an integer $n$, so
that an $A$-brane with support $M_0$ exists.  (Recall from section
3.2 that the space of $(\Bcc,\B_0)$ strings has dimension
$\beta$.) The decomposition $\B'=\B_+\oplus \B_0\oplus \B_-$
further needs a condition on the Chan-Paton bundle of $\B'$. The
monodromy around the point of intersection of $M_+$ and $M_0$ (or
of $M_0$ and $M_-$) must be trivial, or else this monodromy will
link the different components.

Though it is possible for $\B'$ to have a decomposition as an
$A$-brane, this is not the main situation that we wish to discuss.
We will primarily be interested in parameters for which $\B'$ is
irreducible.  Actually, for brevity we will focus on the case of
$s=0$, leading to representations of $SL(2,\R)$ in which the
center acts trivially.  One can achieve $s=0$ for any $\beta$ or
$J^2$ by suitably adjusting the monodromy of the Chan-Paton
bundle.  If $s=0$, then the representation of $SL(2,\R)$ is
irreducible unless $J^2=n(n+1)$ for some integer $n$.  When this
occurs, quantization of $M_0$ gives a Hilbert space of dimension
$2n+1$.  This happens for $\beta=2n+1$.  (For $s=1/2$, the
representation becomes reducible for $J^2=n(n+1)$ with $n\in
\Z+1/2$; this happens for even $\beta$. We leave this case to the
reader.)

Let us discuss the space $\H$ of $(\Bcc,\B')$ strings.  First of
all, $M_+$ and $M_-$ are $SL(2,\R)$-invariant, though $M_0$ is
not. But $M_0$ is compact.  So we are in  a situation similar to
the one we encountered above: since $M$ is $SL(2,\R)$-invariant
away from a compact set, $SL(2,\R)$ acts naturally on $\H$.
Furthermore, this action preserves a hermitian form (though not
necessarily a positive-definite one) since $M$ is
$\tilde\tau$-invariant.  In fact, $\tilde\tau$ acts trivially on
the components $M_\pm$, while rotating the two-sphere $M_0$ by an
angle $\pi$ around its points of intersection with $M_\pm$.  These
statements have essentially been summarized in eqn. \znx,
according to which $\tilde\tau$ acts on the fiber of $\pi:Y\to
\R^3$ by \eqn\xnerf{ \tilde\tau:\chi\to\cases{\chi &
$|w_2|>\beta/2$\cr
                                       \chi+\pi     &
                                       $|w_2|<\beta/2$.\cr}}

We consider first the case that  $\beta$ is large, to get a useful
semiclassical description. Also, we take $s=0$, so as to get a
representation of $SL(2,\R)$ with integer eigenvalues of $J_z$.
All integers will appear, since the moment map of $J_z$, which is
$w_2$, is unbounded above and below on $M$. So $\H$ has a basis
$\psi_n$ with $J_z\psi_n=n\psi_n$. Semiclassically, $M_+$, $M_0$,
and $M_-$ support, respectively, states with $J_z$ (or its moment
map $w_2$) greater than $\beta/2$, between $\beta/2$ and
$-\beta/2$, and less than $-\beta/2$, respectively.  Quantization
of $M_+$ or $M_-$ gives a positive-definite hermitian form, since
$\tilde\tau$ acts trivially, while quantization of $M_0$ gives an
oscillatory quadratic form, as described above.  So the norm of
$\psi_n$ is positive for $|n|>\beta/2$ and oscillates in sign for
$|n|<\beta/2$. Semiclassical reasoning justifies these statements
except near $|n|=\beta/2$, where the fact that $\B'$ is actually
irreducible becomes relevant. But actually, the algebraic analysis
of the hermitian form in section  3.5 shows that the statements
are precisely valid.

Now let us consider the opposite region in which $\beta$ is small,
still keeping $s=0$.  At $\beta=0$, $M$ coincides with the brane
used in section 3.4 to describe the $\gamma=0$ case of the
principal series.  The hermitian form on $\H$ is certainly
positive-definite in that case; in fact, $\tilde\tau$ acts
trivially on $M$ at $\beta=0$.  When we turn on $\beta$,
$J^2=(\beta^2-1)/4$ becomes greater than $-1/4$.  The $SL(2,\R)$
action on $\H$ remains irreducible until we reach $\beta=1$.  In
this entire range, there is no highest or lowest weight vector
(eqns. \noble\ and \tobel\ show that for $s=0$ and $J^2<0$, there
cannot be one), and the hermitian inner product remains
positive-definite, according to eqn. \overto.  What we have found
is the complementary series of unitary representations of
$SL(2,\R)$.

At $\beta=1$, the brane $\B'$ is reducible.  $\H$ splits up
accordingly as $\H_+\oplus\H_0\oplus\H_-$, where $\H_\pm$ are
infinite-dimensional Hilbert spaces that realize the discrete
series representations $D_1^\pm$, and $\H_0$ is a one-dimensional
trivial representation of $SL(2,\R)$.  The hermitian structure on
$\H_+\oplus \H_0\oplus \H_-$ is not unique, as signs can be chosen
independently on the three summands, depending on the lift of
$\tilde\tau$ to act on the Chan-Paton bundles of the three branes.
Continuing past $\beta=1$, the hermitian structure becomes unique
again (up to a multiplicative constant, which we choose to get an
almost positive-definite inner product), with a single state of
negative norm.  This follows from the algebraic analysis of
section 3.5 (see eqn. \overto).  This result means that the
hermitian structure at $\beta=1$ that continues smoothly to
$\beta>1$ corresponds to taking  the hermitian form on $\H_+$,
$\H_0$, and $\H_-$ to be respectively positive, negative, and
positive.

If we continue to increase $\beta$, the brane $\B'$ and the
representation furnished by $\H$ are irreducible, as noted above,
except when $\beta$ is an odd integer $2n+1$, for some $n$.
 For $\beta$ between $2n-1$ and $2n+1$,  the number of
negative norm states is $n$, according to the algebraic analysis
of section 3.5. At $\beta=2n+1$, $\H$ decomposes as $\H_+\oplus
\H_0\oplus\H_-$, where $\H_\pm$ furnish discrete series
representations, and $\H_0$ is a representation of dimension
$2n+1$.  Taking a basis of $\H_0$ with $J_z\psi_k=k\psi_k$,
$|k|\leq n$, the norm of $\psi_k$ has sign $c(-1)^k$, as explained
earlier, where depending on the choice of lift of $\tilde\tau$,
$c$ may be either 1 or $-1$.  One lift gives $n$ states of
negative norm and one gives $n+1$.  The lift that gives $n$ states
of negative norm continues smoothly to $\beta<2n+1$ and the one
that gives $n+1$ states of negative norm continues smoothly to
$\beta>2n+1$.  The fact that one lift extends in one direction and
one in the other can be explained by a topological argument.

\subsec{Relation To ${\cal D}$-Modules}

In section 3.7, we showed how to describe a  Harish-Chandra module
in terms of a curve $\ell$ in the plane $\R^2_b$. We adopted as
much as possible a topological point of view, not attempting to
make a distinguished choice of $\ell$. Instead we required only
that $w_2$ is a good coordinate along $\ell$, so that the space of
$(\Bcc,\B')$ strings in the $A$-model can be related to
quantization of $M$.

\ifig\hermitlafig{A vertical line in the plane $\R^2_b$,
corresponding to a brane of type $(A,B,A)$.}
{\epsfxsize3.0in\epsfbox{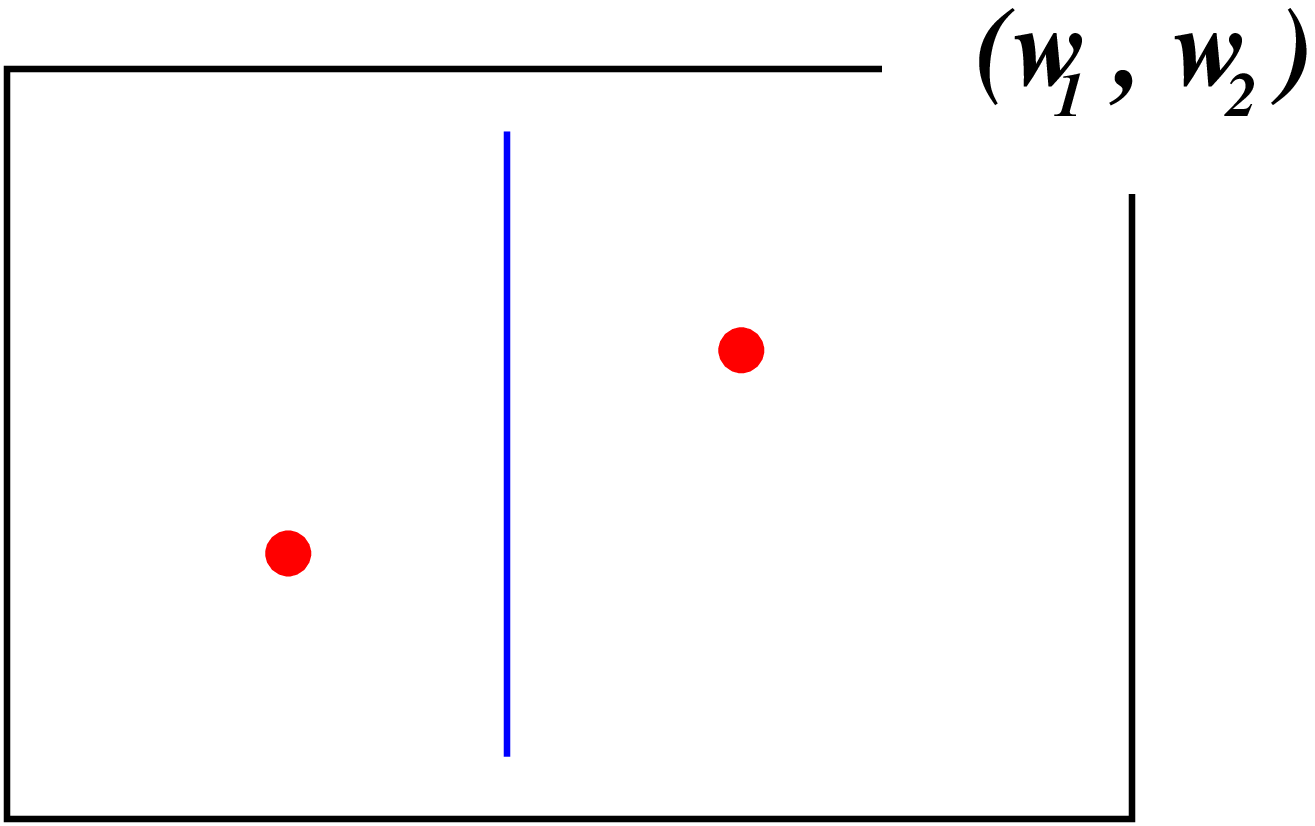}}

However, certain choices of $\ell$ have particularly nice
properties.  One choice that is particularly nice from the point
of view of quantization is to take $\ell$ to be a vertical line
(or part of a vertical line) in the plane $\R^2_b$.  In this case
(\hermitlafig), $M=\pi^{-1}(\ell)$ is a complex manifold in
complex structure $J$. The associated $A$-brane $\B'$ is then a
brane of type $(A,B,A)$ (that is, it is a $B$-brane with respect
to complex structure $J$ and an $A$-brane for any linear
combination of $\omega_I$ and $\omega_K$).  This can be
convenient, for the following reason. The canonical coisotropic
brane $\Bcc$ associated with a choice of hyper-Kahler metric on
$Y$ is also a brane of type $(A,B,A)$.  The space of $(\Bcc,\B')$
strings in the $A$-model can be viewed as the space of string
states of zero energy in the underlying sigma-model of $Y$ with
$\N=4$ supersymmetry, or alternatively as the $(\Bcc,\B')$ strings
in the $B$-model of complex structure $J$.  In other words,
quantization in our sense coincides in this situation with what in
geometric quantization is called quantization using the Kahler
polarization determined by $J$.  We exploited this relationship in
section 2.3.

\ifig\hermit{A horizontal line in the plane $\R^2_b$,
corresponding to a brane of type $(B,A,A)$.}
{\epsfxsize3.0in\epsfbox{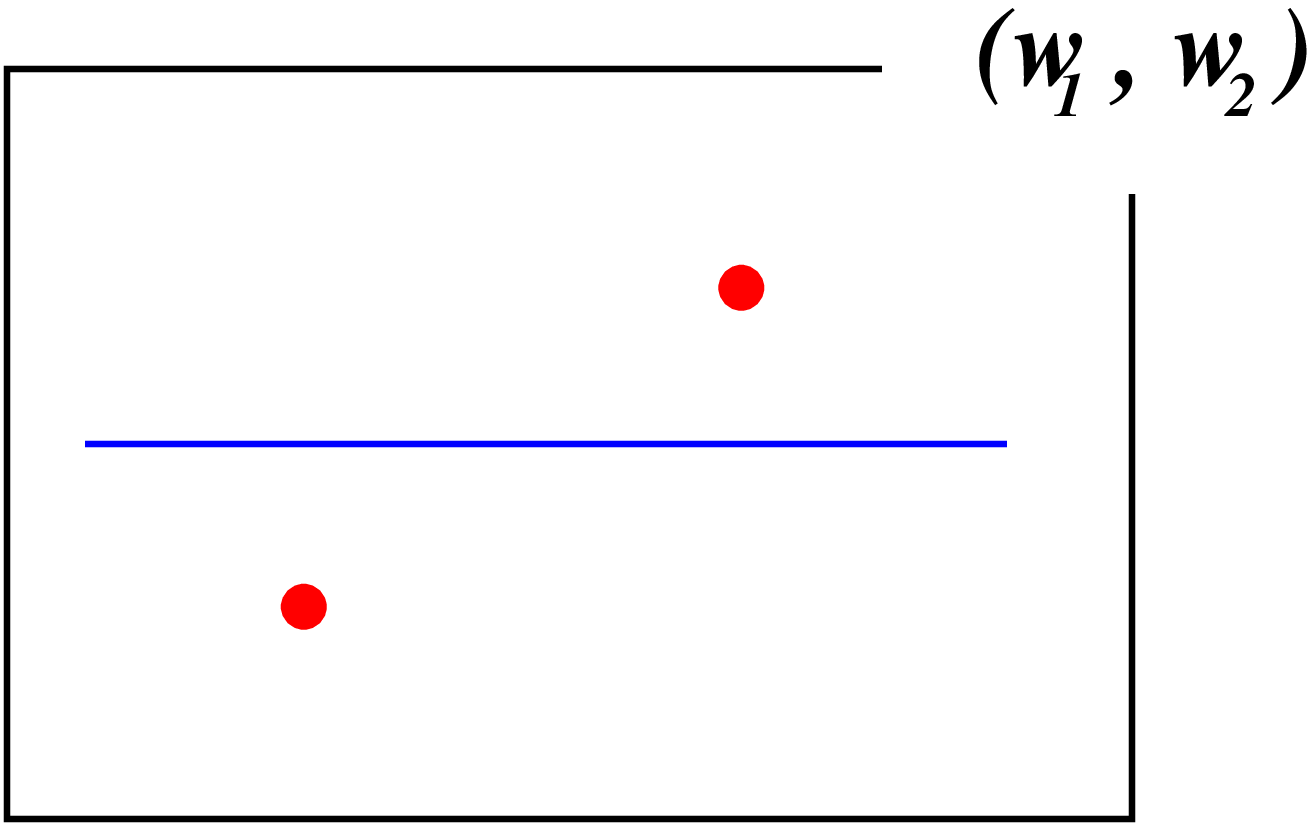}} There is another type of
choice for $\ell$ (\hermit) that is not closely related to
quantization but is interesting from a different point of view. We
take $\ell$ to run horizontally in $\R^2_b$, parallel to the $w_1$
axis. In this case, $M$ is holomorphic in complex structure $I$
and $\B'$ is a brane of type $(B,A,A)$. This choice of $M$ is
maximally unsuitable for an interpretation via quantization,
because $\omega_J$ vanishes when restricted to $M$, rather than
being nondegenerate.  However, it has another virtue: it
simplifies the relation between $A$-branes and ${\cal D}$-modules.

 $Y$ can be interpreted as $T^*\Bbb{CP}^1$ if
$\alpha\not=0$, $\beta=\gamma=0$; more generally if $\beta$ and
$\gamma$ are not zero, $Y$ is an affine deformation of
$T^*\Bbb{CP}^1$.  In any event, $Y$ admits an $SL(2,\C)$-invariant
holomorphic map  $\Psi:Y\to\Bbb{CP}^1$ (in fact two such maps,
related in a sense by a Weyl transformation; explicit formulas are
given below). In this situation, as explained in section 11 of \KW,
the space of $(\Bcc,\B')$ strings can be sheafified over
$\Bbb{CP}^1$ and interpreted as the sheaf of sections of a twisted
${\cal D}$-module over $\Bbb{CP}^1$. The ${\cal D}$-module is
twisted by $K^{1/2}\otimes {\cal O}(1)^{\lambda}$, where $K$ is the
canonical line bundle of $\Bbb{CP}^1$, ${\cal O}(1)$ is the usual
line bundle of degree 1, and $\lambda=\eta+i\gamma$.  (Such twisting
is described in \Ramified, section 4.4.)

The fact that an $A$-brane leads to a twisted ${\cal D}$-module,
as well as an $\frak{sl}(2,\C)$ module, enables us to make contact
with the theory of Beilinson and Bernstein \bb\ relating
$\frak{sl}(2,\C)$ modules to twisted ${\cal D}$-modules on the
flag manifold $\Bbb{CP}^1$. To compare to that theory, we would
like to be able to
 explicitly describe the ${\cal D}$-module corresponding to a
given $A$-brane. In general, this is difficult, but for branes of
type $(B,A,A)$, there is a natural framework for doing so, as
described in section 4.3 of \fw.  In general, this involves
solving Hitchin's equations, but for the branes considered in the
present paper, one can get an explicit answer as the relevant
equations are abelian.

  If $\B'$ is a brane of type $(B,A,A)$ supported on $M\subset
Y$ (so in particular $M$ is holomorphic in complex structure $I$),
then the support of the corresponding ${\cal D}$-module is simply
the projection of $M$ under the holomorphic map $\Psi: Y \to
\Bbb{CP}^1$. In order to describe $\Psi$ explicitly, we recall that,
in complex structure $I$, the complex symplectic manifold $Y$ is
defined by the equation $x^2+y^2+z^2=\mu^2/4$ in complex variables
$x,y,z$. To an element of $Y$, we associated a complex $2 \times 2$
traceless matrix \eqn\refl{ A = \pmatrix{x & y - iz \cr y + i z & -
x}}
 with  determinant $-\mu^2/4$ or, equivalently,
with eigenvalues $\pm \mu/2$.  Because $\mu/2$ is an eigenvalue of
$A$, there exists a nonzero column vector $\Upsilon$, unique up to
scaling, that obeys
\eqn\laeigen{ A \Upsilon = {\mu \over 2} \Upsilon .}
Up to scaling, $\Upsilon$ determines a point in $\Bbb{CP}^1$, and
$\Psi$ is defined to map $A$ to this point. (A second such map can
be defined by taking $\Upsilon$ to obey
$A\Upsilon=-(\mu/2)\Upsilon$.) Setting $\Upsilon = \pmatrix{1 \cr
t}$ gives \eqn\jhsin{ t = {\mu/2 - z \over x - i y} = {x + i y \over
\mu/2 + z}.} This formula explicitly describes the map
$\Psi:Y\to\Bbb{CP}^1$. In terms of new variables $x_{\pm} = \pm x +
i y$, we have
 \eqn\maptocpone{ t = {z - \mu/2
\over x_-} = {x_+ \over \mu/2 + z} } In terms of these variables,
the equation defining $Y$ becomes \eqn\trney{ x_+ x_- =
z^2-\mu^2/4.} The variable $z$ here can be identified with the
complex variable $w_2 + i w_3$ in the Gibbons-Hawking description
of the hyper-Kahler metric on $Y$.

\ifig\baaiibranefig{A brane $\B'$ of type $(B,A,A)$ is located at
a fixed value of $z = w_2 + i w_3$.}
{\epsfxsize2.7in\epsfbox{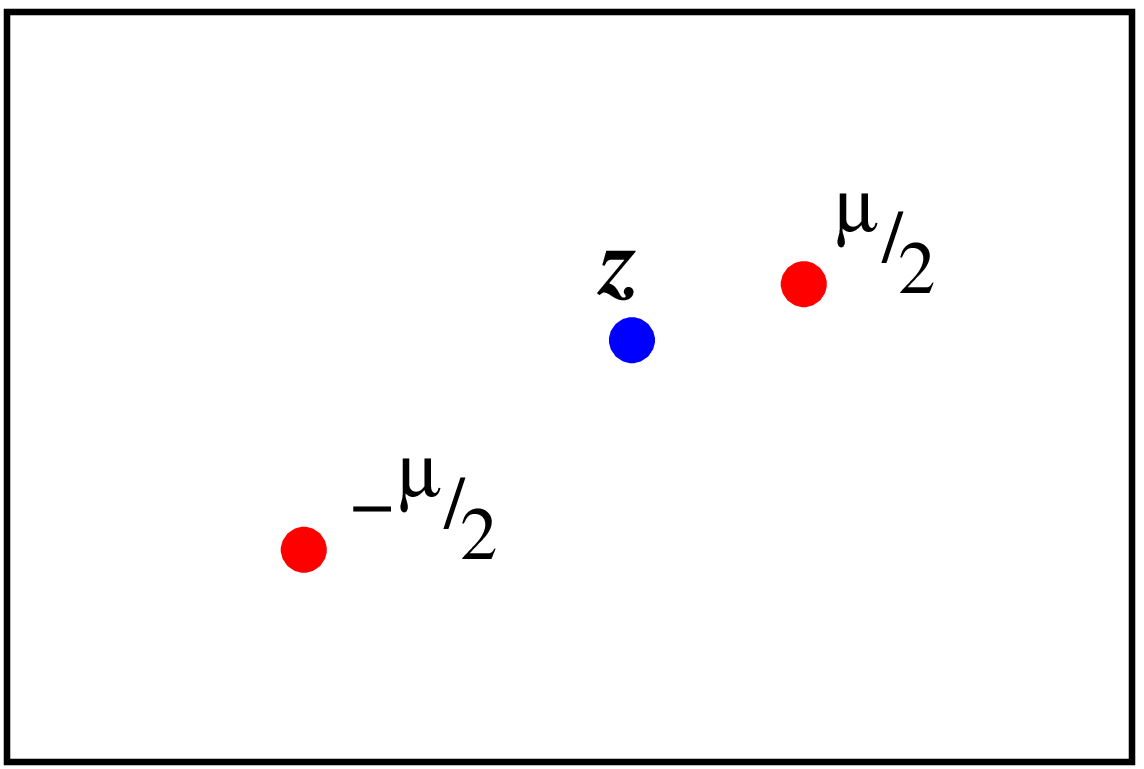}}

Now let us return to branes of type $(B,A,A)$. As we explained in
section 3.7, in the Gibbons-Hawking description of the
hyper-Kahler metric on $Y$, a brane $\B'$ of type $(B,A,A)$ is
represented by a curve $\ell\subset\R^3$ with fixed values of
$w_2$ and $w_3$. In other words, $\ell$ is a line parallel to the
$w_1$ axis, or a part of it. In particular, since the value of $z
= w_2 + i w_3$ is fixed, the support of this brane is a complex
subvariety of $Y$ defined by a constant value of $x^2+y^2$ or
equivalently of $x_+x_-$:
\eqn\xyepsilon{ x_+ x_- = \epsilon. }
Here $\epsilon=-\mu^2/4+z^2$ is a constant that vanishes precisely
if $z=\pm \mu/2$.

\medskip\noindent{{\it Unbounded Modules}}\medskip

We start with the general case $\epsilon \ne 0$. In this case, $M$
is an irreducible algebraic curve defined by eqn. \xyepsilon. It
has the topology of $\R\times S^1$ and corresponds to a principal
series representation of $SL(2,\R)$ or its universal cover, or,
more generally, to a Harish-Chandra module that has $J_z$
eigenvalues unbounded above and below. From \maptocpone\ it
follows that a brane $\B'$ of this type corresponds to a ${\cal
D}$-module supported on \eqn\hyy{ M'= \Bbb{CP}^1 \setminus \{
0,\infty \}.}

Because the map from $M$ to $M'$ is $1:1$, the ${\cal D}$-module
is of rank 1, and is given by a flat bundle with structure group
$\C^*=GL(1,\C)$.  As $\pi_1(M')=\Z$, such flat bundles are
classified by a single element of $\C^*$, which we can take to be
the monodromy around the origin (or the inverse of the monodromy
at infinity).  The logarithm of this monodromy corresponds to the
parameter $s=\delta+ib$ of section 3.7.  If the projection of $M$
to its image in $\Bbb{CP}^1$ were $k:1$, we would have to solve
rank $k$ Hitchin equations to describe the ${\cal D}$-module
explicitly.

$M'$ is one of the $\K_\C$ orbits in $\Bbb{CP}^1$. Indeed, the
complex group $\K_\C = \C^*$ acts on the flag variety
$\Bbb{CP}^1$, which is a union of three $\K_\C$ orbits: two
compact orbits $\{ 0 \}$ and $\{ \infty \}$, and one open orbit
$\C^* \simeq \Bbb{CP}^1 \setminus \{ 0,\infty \}$.

\medskip\noindent{{\it Highest And Lowest Weight Modules}}\medskip

As $z$ approaches one of the points $\pm \mu/2$, say $\mu/2$, we
have $\epsilon \to 0$ in eqn. \xyepsilon, and the brane $\B'$
supported on the curve \xyepsilon\ degenerates into a sum of two
branes, one supported at $x_+=0$, and the other supported at
$x_-=0$. We denote these branes as $\B_+'$ and $\B_-'$,
respectively. (The imaginary part of the condition $z = \mu/2$ is
$b = \gamma/2$, which is precisely the condition \ols\ for $\ell$
to end at a special point.) {}From \maptocpone, it follows that
the supports $M_\pm'$ of the branes $\B_\pm'$ are
\eqn\dfuyd{\eqalign{ M_+' & = \{ 0 \} \cr M_-' & = \Bbb{CP}^1
\setminus \{ \infty \} }}
At the second special point, $z = - \mu/2$, the roles of $\B_+'$
and $\B_-'$ are reversed and the branes $\B_+'$ and $\B_-'$
correspond to ${\cal D}$-modules supported on
\eqn\dfwe{\eqalign{ M_+' & = \Bbb{CP}^1 \setminus \{ 0 \} \cr M_-'
& = \{ \infty \} }}
In both cases, the ${\cal D}$-modules supported on $M_+'$ and
$M_-'$ correspond, respectively, to Harish-Chandra modules with
lowest and highest weight vectors.

 \subsec{Groups Of Higher
Rank}

\def\O{{\cal O}}
We have concentrated on $SL(2,\C)$ in this section to keep the
arguments elementary, but there is a fairly immediate analog for
any complex Lie group $G_\C$, with $Y$ taken to be a coadjoint
orbit of $G_\C$, and $\Omega$ the natural holomorphic symplectic
form of the coadjoint orbit.  The algebra $\CA$ of $(\Bcc,\Bcc)$
strings in that case is ${\cal U}(\frak g_\C)/{\cal I},$ where
${\cal U}(\frak g_\C)$ is the universal enveloping algebra of
$G_\C$, and ${\cal I}$ is a deformation of the ideal that defines
$Y$.

Suppose that $G_\C$ is of rank $r$.  Then the ring of invariant
polynomials on $\frak g_\C$ is a polynomial ring with $r$
generators $\O_1,\dots,\O_r$, which are known as Casimir
operators.  A regular orbit in $\frak g_\C$ is obtained by setting
the $\O_i$ to complex constants $c_i$. So if $Y$ is such a regular
orbit, then the ideal $\I$ is generated by $\O_i-c_i'$,  with some
constants $c_i'$, $i=1,\dots,r$ (which may differ from the
classical values $c_i$, as we have seen for rank 1).

For any Lagrangian $A$-brane $\B'$, the space $\cal H$ of
$(\Bcc,\B')$ strings is a $\frak g_\C$-module in which the values
of the Casimir operators are $\O_i=c_i'$.  Thus, just as we saw
for the rank 1 case, by making different choices of $\B'$, we get
many different $\frak g_\C$-modules with the same values of the
Casimir operators. Moreover, if $\B'$ is invariant under an
antiholomorphic involution of $Y$ (which leaves fixed the support
of $\B'$), then $\cal H$ furnishes a unitary representation of the
corresponding real form of $G_\C$.

With suitable choices of $\B'$, one can construct unitary
representations of the compact form of $G_\C$ and analogs of the
discrete and principal series for noncompact forms, as well as
mixtures of these.  The greatest novelty in going to rank greater
than 1 is, however, that there are also interesting
representations associated with non-regular orbits.  For
$SL(2,\C)$, the only nonregular coadjoint orbit is the orbit of
the zero element of $\frak{sl}(2,\C)$; this orbit is a point,
leading to the trivial representation.  Groups of higher rank have
a greater  variety of non-regular orbits, leading to non-trivial
but in some sense small representations.

\newsec{Quantization Of Chern-Simons Gauge Theory}

\def\CA{{\eusm A}}
Finally, we conclude  by examining one of the few known examples of
a quantum field theory in which the subtleties of quantization
actually play an important role.

The relevant quantum field theory is three-dimensional
Chern-Simons gauge theory.  It was analyzed in \witten\ and in
more detail in \refs{\elitz,\hitch,\ADW} from the viewpoint of
geometric quantization, and has been studied from numerous other
points of view ranging from conformal field theory to algebraic
geometry and deformation quantization (references include \TUY,
\faltings, and \anderson, respectively).  See \atiyah\ for an
introduction to the theory.

\def\k{\hat k}
The aspect of interest to us is to construct the space of physical
states of Chern-Simons theory with compact gauge group $G$ on an
oriented two-manifold without boundary that we call $C$.
 For brevity, we take $G$ connected and
simply-connected. We define $M$ to be the moduli space of
homomorphisms from $\pi_1(C)$ into $G$, up to conjugation, of a
given topological type. $M$ (which is the classical phase space of
the Chern-Simons theory) has a natural symplectic form $\omega_*$,
inherited \abott\ from a symplectic structure on the
infinite-dimensional linear space $\CA$ of all connections on a
$G$-bundle $E\to C$ of the appropriate topological type.  In gauge
theory, letting $A$ denote a connection on $E$, the symplectic
form can be described explicitly by \eqn\tonx{\omega_*={1\over
4\pi}\int_C\Tr \,\delta A\wedge \delta A.} The symbol $\Tr$
represents an invariant quadratic form on the Lie algebra $\frak
g$ of $G$, which we normalize so that $\omega_*/2\pi$ is the image
in de Rham cohomology of a generator of $H^2(M,\Z)\cong\Z$. (For
$G=SU(n)$, $\Tr$ is the trace in the $n$-dimensional
representation.) A line bundle $\L_*$ with curvature $\omega_*$
can be naturally constructed using gauge theory \RSW. Then,
picking a positive integer $\k$, we seek to quantize $M$ with
symplectic form $\omega=\k\omega_*$ and prequantum line bundle
$\L=\L_*^{\k}$.

To find a sigma-model that will  do this, we need a natural
complexification $Y$ of $M$, with certain properties. We simply
take $Y$ to be the moduli space of homomorphisms from $\pi_1(C)$
to $G_\C$, the complexification of $G$, again up to conjugation.
$Y$ is naturally a complex manifold. Indeed, $Y$ can be defined by
giving the holonomies $U_i$ and $V_j$, $i,j=1,\dots,g$ around a
complete set of $a$-cycles and $b$-cycles in $C$.  These are
defined up to conjugation by an element of $G_\C$, and obey a
single relation \eqn\obrel{U_1V_1U_1^{-1}V_1^{-1}\dots
U_gV_gU_g^{-1}V_g^{-1}=1.} This description makes it clear that
$Y$ is a complex manifold, and in fact an affine variety, with a
great deal of holomorphic functions.  We can describe these
functions explicitly.  Let $R$ be a finite-dimensional
representation of $G_\C$.  For $S$ an oriented closed loop on $C$,
let ${\rm Hol}(S)$ be the holonomy of a flat connection around
$S$. Then $W_R(S)=\Tr_R\,{\rm Hol}(S)$ is a holomorphic function
on $Y$. (We took $R$ to be finite-dimensional to ensure that the
trace always converges; for suitable infinite-dimensional $R$, the
same formula gives a meromorphic function on $Y$.) In gauge
theory, with $A$ now understood as a $\frak g_\C$-valued
connection, we can write
\eqn\yto{W_R(S)=\Tr_R\,P\exp\biggl(-\oint_SA\biggr).}
Alternatively, if we write $S$ as a word in the $a$-cycles and
$b$-cycles (regarded now as generators of $\pi_1(C)$), then
$W_R(S)$ is the trace of the corresponding word in the $U_i$ and
$V_j$.  The holomorphic functions $W_R(S)$ restrict on $M$ to
holonomy functions that we define and denote in the same way (the
holonomies are now $G$-valued rather than $G_\C$-valued). Thus the
restrictions of holomorphic functions on $Y$ give a dense set of
functions on $M$.

$Y$ has a nondegenerate holomorphic two-form $\Omega_*$ that is
defined by the same formula as \tonx, with $A$ now understood as a
$\frak g_\C$ valued connection.  We consider $Y$ as a complex
symplectic manifold with holomorphic symplectic form
$\Omega=\k\Omega_*$.  Clearly, the restriction of $\Omega$ to $M$
coincides with $\omega$. The line bundle $\L\to M$ extends to a
unitary line bundle $\L\to Y$ with a connection of curvature ${\rm
Re}\,\Omega$.

As usual, we want to consider the $A$-model of $Y$ with symplectic
structure $\omega_Y={\rm Im}\,\Omega$.  (This is the same
$A$-model that is studied in the gauge theory approach to
geometric Langlands \KW, though the motivation there is a little
different.)
 $Y$ is a space with a very
good $A$-model, since in fact \hitchin\ it can be endowed with a
complete hyper-Kahler metric extending its structure as a complex
symplectic manifold.    To find such a hyper-Kahler metric, one
picks a complex structure on $C$, which enables one to write down
Hitchin's equations; one then endows $Y$ with a complete
hyper-Kahler metric by interpreting it as the moduli space of
solutions of those equations.  An important point here is that the
structure of $Y$ as a complex symplectic manifold is completely
natural (requiring no structure on $C$ except an orientation). But
to endow $Y$ with a hyper-Kahler structure, which is useful for
making the $A$-model concrete, we have to pick a complex structure
on $C$.  The choice of such a hyper-Kahler metric is a
hyper-Kahler polarization of $(Y,M)$ in a sense described in
sections 1.3 and 2.3.

$G_\C$ has an antiholomorphic involution that keeps $G$ fixed; we
write it as $U\to \bar U$ and call it complex conjugation. We define
an antiholomorphic involution $\tau:Y\to Y$ that acts by complex
conjugating all monodromies.  $M$ is a component of the fixed point
set of $\tau$, since by definition, $M$ is the locus in $Y$ with
$G$-valued monodromies.  (The fixed point set of $\tau$ has other
components, as explained in \hitchin.)

To place quantization of $M$ in the framework of this paper, we
must as usual introduce two branes in the $A$-model of $Y$. One
brane is the canonical isotropic brane $\Bcc$, whose support is
all of $Y$ and whose curvature form is ${\rm Re}\,\Omega$.  This
brane exists and is unique up to isomorphism because we have taken
$G$ to be simply-connected and $\hat k$ to be an integer.
Restricted to $M$, ${\rm Re}\,\Omega$ is the symplectic form
$\omega=\k\omega_*$ of $M$, which we wish to quantize. We let
$\B'$ be a rank 1 $A$-brane supported on $M$; it exists and is
unique up to isomorphism as $M$ is a simply-connected spin
manifold.

The space $\H$ of $(\Bcc,\B')$ strings in the $A$-model gives a
quantization of $M$ with symplectic structure $\omega$.
Diffeomorphisms of $C$ that are continuously connected to the
identity act trivially on $Y$ and on its $A$-model.  (They do not
preserve a hyper-Kahler metric on $Y$ that we may use to facilitate
computations in the $A$-model, but $A$-model observables do not
depend on this hyper-Kahler metric.)   However, the mapping class
group ${\eusm M}_C$ of $C$ acts on $Y$ in a way that is nontrivial
(and not isotopic to the identity) so it can act non-trivially  on
$\H$. It is useful to introduce the Teichmuller space $\T$ of $C$.
Any point $t\in\T$ determines a complex structure on $C$ (unique up
to isotopy) and hence a hyper-Kahler polarization of $(Y,M)$.  We
denote as $\H_t$ the space of $(\Bcc,\B')$ strings constructed with
this polarization.  It is locally independent of $t$, since the
$A$-model is invariant under a local change in the hyper-Kahler
polarization, so the $\H_t$ fit together as fibers of a flat vector
bundle over $\T$.  Taking the monodromy of the flat connection, we
get an action of ${\eusm M}_C$ on $\H_t$ (for any choice of $t$).
Actually, to be more precise, as is known from other approaches
cited at the beginning of this section, what acts on $\H$ is a
central extension of ${\eusm M}_C$. Though the occurrence of a
central extension is not surprising in quantum mechanics, to compute
the central extension from the present point of view, we would need
a better understanding of how to explicitly construct the flat
connection.

As explained in general in section 2.3, after a choice of
hyper-Kahler polarization corresponding to a point $t\in\T$, $\H$
can be computed explicitly as a vector space by taking the space
of holomorphic sections of the appropriate line bundle,
\eqn\zolf{\H= H^0(M,\L_*^{\k}\otimes K^{1/2}),} but  the proper
Hilbert space structure of the $A$-model is not given by an
elementary formula in terms of the Kahler geometry of $M$.  That
is actually the standard result in Chern-Simons gauge theory; $\H$
can be constructed as a vector space by taking a suitable space of
holomorphic sections, but that does not lead to a simple
expression for the Hilbert space structure. \zolf\ can be slightly
simplified using the fact that $K^{1/2}\cong \L_*^{-h}$, where $h$
is the dual Coxeter number of $G$. (This fact can be proved using
the index theorem for a family of Dirac operators.)  The standard
algebro-geometric description of the physical Hilbert space of
Chern-Simons theory at level $k$ is $\H=H^0(M,\L_*^k)$.  So $k$ as
conventionally defined in gauge theory or two-dimensional current
algebra is related to $\k$ in the $A$-model by
\eqn\snur{k=\k-h,~~\k=k+h.} Many formulas in Chern-Simons gauge
theory are most simply written not in terms of the underlying
coupling $k$ but in terms of $k+h$, which, as we now see, is the
natural $A$-model parameter $\k$.

\def\EUBB{\cmmib B}
The holonomy functions $W_R(S)$ generate, classically, a
commutative algebra.  In the $A$-model, this commutative algebra
is deformed as usual to a noncommutative algebra $\CA$, the space
of $(\Bcc,\Bcc)$ strings. This algebra will act on $\H$. In
Chern-Simons gauge theory, what this means is simply that Wilson
loops on $C$, upon quantization, become operators that act on the
quantum Hilbert space.  Some aspects of this were described in
\witten; for more from the point of view of deformation
quantization, and additional references, see \anderson.

The greatest novelty of the present approach to this much-studied
subject is probably that it is clear that the very same algebra
$\CA$ acts on the space of $(\Bcc,\B)$ strings for any other choice
of $A$-brane $\B$.  Quite a few interesting choices can be
contemplated. We conclude by mentioning some illustrative examples.

We have already considered one case, in which $\B$ is a rank 1
$A$-brane supported on $M$, and the space $\H$ of $(\Bcc,\B)$
strings can be interpreted as a quantization of $M$.

Alternatively, since the $A$-model we are considering here is the
same one that is considered in the gauge theory approach to
geometric Langlands, we can consider $A$-branes that are important
in that context. These are rank 1 $A$-branes supported on a fiber of
the Hitchin fibration. The Hitchin fibration is a map $\pi:Y\to
\EUBB$, where $\EUBB$ is an affine space of half the dimension of
$Y$.  The map is holomorphic not in the natural complex structure of
$Y$ with which we began the discussion but in another complex
structure \foot{This complex structure is called $I$ in \hitchin.
There is a difficult-to-avoid clash in notation with section 2,
where the analogous complex structure was  called $J$. The notation
there was motivated by compatibility with the gauge theory approach
to geometric Langlands.}  that is found by solving Hitchin's
equations. A generic fiber $\cmmib F$ of the Hitchin fibration is a
torus that is a complex abelian variety from the point of view of
this other complex structure. More relevant for our present
purposes, $\cmmib F$ is a Lagrangian submanifold from the point of
view of $\omega_Y={\rm Im}\,\Omega$, so it can be the support of a
rank 1 $A$-brane $\B^*$. Moreover, ${\rm Re}\,\Omega$ is
nondegenerate when restricted to $\cmmib F$, so the space $\H^*$ of
$(\Bcc,\B^*)$ strings can be regarded as a quantization of $\cmmib
F$.

Since $\cmmib F$ is a torus, one would naively expect quantization
of $\cmmib F$ to be related to abelian current algebra, not
nonabelian current algebra. However, there should be a close
relation between $\H$ and $\H^*$, because a certain very singular
special fiber of the Hitchin fibration (the fiber at the
``origin'') has $M$, taken with multiplicity greater than 1, as
one of its components.  This fact should lead to an embedding of
$\H$ (or possibly of the direct sum of several copies of $\H$) in
$\H^*$, something that is very likely related to various results
in conformal field theory in which current algebra of the
nonabelian group $G$ is expressed in terms of an abelian current
algebra.

For a quite different kind of example, let $G_\R$ be an arbitrary
real form of the complex Lie group $G_\C$.  For any such real form,
there is an antiholomorphic involution $\phi:G_\C\to G_\C$ that
leaves $G_\R$ fixed.  Mapping the holonomies $U_i,V_j$ to their
transforms by $\phi$ gives an antiholomorphic involution of $Y$ that
we call $\tau_\phi$. The fixed point set of $\tau_\phi$ has a
component that is the moduli space $M_\phi$ of $G_\R$-valued flat
connections on $C$.  If $G_\R$ is compact, then $M_\phi$ simply
coincides with the space $M$ whose quantization we have already
discussed.

\def\tt{\tilde}
In general, $M_\phi$ is the phase space of three-dimensional
Chern-Simons theory with gauge group $G_\R$, compactified on a
two-manifold $C$. It is a Lagrangian submanifold of $Y$ with respect
to $\omega_Y={\rm Im}\,\Omega$ and supports a rank 1 $A$-brane
$\tt\B$. The space $\tt\H$ of $(\Bcc,\tt\B)$ strings can be
interpreted as the space of physical states in quantization on $C$
of Chern-Simons theory with gauge group $G_\R$.  One interesting
consequence of the present point of view is that the same algebra
$\CA$ of quantized holonomies that acts in the compact case also
acts on the space of physical states for any noncompact real form.
This is almost clear perturbatively (except for a subtlety at the
one-loop level \BW, which we explain shortly), but is less obvious
nonperturbatively.

To describe $\tt\H$ explicitly, we pick again a point $t\in\T$,
giving a hyper-Kahler polarization of $(Y,M_\phi)$. In other
words, after picking $t$, we solve Hitchin's equations to get a
hyper-Kahler metric on $Y$, which determines a complex structure
on $M_\phi$. Then in this hyper-Kahler polarization, $\tt\H_t$ is
explicitly $H^0(M_\phi,\L_*^{\k}\otimes K^{1/2})$, where now $K$
is the canonical bundle of $M_\phi$. However, it is not true that
$K^{1/2}\cong \L_*^{-h}$. Rather, $K^{1/2}\cong \L_*^{-h_\phi}$,
where $h_\phi$ is defined as follows. Decompose the Lie algebra of
$G_\R$ as ${\frak g}={\bf k}\oplus {\frak p}$, where $\bf k$ is
the Lie algebra of a maximal compact subgroup of $G_\R$ and $\frak
p$ is its orthocomplement. After expressing $h$ in terms of the
trace in $\frak g$ of the square of a suitable element of $\frak
g$, write $h=h_++h_-$, where $h_+$ and $h_-$ come from traces in
$\bf k$ and $\frak p$, respectively.  Then (as one can again prove
using the families index theorem) $h_\phi=h_+-h_-$. We thus have
\eqn\nolgo{\tt\H_t=H^0(M_\phi,\L_*^{\k-h_\phi}).}

We would like to compare this result to Chern-Simons gauge theory,
but it is difficult to do so because Chern-Simons gauge theory of
a non-compact gauge group is not well-understood.  However, it is
a reasonable conjecture that the space of physical states is
$\tt\H_t=H^0(M_\phi,\L_*^{k_\phi})$, where $k_\phi$ is the
Chern-Simons coupling. If so, the relation between $k_\phi$ and
the $A$-model parameter $\k$ is
\eqn\zolgo{k_\phi=\k-h_\phi,~~\k=k_\phi+h_\phi.}
   Equivalently, for the same
algebra $\CA$ to act in Chern-Simons theory with compact gauge
group and coupling $k$ as in Chern-Simons gauge theory with gauge
group $G_\R$ and coupling $k_\phi$, the relation between $k$ and
$k_\phi$ must be \eqn\olgo{k+h=k_\phi+h_\phi.} The couplings are
here defined so that the spaces of physical states are
$H^0(M,\L_*^k)$ and $H^0(M_\phi,\L_*^{k_\phi})$, respectively.
These formulas are consistent with a computation \BW\ of the
one-loop quantum correction in Chern-Simons theory with a
noncompact gauge group.

Either the topological invariance of Chern-Simons gauge theory or
the fact that the $A$-model is independent of a choice of
hyper-Kahler polarization implies that there should be a natural
projectively flat connection governing the dependence of $\tt\H_t$
on $t$.  Except in the compact case, it is not known how to
explicitly construct such a connection.  However, for $G_\R$ a
split real form, and a particular component of $M$ (the one that
is contractible topologically), the appropriate representation of
the mapping class group has been constructed by another method
based on real polarizations \gonch.

\vskip 30pt

\centerline{\bf Acknowledgments}

We would like to thank D.~Kazhdan, M.~Kontsevich, N. Hitchin,
and  P. Sarnak  for valuable discussions.
Research of SG is supported in part by NSF Grants DMS-0635607 and
PHY-0757647, in part by RFBR grant 07-02-00645, and in part by the
Alfred P. Sloan Foundation. Research of EW is supported in part by
NSF Grant PHY-0503584. Conclusions reported here are those of the
authors and not of funding agencies.

\listrefs
\end

\subsec{Hamiltonian Isotopy Of The $A$-Model }

Now we will discuss a crucial link in the relation between the
$A$-model and quantization.

We begin as above with a rank 1 $A$-brane $\cal B'$ supported on a
submanifold $M\subset Y$, with the property that $\omega_K$
vanishes on $M$ and $\omega_J$ is nondegenerate when restricted to
$M$.  The definition of $\cal B'$ also depends on a flat
Chan-Paton bundle on $M$; as this contains only global topological
information, we can keep it fixed and suppress it in the following
analysis.

Now we want to change $\cal B'$ by applying a Hamiltonian isotopy
(or diffeomorphism) with respect to the symplectic structure
$\omega_K$ of the $A$-model.  For some Hamiltonian function $f$,
such a diffeomorphism is generated by a vector field
\eqn\zelo{\delta x^\mu=\omega_K^{\mu\nu}{\partial f\over\partial
x^\nu}.}   Such a transformation applied to $M$ does not change
the $A$-brane determined by $M$.  We want to discuss what this
means for quantization of $M$.

We define quantization by saying that the quantum Hilbert space of
$M$ is the space of $(\Bcc,\B')$ strings, where $\Bcc$ is another
$A$-brane.  In the $A$-model, the space of $(\B_1,\B_2)$ strings
only depends on $\B_1 $ and $\B_2$ as $A$-branes, so it is
invariant, up to a natural isomorphism, under {\it independent}
Hamiltonian isotopies applied to $\B_1$ and $\B_2$.  So keeping
$\Bcc$ fixed and applying the Hamiltonian isotopy only to $\B'$,
the space of $(\Bcc,\B')$ strings must be unchanged up to natural
isomorphism.

On the other hand, we claim that the space of $(\Bcc,\B')$ strings
is the result of quantization of the symplectic manifold
$(M,\omega_J)$. It must then be that the pair $(M,\omega_J)$ is
invariant, up to natural isomorphism, when we apply to $X$ a
Hamiltonian isotopy with respect to the symplectic structure
$\omega_K$.

Since this is a crucial statement, we pause to amplify on it.
First let us discuss how in general one could perturb a compact
symplectic manifold $(N,\omega)$.  Of course, one could perturb it
by adding a cohomologically non-trivial term to $\omega$; the
result is a symplectic manifold $(N,\omega+\delta\omega)$ that is
not isomorphic to $(N,\omega)$. Alternatively, suppose that we add
to $\omega$ a cohomologically trivial term $\delta\omega=d\alpha$
for some one-form $\alpha$. Then $(N,\omega+d\alpha)$ is
isomorphic, as a symplectic manifold, to $(N,\omega)$.  In fact,
the change in $\omega$ under a symmetry generated by a vector
field $V$ is $\delta\omega=d(\iota_V\omega)$, where $\iota_V$ is
the operation of contraction with $V$.  If we set
$V=\omega^{-1}\alpha$, we get $\iota_V\omega=\alpha$.  So to first
order in $\alpha$, $(N,\omega)$ can be transformed to
$(N,\omega+\delta\omega)$ using the vector field $V$.  This
extends to a more complete statement: any continuous change in
$\omega$, keeping its cohomology class fixed, can be obtained by a
diffeomorphism of $N$.

But this argument does  not give a canonical isomorphism.  The
reason for this is that if there exists $\alpha$ such that
$\delta\omega=d\alpha$, then $\alpha$ is not unique; we could
substitute $\alpha\to\alpha+df$ for any Hamiltonian function $f$.
Then the vector field $V$ used in the last paragraph is replaced
by  $V'= V+\omega^{-1}df$, where $\omega^{-1}df$ is a completely
arbitrary Hamiltonian vector field.  So the isomorphism we defined
between $(N,\omega)$ and $(N,\omega+\delta \omega)$ is only unique
up to a canonical transformation of $N$.

Now, let us go back to our problem with the pair $(M,\omega_J)$. A
Hamiltonian isotopy with respect to $\omega_K$ moves $M$ to a
nearby manifold $M'$.  $\omega_J$ is still nondegenerate when
restricted to $M'$, since nondegeneracy is a generic or  ``open''
condition. So $(M',\omega_J)$ is a symplectic manifold.
Topologically, we can identify $M'$ with $M$; pick a smoothly
varying way to do so. Since $\omega_J$ is closed, its periods are
constant when $M$ is perturbed to $M'$.  Given this, it follows
from the remarks in the last paragraph that $(M',\omega_J)$ is
equivalent to $(M,\omega_J)$ as a symplectic manifold.

For it to make sense to interpret the $A$-model in terms of
quantization, we need a much more precise result than this.  A
noncanonical equivalence would tell us that the Hilbert spaces
$\H$ and $\H'$ are unitarily equivalent, but with no distinguished
unitary map between them.  All that we would really learn from
this is that they have the same dimension. An equivalence between
two quantum systems should be a natural unitary transformation (up
to a complex scalar of modulus one) from states of one to states
of the other, intertwining the observables of the two systems. The
classical analog of a distinguished unitary transformation between
the two quantum systems is a distinguished isomorphism between the
two classical phase spaces. So we need a {\it canonical}
equivalence between $(M,\omega_J)$ and $(M',\omega_J)$, not just
the above weak argument showing that on abstract grounds they are
equivalent.